\documentclass[aps,prd,twocolumn,superscriptaddress,showpacs,longbibliography]{revtex4-1}
\usepackage{amsmath}
\usepackage{amssymb}
\usepackage{natbib}
\usepackage{graphicx}
\usepackage{appendix}
\usepackage{hyperref}
\usepackage{url}
\usepackage{longtable}

\begin{document}


\title{Resampling to accelerate cross-correlation searches for continuous gravitational waves from binary systems}


\author{G.D. Meadors}
\email[]{grant.meadors@ligo.org}
\affiliation{Albert-Einstein-Institut, Max-Planck-Institut f\"ur Gravitations\-physik, D-14476 Potsdam-Golm, Germany}
\affiliation{Albert-Einstein-Institut, Max-Planck-Institut f\"ur Gravi\-ta\-tions\-physik, D-30167 Hannover, Germany }

\author{B. Krishnan}
\affiliation{Albert-Einstein-Institut, Max-Planck-Institut f\"ur Gravitations\-physik, D-14476 Potsdam-Golm, Germany}

\author{M.A. Papa}
\affiliation{Albert-Einstein-Institut, Max-Planck-Institut f\"ur Gravitations\-physik, D-14476 Potsdam-Golm, Germany}
\affiliation{Albert-Einstein-Institut, Max-Planck-Institut f\"ur Gravi\-ta\-tions\-physik, D-30167 Hannover, Germany }
\affiliation{University of Wisconsin-Milwaukee, Milwaukee, Wisconsin 53201, USA}

\author{John T. Whelan}
\affiliation{Rochester Institute of Technology, Rochester, New York 14623, USA}

\author{Yuanhao Zhang}
\affiliation{Rochester Institute of Technology, Rochester, New York 14623, USA}
\affiliation{Albert-Einstein-Institut, Max-Planck-Institut f\"ur Gravitations\-physik, D-14476 Potsdam-Golm, Germany}
\affiliation{Albert-Einstein-Institut, Max-Planck-Institut f\"ur Gravi\-ta\-tions\-physik, D-30167 Hannover, Germany }



\date{\today}

\begin{abstract}
Continuous-wave (CW) gravitational waves (GWs) call for computationally-intensive methods.
Low signal-to-noise ratio signals need templated searches with long coherent integration times and thus fine parameter-space resolution.
Longer integration increases sensitivity.
Low-Mass X-ray Binaries (LMXBs) such as Scorpius X-1 (Sco X-1) may emit accretion-driven CWs at strains reachable by current ground-based observatories.
Binary orbital parameters induce phase modulation.
This paper describes how resampling corrects binary and detector motion, yielding source-frame time series used for cross-correlation.
Compared to the previous, detector-frame, templated cross-correlation method, used for Sco X-1 on data from the first Advanced LIGO observing run (O1), resampling is about $20\times$ faster in the costliest, most-sensitive frequency bands.
Speed-up factors depend on integration time and search set-up.
The speed could be reinvested into longer integration with a forecast sensitivity gain, $20$ to $125$ Hz median, of approximately $51\%$, or from $20$ to $250$ Hz, $11\%$, given the same per-band cost and set-up.
This paper's timing model enables future set-up optimization.
Resampling scales well with longer integration, and at $10\times$ unoptimized cost could reach respectively $2.83\times$ and $2.75\times$ median sensitivities, limited by spin-wandering.
Then an O1 search could yield a marginalized-polarization upper limit reaching torque-balance at 100 Hz.
Frequencies from 40 to 140 Hz might be probed in equal observing time with $2\times$ improved detectors.
\end{abstract}


\maketitle



\section{Introduction\label{introduction}}

New gravitational-wave (GW) source types await sensitive analyses.
Transient signals such as GW150914~\cite{GW150914LIGO} can reach strain amplitudes $h_0$ of approximately $10^{-21}$.
Yet-unseen continuous-wave (CW) signals, from sources such as non-axisymmetric neutron stars (NSs)~\cite{Brady1998}, are constrained to be significantly weaker: for Scorpius X-1 (Sco X-1), the brightest Low Mass X-ray Binary (LMXB), the best $95\%$-confidence marginalized-polarization upper limit reaches $2.3\times10^{-25}$~\cite{ScoX1CrossCorr2017ApJO1}.
Accretion-driven torque-balance could drive GW emission from LMXBs: infalling matter's angular momentum is predicted to be balanced by that radiated gravitationally~\cite{PapaloizouPringle1978,Wagoner1984}.
Sco X-1 attracts attention~\cite{Bildsten1998} as the brightest persistent X-ray source~\cite{Giacconi1962}.
Emission might be expected at a GW frequency $f_0$ equal to $2 \nu$, for an NS spin frequency $\nu$, assuming that the compact object in the system is an NS radiating \textit{via} the $l=m=2$ mass quadrupole moment.
An NS could also emit \textit{via} $r$-mode (Rossby) oscillations~\cite{Shawhan2010,Owen2010}, depending on the equation of state and dissipative mechanisms~\cite{Andersson1998,Friedman1998,Owen1998}.
Its spin frequency is unknown, so an $f_0$ range must be searched.

In this paper, we discuss how to accelerate and increase the sensitivity of a broadband search for Sco X-1.
CW analyses are computationally demanding.
Long coherent integration times $T_\mathrm{coh}$, for low signal-to-noise ratio (SNR) signals, induce a steep metric~\cite{Brady1998} on the parameter space, increasing the matched-filtering template density.
While an optimal statistic~\cite{Jaranowski1998} can maximize out \textit{amplitude parameters}, the \textit{Doppler parameters} need explicit templating.
Sensitivity, which grows from longer integration, conflicts with computational cost, which grows faster.
Semicoherent methods~\cite{HierarchicalBrady2000} tune this balance: an observing run of data is subdivided into coherent segments.
Summing statistics from segments increases total sensitivity, while the metric depends mainly on the coherent-segment length.
Sensitivity benefits from both total observing time $T_\mathrm{obs}$ and $T_\mathrm{coh}$.
Whole observation runs are typically used, with coherent segments as long as resources permit.
Speed frees resources to be invested in coherent integration time.
\textit{Resampling}~\cite{Jaranowski1998,Patel:2009qe} techniques can accelerate the cross-correlation methods (CrossCorr)~\cite{Dhurandhar2008,Chung2011,ScoX1CrossCorr2015PRD} that have to date shown the most sensitive results for Sco X-$1$ in simulation~\cite{ScoX1MDC2015PRD} and Advanced LIGO data~\cite{ScoX1CrossCorr2017ApJO1}.
After over a decade of GW investigations into Sco X-$1$~\cite{AbbottScoX12007,AbadieStoch2011,GoetzTwoSpectResults2014,Sammut2014PRD,Sideband2015,MeadorsS6ScoX1PRD2017,O1Radiometer2017,O1Sideband2017}, the nominal torque-balance level is near.
Discovery may yield new astrophysics.

Detection becomes more likely as the GW strain amplitude $h_0$ sensitivity approaches torque-balance (TB).
LMXB accretion torque could recycle NSs to higher $\nu$~\cite{PapaloizouPringle1978}.
If spin-up torque balances GW spin-down~\cite{Wagoner1984}, the apparent speed limit on millisecond pulsars slightly over 700~Hz~\cite{Chakrabarty2003} may be explained.
Sco X-1 and similar LMXBs could radiate GWs from NS asymmetries.
By Bildsten Equation~4~\cite{Bildsten1998}, with characteristic strain $h_c$ related by $h_c/h_0 = 2.9/4.0$, for an LMXB with flux $\mathcal{F}_\mathrm{X-ray}$,

\begin{equation}
h_c \approx 4\times10^{-27}\left[\frac{300\mathrm{~Hz}}{\nu_s} \times \frac{\mathcal{F}_\mathrm{X-ray}}{10^{-8}\mathrm{~erg~cm}^{-2}\mathrm{~s}^{-1}}\right]^{1/2}.
\label{torque_bal_eq}
\end{equation}

\noindent High X-ray flux ($3.9 \times 10^{-7}$ erg cm$^{-2}$ s$^{-1}$~\cite{Watts2008}), assuming a nominal 1.4 solar mass, 10 km radius NS of unknown spin frequency, implies Sco X-1's $h_0$:

\begin{equation}
h_0 \approx 3.5\times 10^{-26} [600~\mathrm{Hz}]^{1/2} f_0^{-1/2}.
\end{equation}

\noindent Advanced LIGO Observing Run 1 (O1) data was searched~\cite{ScoX1CrossCorr2017ApJO1} with the cross-correlation method~\cite{ScoX1CrossCorr2015PRD}, setting a 95\%-confidence marginalized-polarization upper limit at 175 Hz of $2.3\times 10^{-25}$, or $8.0\times 10^{-26}$ assuming optimal, circular polarization, respectively $3.5\times$ and $1.2\times$ above torque-balance.
This analysis spanned 25 to 2000 Hz, with the detector noise curve and computational cost reducing the depth of the upper limits.

As Advanced LIGO~\cite{ALIGOStandardRef}, Advanced Virgo~\cite{AVirgoStandardRef}, and KAGRA~\cite{KAGRAStandardRef} observatories improve, sensitivity varies linearly with noise amplitude spectral density(ASD), $S_H^{1/2}(f)$, for fixed $T_\mathrm{obs}$. 
\textit{Sensitivity depth} $D^C(f_0)$~\cite{BehnkeGalacticCenter2015,LeaciPrixDirectedFStatPRD} factors away this noise floor, to characterize analyses:

\begin{equation}
D^C(f_0) \equiv S_H^{1/2}(f_0)[h_0^C(f_0)]^{-1}.
\end{equation}

\noindent Depth should be specified at a confidence $C$, such as $D^{95\%}(f_0)$, based on a strain upper limit $h_0^C$.
\textit{Coherent SNR} is proportional to $h_0\sqrt{T_\mathrm{obs} / S_H}$; deeper methods find lower SNR signals.

Methods vary~\cite{Riles2013} for finding CWs from NS in binary systems. 
Isolated CW techniques~\cite{Jaranowski1998,HoughTransformKrishnan2004,LSCPulsarS4,LSCPowerFlux2009,PowerFluxMethod2010,PowerFluxAllSky2012} inform searches at unknown sky location, as well as for known ephemerides~\cite{DupuisWoan2005,AasiPulsarInitialResults2014}, and also for the \textit{directed} case of known sky location but uncertain ephemerides.
Sco X-1 searches are directed (Table~\ref{scox1_table_params}).
Five Doppler parameters arise from the binary orbit.
New techniques address these parameters' computational cost~\cite{Messenger2007CQG,Sammut2014PRD,SidebandMarkovModelSuvorova2016,GoetzTwoSpectMethods2011,MeadorsDirectedMethods2016,Ballmer2006CQG,AbadieStoch2011,2010JPhCS.228a2005V,Dhurandhar2008,ScoX1CrossCorr2015PRD}, with more in development~\cite{LeaciPrixDirectedFStatPRD}.
The cross-correlation method found all simulations in a 2015 Mock Data Challenge (MDC)~\cite{ScoX1MDC2015PRD} and sets O1 upper limits 3 to 4 times more stringent than others~\cite{O1Radiometer2017,O1Sideband2017}.
The \textit{resampled} cross-correlation method could surpass these limits.

Resampling was proposed~\cite{Jaranowski1998} and detailed~\cite{Patel:2009qe} for isolated-star $\mathcal{F}$-statistic calculations.
Strain $h_0(t)$ is interpolated from the detector frame, where Earth and source motion introduce phase modulation, to the source frame.
In the source frame, the statistic simplifies (with normalization factors determined by the detector antenna functions) to frequency bin power.
Although interpolation is costly, subsequent computations can be faster than interpolating across time-varying frequency bins.
We adapt the cross-correlation method for resampling.
Speed-up and sensitivity performance projections are estimated from implemented code tested with simulated data.  
Deeper, resampled cross-correlation methods could bring CW analyses of Sco X-1 and similar LMXBs to the brink of detection.

Section~\ref{crosscorr_method} details the cross-correlation method, Section~\ref{resampling} explains resampling, and Section~\ref{resamp_cost} measures the cost and benefits.
Figures~\ref{projected-ul} and~\ref{projected-sens-depth} show predicted astrophysical reach.
Section~\ref{conclusion} concludes.

\begin{table*}[t]
\begin{tabular}{r r r r r}
Sco X-1 parameter & Ref. & Value & Uncertainty & Units\\
\hline \\
Right ascension ($\alpha$) &~\cite{2mass06} & 16:19:55.067  & $\pm 0.06'' $ & --- \\
Declination ($\delta$) &~\cite{2mass06} & $-15^\circ 38'25.02''$ & $\pm 0.06''$ & ---\\
Distance ($d$) &~\cite{Bradshaw1999} & $2.8$ & $\pm0.3$ & kpc\\
X-ray flux at Earth ($\mathcal{F}_\mathrm{X-ray}$) &~\cite{Watts2008} & $3.9\times10^{-7}$ & --- &  erg cm$^{-2}$ s$^{-1}$\\
Orbital eccentricity ($e$) &~\cite{ScoX1MDC2015PRD,Galloway2014} & $< 0.068$  & $(3 \sigma)$ & ---\\
Orbital period ($P$) &~\cite{Galloway2014} & $68023.70 $ & $\pm 0.04$ & s\\
Orbital projected semi-major axis ($a_p$) &~\cite{WangSteeghsGalloway2016,ScoX1CrossCorr2017ApJO1} & $1.805$ & $\pm 1.445$ & s\\
Compact object time of ascension ($T_\mathrm{asc}$) &~\cite{Galloway2014,ScoX1MDC2015PRD} & $897753994$ & $\pm100$ & s\\
Companion mass ($M_2$) &~\cite{2002ApJ...568..273S} & $0.42$ & --- & $M_\mathrm{sol}$ \\
\end{tabular}
\caption{
Sco X-1 prior measured parameters from electromagnetic observations (compare~\cite{MeadorsDirectedMethods2016,ScoX1CrossCorr2017ApJO1}).
Uncertainties are $1 \sigma$, except for $a_p$, explained below.
The projected semi-major axis is in time units, $a_p = (a \sin i)/c$; it depends on velocity $K_1=[10,90] {\rm~km\,s^{-1}}$ with approximately-uniform uncertainty distribution (previously~\cite{AbbottScoX12007,ScoX1MDC2015PRD,Galloway2014}, increased uncertainty after private communication~\cite{WangSteeghsGalloway2016}).
Time of ascension $T_\mathrm{asc}$ is defined here as the time when the compact object crosses the ascending node, heading away from the observer in the solar system barycenter (SSB).
Because the companion's inferior conjugation $T_0$ is measured~\cite{ScoX1MDC2015PRD}, we calculate $T_\mathrm{asc} = T_0 - P/4$~\cite{Galloway2014}.
\label{scox1_table_params}
}
\end{table*}

\section{Cross-Correlation Method\label{crosscorr_method}}

Detecting a sinusoid should be simple.
Low SNR, amplitude- and phase-modulated sinusoids are hard.
The `CrossCorr' cross-correlation method~\cite{Dhurandhar2008,ScoX1CrossCorr2015PRD} intersects two paths to this problem: the stochastic radiometer~\cite{Allen1999,Ballmer2006CQG} and the multi-detector $\mathcal{F}$-statistic~\cite{Jaranowski1998,CutlerMulti2005,BStatPrix2009}. 
This cross-correlation method computes a statistic, $\rho$, which approaches the others in limiting cases.
We summarize $\rho$ to clarify, and to explain how resampling~\cite{Jaranowski1998,Patel:2009qe}, designed for the $\mathcal{F}$-statistic, is transferable.
The principle remains -- a semicoherent matched filter using a signal model for continuous, modulated GWs, then a frequentist statistic proportional to the power,~$(h_0)^2$.

\subsection{Signal model\label{signal_model}}

Continuous waves from NS in binary systems are defined by a signal model in amplitude and Doppler parameters.
Amplitude parameters $\bar{\mathcal{A}}^i$~\cite{Jaranowski1998}, are factored out: reference phase $\Phi_0$, polarization angle $\psi$, NS inclination angle $\iota$ (with respect to the line of sight), and strain amplitude $h_0$.
Sky location is in right ascension $\alpha$ and declination $\delta$.

The Doppler parameters $\lambda$ for an isolated system include frequency $f_0$ and higher-order Taylor-expanded \textit{spindown} (or \textit{spinup}) terms $f^{(1)}$, $f^{(2)}$, \textit{etc}.
Assuming an NS source spinning at frequency $\nu$, GW $f^{(k)} \equiv (\sigma d^k \nu(\tau)/d\tau^k | \tau = t_\mathrm{ref})$, with emission time in the source frame $\tau$, evaluated at arbitrary reference time $t_\mathrm{ref}$ (conventions follow~\cite{LeaciPrixDirectedFStatPRD}).
For quadrupole emission, $\sigma = 2$.
Assuming torque-balance, LMXB searches have set spindown terms to zero and instead consider \textit{spin-wandering}~\cite{MukherjeeSpinWandering2016}, an unmodeled stochastic drift about $f_0$.
For an isolated system without spindown, the measured frequency in the solar system barycenter (SSB) will be constant.

For a binary system, the $\lambda$ parameters further include $(a_P,P,T_\mathrm{asc},T_\mathrm{p},e)$.
Orbital projected semi-major axis, in time units, is $a_p \equiv (a \sin i) / c$ ($a$ measured in light-seconds).
Orbital period is $P$.
Time of ascension is $T_\mathrm{asc}$, when the compact object crosses the ascending node, heading away from an SSB observer.
Because only the companion's inferior conjunction time $T_0$~\cite{Galloway2014} is known, the compact object $T_\mathrm{asc} = T_0 - P/4$~\cite{ScoX1MDC2015PRD} (stated in SSB GPS seconds).
Time of periapsis passage is $T_\mathrm{p}$.
Orbital eccentricity is $e$.
When $e = 0$, $T_\mathrm{p} = T_\mathrm{asc}$ by convention.
For Sco X-1, $\alpha$ and $\delta$ are precise enough that one point covers uncertainty.

\subsubsection{Strain and amplitude parameters}

Strain amplitude $h(t)$ is measured; GW phase $\Phi(t;\lambda)$ is key to its signal model:

\begin{eqnarray}
h(t) 
&=& 
\left[F_+ (t; \alpha, \delta), F_\times (t; \alpha, \delta) \right]
  \left[ \begin{array}{c} A_+ \cos \Phi(t;\lambda) \\ A_\times \sin \Phi(t;\lambda) \end{array}\right],
\end{eqnarray}

\noindent where $t$ is detector GPS time measured; $F_+$ and $F_\times$ are called \textit{beam-pattern functions}.
The amplitude model factors loosely depend on time and sky location \textit{via} the detector response in the \textit{antenna functions} $a(t; \alpha, \delta)$ and $b(t; \alpha, \delta)$~\cite{Jaranowski1998,Dhurandhar2008}.
Since we discuss known sky location targets, $(\alpha,\delta)$ will be implicit in $a(t)$, $b(t)$:

\begin{eqnarray}
\left[ \begin{array}{c} F_+ (t) \\ F_\times (t) \end{array} \right] &=& \left[ \begin{array}{c} a(t) \cos 2\psi + b(t) \sin 2 \psi \\ b(t) \cos 2 \psi - a(t) \sin 2\psi \end{array} \right],\\
\left[ \begin{array}{c} A_+ \\ A_\times \end{array} \right] &=& h_0 \left[ \begin{array}{c} \frac{1+\cos^2 \iota}{2} \\ \cos \iota \end{array} \right].
\end{eqnarray}

\noindent
Amplitude parameters can be projected into four new coordinates which affect the waveform linearly~\cite{Jaranowski1998,CutlerMulti2005,PrixMultiMetric2007,BStatPrix2009,WhelanNewAmplitude2014CQG}.
These \textit{canonical} coordinates $\mathcal{A}^\mu$ satisfy, for basis functions $h_\mu(t; \lambda)$,

\begin{eqnarray}
h (t;\lambda) &=& \sum_{\mu=1}^{4} \mathcal{A}^\mu h_\mu (t;\lambda).
\label{decomposition-projection-f}
\end{eqnarray}

\noindent Maximization, or marginalization, over $\mathcal{A}^\mu$ leads to approximately Neyman-Pearson optimal statistics (respectively $\mathcal{F}$, $\mathcal{B}$)~\cite{BStatPrix2009}.
The cross-correlation method obtains a similar statistic $\rho$ with different motives~\cite{Dhurandhar2008} (see Section~\ref{crosscorr-stat}).

\subsubsection{Doppler parameters}

The $\Phi(t;\lambda)$ model is defined with source time $\tau$ as a function of $t$, \textit{via} SSB time $t_\mathrm{SSB}$.
In the spindown-free source frame,  $\Phi(\tau) = 2\pi f_0 \tau$, so $\Phi(t;\lambda) = 2\pi f_0 \tau(t_\mathrm{SSB}(t;\alpha, \delta);\lambda)$.
One can find the barycentric time $t_\mathrm{SSB}$ from $(\alpha, \delta)$, \textit{via} the vector $\vec r(t)$ pointing from the SSB to the detector and the unit vector $\vec n$ pointing from the SSB to the source.
The latter vector is defined as $\vec n(\alpha, \delta) = (\cos \alpha \cos \delta, \sin \alpha \cos \delta, \sin \delta)$~\cite{LeaciPrixDirectedFStatPRD}.
With GW unit wavevector $\vec k = -\vec n$ in the far-field approximation,

\begin{equation}
t_\mathrm{SSB}(t; \alpha,\delta) = t + \frac{\vec r(t) \cdot \vec n(\alpha, \delta) }{c}.
\label{t-ssb-equation}
\end{equation}

\noindent
The relativistic $t_\mathrm{SSB}$ is corrected for Shapiro and Einstein delays, in addition to the Earth orbital and rotational Roemer delays encoded by $\vec r$~\cite{LeaciPrixDirectedFStatPRD}.
The binary orbital Roemer delay comes from the projected radial distance $R$ along the line of sight.
Following conventions~\cite{BlandfordBinary1976,LeaciPrixDirectedFStatPRD},

\begin{equation}
\tau(t_\mathrm{SSB};\lambda) = t_\mathrm{SSB} - \frac{d}{c} - \frac{R(t_\mathrm{SSB};\lambda)}{c},
\label{tau-equation}
\end{equation}

\noindent
wherein larger $R$ signifies greater distance from the binary barycenter (BB) along the line of sight, away from the observer.
Source distance will affect $h_0$ and cause an overall time shift $d/c$ equivalent to changing $\Phi_0$, and inertial motion effects an overall constant Doppler shift to $f_0$.
As $d$ would also affect electromagnetic observations and is indistinguishable from other parameters, we now drop $(d/c)$, in effect equating the SSB with the BB.

Kepler's equations involve a constant argument of periapse $\omega$ (the angle from the ascending node to periapsis in the direction of motion, dependent on $T_\mathrm{p}$ and $T_\mathrm{asc}$) and a time-varying eccentric anomaly $E$ (implicit in $\tau$~\cite{LeaciPrixDirectedFStatPRD}).
These equations describe system dynamics:

\begin{eqnarray}
\tau &=& T_\mathrm{p} + \frac{P}{2 \pi} (E - e \sin E),\label{solve-for-E}\\
\frac{R}{c} &=& a_p \left[ \sin  \omega (\cos E - e) + \cos \omega \sin E \sqrt{1-e^2}\right].\label{Kepler-equation}
\end{eqnarray}

\noindent
Sco X-1's orbit is near-circular ($e < 0.068$ at $3\sigma$), so we will focus on $e=0$, though resampling can handle elliptical orbits.
Sco X-1's $a_p$ is four orders of magnitude less than $P$, so we approximate $E(\tau) = E(t)$.
Let $\Omega \equiv 2 \pi/P$.
In this circular case~\cite{ScoX1CrossCorr2015PRD},

\begin{eqnarray}
\frac{R(t;\lambda)}{c} &=& a_p \sin \left(\Omega [t - T_\mathrm{asc}] \right),\\
\phi(t;\lambda) &=& 2 \pi f_0 \left[ t_\mathrm{SSB}(t;\alpha, \delta) - \frac{R(t;\lambda)}{c} \right],\label{time-varying-phase-eq}\\
\Phi (t; \lambda) 
&=& \Phi_0 + \phi(t;\lambda). \label{phase-model-eq}
\end{eqnarray}

Phase modulation induces an effective frequency modulation depth, $\Delta f_\mathrm{obs}$.
This modulation adds to Doppler shift from detector velocity, $\vec v = d \vec r/dt$ (dominated by Earth's orbit $v_\mathrm{Earth}$), when calculating the total physical frequency bandwidth $\Delta f_\mathrm{drift}$ through which the signal can drift:

\begin{eqnarray}
\Delta f_\mathrm{drift} &=& 2\times\left(\frac{\max{(\vec v \cdot \vec n)}}{c} + \Delta f_\mathrm{obs} \right), \label{f-drift-eq}\\
\Delta f_\mathrm{obs} &=& a_p \Omega f_0.\label{delta-f-obs}
\end{eqnarray}

\noindent With $|\vec v| \approx v_\mathrm{Earth}$, $\mathrm{max}(v / c) \approx 10^{-4}$ (lower off-ecliptic).

For an unmodulated signal, $d\Phi/dt = 2\pi f_0$, reducing to a Fourier transform~\cite{Brady1998}.
For modulated signals, $\phi(t)$ must be tracked to maintain coherence.
Given Equation~\ref{phase-model-eq}, the cross-correlation method tracks a CW signal as the signal changes instantaneous frequency.

Mismatch in Doppler parameters can lead to false dismissal.
The phase mismatch metric~\cite{Brady1998} (\cite{ScoX1CrossCorr2015PRD} for the cross-correlation method) sets the parameter-space density required for Doppler parameters.
A mismatch in the phase-model Roemer delay of about a half-cycle of $f_0$ between the beginning and end of each integration time $T_\mathrm{coh}$ will lose the signal.
(A $100$~Hz signal accumulates $\mathcal{O}(10^2)$ cycles over $a_p$ of Sco X-1 and $\mathcal{O}(10^5)$ cycles over 2 AU).
The computational cost stems from the parameter-space density needed for the long $T_\mathrm{coh}$ that low-SNR signals require. 

We define detection statistics for these signals.
This paper will show that resampling is a more efficient way to compute the cross-correlation method's $\rho$ statistic.

\subsection{Cross-correlation statistic\label{crosscorr-stat}}

\begin{figure*}
\includegraphics[width=0.9\paperwidth, trim={0 30 0 0}, clip]{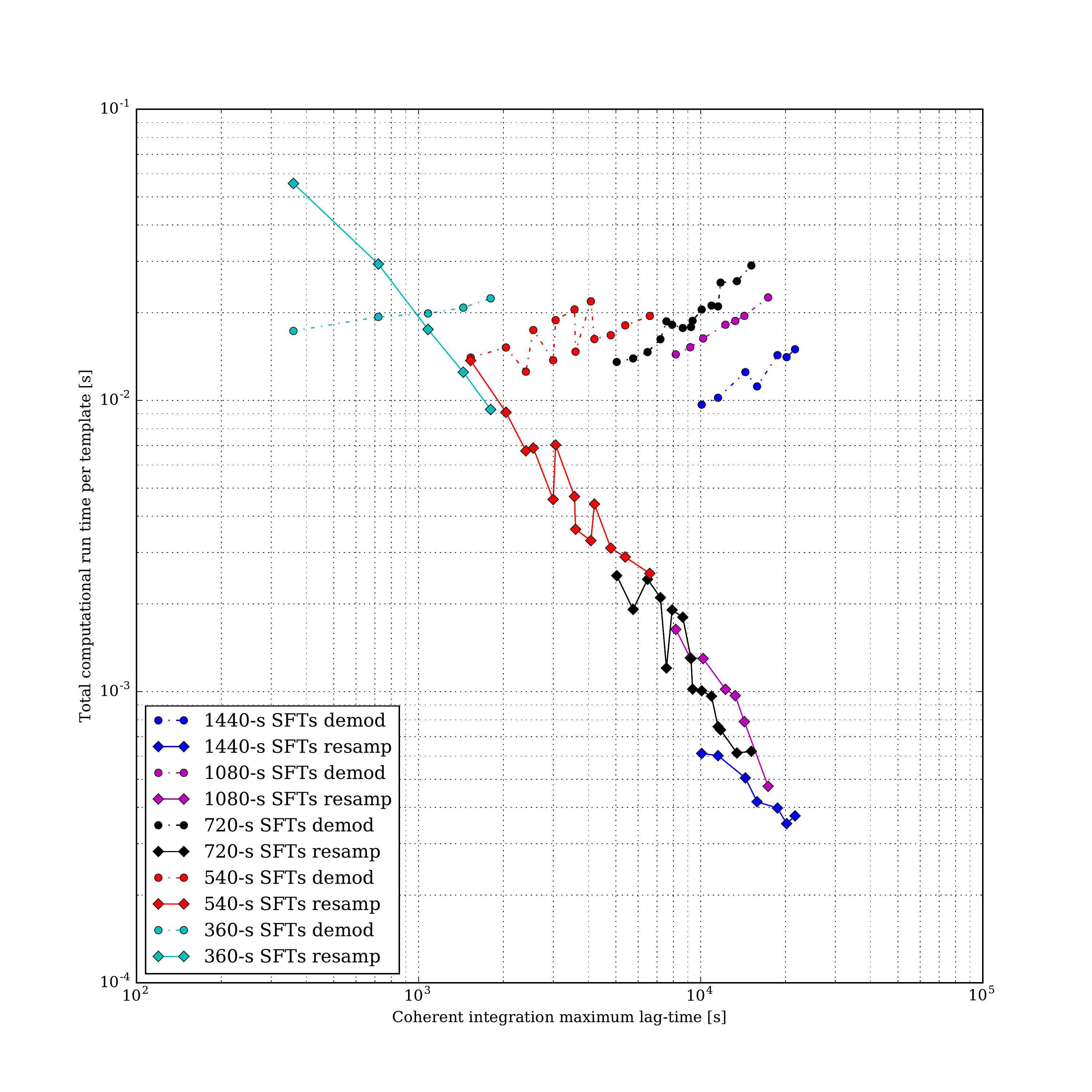}
\caption{
Measured demodulation \textit{(dotted lines, cirles)} and resampling \textit{(solid lines, diamonds)} computational cost-per-template \textit{vs} lag-time $T_\mathrm{max}$.
Lag-time is the maximum difference in start times between the cross-correlation method's data segments.
Results plotted for various Short Fourier Transform (SFT)~\cite{AllenMendellSFT2004} data durations $T_\mathrm{sft}$ \textit{(colors)}.
Average run time for a given lag-time is shown; roughly 190 `demod' and 180 `resamp' measurements done on the Atlas cluster.
Run time shown is total (not per-segment).
The overhead costs for resampling include the barycentering interpolation and FFT (with Doppler wings).
Doppler wings large in proportion to $f_\mathrm{band}$ reduce resampling efficiency; these tests use $0.050$-Hz $f_\mathrm{band}$.
At longer lag-times, these costs are proportionally less, so resampling is faster.
Demodulation's cost increases with $T_\mathrm{max}$ for fixed $T_\mathrm{sft}$, but because the setup uses longer $T_\mathrm{sft}$ with longer $T_\mathrm{max}$ (such as 360-s SFTs at 1000-s lag-time but 1440-s SFTs at 20000-s lag-time), the cost per-template is kept between about~10 and~30 ms.
Per-template costs for demodulation jump when comparing one SFT duration to another.
This effect is because longer SFTs require fewer pairing operations, but length is limited by source and detector acceleration timescales.
At fixed $T_\mathrm{sft}$, demodulation run time increases with lag-time.
Resampling allows pairing to be done in the source frame regardless of acceleration, with a flexible timescale $T_\mathrm{short}$ instead of $T_\mathrm{sft}$.
Fewer pairs and use of the FFT make resampling better suited to long $T_\mathrm{max}$.
From one SFT duration to another, resampling per-template costs are usually consistent.
At long $T_\mathrm{max}$, resampling per-template costs are lower than demodulation.
}
\label{cost_per_template_figure}
\end{figure*}

The goal is to calculate the statistic, $\rho$, as efficiently as possible.
See Figure~\ref{cost_per_template_figure} for a cost per template comparison of the previous `demodulation' and resampled methods.

Let us define $\rho$ as in Whelan \textit{et al}~\cite{ScoX1CrossCorr2015PRD}.
(In Appendix~\ref{relationships-to-other-optimal-statistics} we compare $\rho$, like Dhurandhar \textit{et al}~\cite{Dhurandhar2008}, with the $\mathcal{F}$-statistic~\cite{Jaranowski1998} and radiometer~\cite{Allen1999,Ballmer2006CQG}).
Data start by being parcelled into short Fourier transforms (SFTs)~\cite{AllenMendellSFT2004}, each of duration $T_\mathrm{sft}$.
The total data set spanning $T_\mathrm{obs}$ for $Q$ detectors may contain up to $N_\mathrm{sft} \leq Q T_\mathrm{obs}/T_\mathrm{sft}$ SFTs.

The \textit{cross-correlation} in our method is made between pairs of SFTs: the first component of the pair is indexed by $K$, the second component by $L$.
In the Whelan \textit{et al} construction, $K$ and $L$ span all detectors, meaning they both can range from $0$ up to $N_\mathrm{sft}$.
(Particulars are discussed in Section~\ref{pair-selection-for-resampling}, where $K$ and $L$ are redefined).
The sets of SFTs $\{K\}$ and $\{L\}$ are defined by an allowable lag-time, $T_\mathrm{max}$, the difference between start times of given SFTs $K$ and $L$.
It is common to require $K \neq L$ (to avoid auto-correlation).
SFT pairs $KL$ in the set $\mathcal{P}$ are \textit{cross-correlated.}

A time series $x_K(t) = h_K(t) + n_K(t)$, signal $h$ and noise $n$, has one-sided power spectral density (PSD) $S_K$.
Analyze the Fourier transform $\tilde{x}$ (using Equation 2.1~\cite{ScoX1CrossCorr2015PRD} conventions), with sampling time $\delta t$, SFT mid-time $t_K$:

\begin{eqnarray}
\tilde{x}_{Km} = \sum_{j=0}^{N-1} x_K (t_K - T_\mathrm{sft}/2 + j \delta t) e^{-\mathrm{i} 2\pi j m \delta t / T_\mathrm{sft}} \delta t,
\label{fourier-transform-def}
\end{eqnarray}

\noindent so normalized data $z_{K m}$ in frequency bin $m$ ($k$ in~\cite{ScoX1CrossCorr2015PRD} and our appendices) is,

\begin{equation} 
z_{K m} = \tilde{x}_{K m} \sqrt{\frac{2}{T_\mathrm{sft} S_K}}.
\label{normalized-z-bin}
\end{equation}

SFT bin frequency is $f_m$, but the signal instantaneous frequency is $f_K$; these must not be confused.
The discrepancy, to the nearest bin from the instantaneous frequency, is $\kappa_{K m}$,

\begin{eqnarray}
f_m &=& \frac{m}{T_\mathrm{sft}},\\
\kappa_{K m} &=& m - f_K T_\mathrm{sft}.
\end{eqnarray}

\noindent
Multiple bins in a set $\mathcal{K}_K$ are part of the \textit{Dirichlet kernel}, discussed around Equation 6.5 of~\cite{Allen2002}.
The signal contribution to each bin is found by the normalized \textit{sinc} function, $\mathrm{sinc} \alpha = \frac{\sin{\pi \alpha}}{\pi \alpha}$.
The total data vector $\textbf{z}$ has elements $z_K$, which are the Fourier-transformed data.
Each element is summed from all bins that could contain a signal at a frequency $f$ (implicitly specified by the set $\mathcal{K}_K$ and the $f_K$ model in $\kappa_{K m}$), then indexed by SFT $K$,

\begin{eqnarray}
\Xi_K &\equiv& \sqrt{\sum_{m'\in \mathcal{K}_K} \mathrm{sinc}^2(\kappa_{K m'})},\\
z_K &=& \frac{1}{\Xi_K} \sum_{m\in\mathcal{K}_K} (-1)^m \mathrm{sinc} (\kappa_{K m}) z_{K m} . \label{total-data-eq}
\end{eqnarray}

\noindent The cross-correlation method constructs $\rho$ with a \textit{filter}, Hermitian weighting matrix $\textbf{W}$.
It uses the conjugate transpose $\dag$.
With matrix entries $KL$ that correlate elements $K$ from the SFT vector $\textbf{z}$,

\begin{equation}
\rho = \textbf{z}^\dag \textbf{W} \textbf{z},
\label{abstract-rho-matrix}
\end{equation}

\noindent
or in explicit notation, $\rho = \sum_K \left(\sum_L z_L^* W_{KL}\right) z_K$.
Equation~\ref{abstract-rho-matrix} depends, \textit{via} $\textbf{W}$, on the point in $\lambda$ parameter space (including frequency $f$) of the signal model. 

A near-optimal $\textbf{W}$ is the geometrical factor $\hat \Gamma^\mathrm{ave}_{KL}$ (chosen for $\psi$-independence, Whelan \textit{et al} Equation 2.33~\cite{ScoX1CrossCorr2015PRD}).
Let a hat symbol indicate noise-weighted normalization, \textit{e.g.,} $\hat a^K \equiv \sqrt{2 T_\mathrm{sft}/S_K} a^K$.
Taking $a^K$ ($a(t)$ at the (mid-)time of SFT $K$) and likewise $a^L$, $b^K$, $b^L$, we can find $\hat \Gamma^\mathrm{ave}_{KL}$.
With overall normalization $N$ (\textit{ibid.} Equation 3.6),

\begin{eqnarray}
\hat \Gamma^\mathrm{ave}_{KL} &=& \frac{1}{10}\left(\hat a^K \hat a^L + \hat b^K \hat b^L \right), \label{geometric-filter}\\
N &=& \left(2 \sum_{KL\in\mathcal{P}} \Xi_K^2 \Xi_L^2 (\hat \Gamma_{KL}^\mathrm{ave})^2\right)^{-1/2}.\label{norm-cc}
\end{eqnarray}

\noindent Another weight, $\hat \Gamma^\mathrm{circ}_{KL} = \frac{1}{10}(\hat a^K \hat b^L - \hat b^K \hat a^L)$, is also $\psi$-independent.
Combining $\hat \Gamma^\mathrm{ave}$ and $\hat \Gamma^\mathrm{circ}$ can fix $\iota$.

To obtain $\rho$, Equation 2.36 of~\cite{ScoX1CrossCorr2015PRD} (analogous to Equation 4.11 of~\cite{Dhurandhar2008}), we cross-correlate with the paired data in SFTs $L$ indexed by bin $n$.
We unite Fourier bins using the filter, complex conjugation $*$, and the signal model phase difference between SFTs, $\Delta \Phi_{KL} = \Phi_K - \Phi_L$:

\begin{eqnarray}
\rho = &N& \sum_{KL\in\mathcal{P}} \hat \Gamma^\mathrm{ave}_{KL} \sum_{m\in\mathcal{K}_K} \sum_{n\in\mathcal{K}_L} (-1)^{m-n} \label{textbook-cc-rho}\\ 
 &\times& \mathrm{sinc}(\kappa_{K m}) \mathrm{sinc}(\kappa_{L n}) \nonumber\\
 &\times& (e^{\mathrm{i}\Delta\Phi_{KL}} z^*_{K m} z_{L n} + e^{-\mathrm{i}\Delta\Phi_{KL}} z_{K m} z^*_{L n} ). \nonumber
\end{eqnarray}

\noindent Implicit in $\Phi_K$ is $f_K$, hence all Doppler parameters: $\rho$ must be calculated for each $\lambda$ template.
Since billions~\cite{ScoX1CrossCorr2017ApJO1} of templates are common, efficiency is paramount.
Thanks to links between $\rho$ and the $\mathcal{F}$-statistic that are explored in Appendix~\ref{relationships-to-other-optimal-statistics}, resampling speeds the search.

\section{Resampling\label{resampling}}

Many signals can be resampled into the source frame.
(Compact binary coalescences were contemplated first~\cite{SchutzCBCResamp2017}).
This paper focuses on CWs~\cite{Jaranowski1998} and adheres in notation to code documentation~\cite{LALAppsRepo,PrixTimingModel2017}.
Resampling abstractly moves phase demodulation from \textbf{W} onto \textbf{z}.

Delay causes phase modulation: Equation~\ref{time-varying-phase-eq}  is Roemer-delayed by Earth and source binary motion.
We want to sample $\phi(t;\lambda) = 2\pi f_0 \tau$, but
 in equally-spaced $\tau$ (source frame) instead of equally-spaced $t$ (detector frame).
Although they consider spindown rather than binary parameters, $\tau \sim t_b$ in~\cite{Patel:2009qe}; calculating Equation~\ref{t-ssb-equation} and Equation~\ref{tau-equation} (with numerical solutions to Equations~\ref{solve-for-E} and~\ref{Kepler-equation}), $\tau(t;\lambda)$ can be found.

Because $x(t)$ is discrete, sampling $x(\tau)$ requires interpolation. 
The sinc function interpolates between time-domain samples, paralleling frequency-domain use~\cite{Allen2002}.
As it is computationally-prudent to analyze small frequency bands $f_\mathrm{band}$ independently, data are heterodyned, by selecting the band of interest from a Fourier transform, then inverse Fourier transforming into a downsampled, complex time series, then interpolating.
Since this procedure differs from~\cite{Patel:2009qe}, we describe it.

A time series $x(t)$ sampled at $\delta t$ has a Nyquist frequency of $f_N = 1/(2\delta t)$.
Each SFT $K$ contains its own set of time indices $j$ ranging from $0$ to $N-1$, so $j$ implicitly refers to $K$.
With respect to an arbitrary reference time, $t = t_K - T_\mathrm{sft}/2 + j \delta t$.
Given a set of $M = T_\mathrm{obs}/T_\mathrm{sft}$ SFTs indexed by $K$, each with frequency bins $k$, spaced by $\delta f = 1/T_\mathrm{sft}$, with $N = T_\mathrm{sft}/(\delta t)$ samples, $x(t)$ can be reconstructed by the inverse FFT:

\begin{eqnarray}
x_K(t_K - T_\mathrm{sft}/2 + j\delta t) &=& \sum_{k=0}^{N-1} z_{Kk} e^{i 2 \pi j k \delta t / T_\mathrm{sft}} \delta f.
\label{inverse-fft-eq}
\end{eqnarray}

\noindent Time series segments and frequency bands can be selected by indices.

Equation~\ref{inverse-fft-eq} can be simplified by using the index $q_K \equiv (t_K - T_\mathrm{sft}/2)/(\delta t) + j$ in its argument.
The $q_K$ is the index with respect to the start of $T_\mathrm{obs}$.
A new sampling interval $\delta t'$ and corresponding index $q_K'$ for some time $t = q_K' \delta t'$ can define a downsampled time series.
This time series (heterodyne frequency $f_h$) is $x' (q' \delta t')$ and is produced as in Appendix~\ref{downsampling-and-heterodyning}.

\subsection{Resampling theory\label{resampling-theory}}

\subsubsection{Interpolation}

When data is unaliased and approximately stationary during each SFT, $x'(q'\delta t')$ is a complete representation.
Sinc-interpolation allows us to interpolate $x'(\tau)$. 
The Shannon formula as implemented~\cite{LALAppsRepo} states that for integer $D$ Dirichlet elements, integer index $j$ and $j^* \equiv \mathrm{round}(t/(\delta t))$, $j0 \equiv j^* - D$, and a window $w_j$ (here, Hamming with length $2D + 1$),

\begin{eqnarray}
\delta_j &\equiv& \frac{t-t_j}{\delta t},\\
x(t) &\approx& 
\frac{\sin{\left( \pi \delta_{j0} \right)}}{\pi} \sum_{j=j^* - \mathrm{D}}^{j^* + \mathrm{D}} (-1)^{(j-j0)} \frac{x_j w_j}{\delta_j},\label{shannon-interp-formula}
\end{eqnarray}

\noindent converging when $D\rightarrow \infty$.
A typical $D=8$, minimizing costs of sinc-interpolation (linear in $D$) plus subsequent FFTs (linear in Appendix~\ref{downsampling-and-heterodyning}'s $\Delta f_\mathrm{load}$).

\subsubsection{Resampling into the source frame}

Let our source-frame time series be indexed by $r$ with constant spacing $\delta t'$: $\tau = r \delta t'$.
We use the function $t(\tau;\lambda)$, the functional inverse of the function $\tau(t;\lambda)$ from Equation~\ref{tau-equation}.
Over timescales $T_\mathrm{sft}$ when the signal stays in one frequency bin, $(d^2\tau/dt^2) T_\mathrm{sft}^2 f_0 \ll 1$, Taylor approximation is valid around $t_0$: 

\begin{eqnarray}
\tau(t;\lambda) &\approx& \tau(t_0;\lambda) + \left[\frac{d\tau(t)}{dt}|_{t=t_0}\right] t,\\
t(\tau;\lambda) &\approx& t(\tau_0;\lambda) + \left[\frac{d\tau(t)}{dt}|_{t=t_0}\right]^{-1} \tau,
\end{eqnarray}

\noindent making computations practical.

Translating from detector time to source time introduces a timeshift $\Delta t^* = r \delta t' - t(r \delta t';\lambda)$ to $x(t)$.
The discrete source-frame time series is $x'(r\delta t') = \exp{(-i2\pi f_h \Delta t^*)} x'(t(r \delta t';\lambda))$:

\begin{eqnarray}
\delta_{q'} &\equiv& \frac{t(r\delta t';\lambda)-q' \delta t'}{\delta t'},\label{delta-q-eq}\\
{r}^*  &\equiv& \mathrm{round}\left(\frac{t(r\delta t';\lambda)}{\delta t'}\right),\label{q-prime-eq}\\
x'(r \delta t') &\approx& 
\frac{\sin{\left( \pi \delta_{q0'} \right)}}{\pi} e^{-2\pi f_h [r \delta t' - t(r \delta t';\lambda)]} \nonumber\\ 
  &&\times \sum_{q'={r}^* - \mathrm{D}}^{{r}^* + \mathrm{D}} (-1)^{(q'-q0')} \frac{x_{q'}' w_{q'}}{\delta_{q'}}.
\label{resampled-time-series-q-eq}
\end{eqnarray} 

\noindent Then $x_r' \equiv x'(r\delta t')$ is the complex, heterodyned, downsampled, discrete time series that equally samples the source frame $x(\tau)$.

Roemer delays vanish in $x(\tau)$, if the Doppler parameters $\lambda$ are accurate.
Mismatch results in residual phase modulation.
No finite lattice of $\lambda$ can perfectly sample the space.
The required resolution is determined by the phase mismatch metric $g$~\cite{Brady1998}.

Derivatives $d/d\lambda$  for $\lambda \in (f, a_p, T_\mathrm{asc}, P)$ have been calculated for the cross-correlation method's metric~\cite{ScoX1CrossCorr2015PRD}.
In the similar $\mathcal{F}$-statistic metric~\cite{LeaciPrixDirectedFStatPRD}, $e$ and $T_\mathrm{p}$ are discussed.
The metric is computed in software over the phase mismatch $\Delta \Phi_{\alpha,i}$ for the cross-correlation method's pairs indexed by $\alpha = KL$ and Doppler parameters indexed by $i$,

\begin{eqnarray}
g_{ij} &\approx& 
\frac{1}{2} \left\langle \left(\frac{\partial (\Phi_K - \Phi_L)}{\partial \lambda^i} \right)\left( \frac{\partial(\Phi_K - \Phi_L)}{\partial \lambda^j}  \right) \right\rangle_\alpha,
\label{phase-derivs-metric}
\end{eqnarray}

\noindent extending to any Doppler parameters in the phase model.
(Metric vielbeins represent the natural units of distance for a parameter-space vector).

Given the metric, a lattice is calculated with the spacing in each dimension set by the allowed mismatch, $\lambda_\mu$.
Mismatch is a tunable choice about the statistic's acceptable \textit{fractional loss:} $\mu_\lambda = (\textrm{max}(\rho) - \rho)/\textrm{max}(\rho)$.
A simple cubic lattice grid for a diagonal metric has spacings $\delta \lambda^i$,

\begin{equation}
\delta \lambda_i = \sqrt{\frac{\mu_{\lambda_i} }{g_{ii}} }.\label{metric-spacing-eq}
\end{equation}

\noindent
However, the metric is only a local approximation~\cite{Brady1998}.
The total derivative $d\tau$ contains many approximate degeneracies, for example when frequency mismatch $df$ equals modulation depth mismatch $d\Delta f_\mathrm{obs}$ arising from offset $a_p$ or $T_\mathrm{asc}$ (see Appendix~\ref{stat-interp-rho}).
Mismatch studies are thus needed to verify the loss and chose spacings.
Each lattice point in orbital parameter space must have its own resampled $x(\tau)$.

Resampling interpolation yields $x(\tau)$ so that a putative signal is concentrated at a single frequency $f_0$.
Next, taking the Fourier transform~\cite{Jaranowski1998,Patel:2009qe} generates $\rho$.

\subsection{Resampled cross-correlation method implementation\label{implementation}}

Source-frame $x(\tau)$ speeds Section~\ref{crosscorr-stat}'s $\rho$ calculation.
Supplied with $T_\mathrm{obs}$, we divide data into semicoherent segments with a shortest timescale of $T_\mathrm{short}$, replacing $T_\mathrm{sft}$.
This $T_\mathrm{short}$ is the duration we will take from each $K$ side of a pair of the cross-correlation method.
The $L$ side of the pair will be composed of all other $T_\mathrm{short}$ intervals with start times up to a \textit{maximum lag-time} $T_\mathrm{max}$ before or after.
A total, cross-detector, coherent integration duration of $T_\mathrm{coh}$ includes a central $T_\mathrm{short}$ plus $T_\mathrm{max}$ on both sides:

\begin{eqnarray}
|\tau_K - \tau_L| &\leq& T_\mathrm{max},\label{lag-time-constraint}\\
T_\mathrm{coh} &=& 2 T_\mathrm{max}+T_\mathrm{short}.
\end{eqnarray}

\noindent
For same-detector correlations, only $T_\mathrm{max}$ on one~side is typically used, to avoid auto-correlation and double-counting, but we preserve the above definition of $T_\mathrm{coh}$ to keep frequency resolution the same.

Times $\tau_K$ and $\tau_L$ evenly divide the resampled time series if calculated in the source frame, though this means that slightly unequal amounts of detector data go into $T_\mathrm{short}$.
As $|\tau_K - t_K| \leq |\vec r \cdot \vec n /c + a_p|$, the difference between an interval start time in detector and source frame is bounded by the Roemer delay.
We neglect these effects because relative inequality from one interval to the next is proportional to $d\tau/dt \leq 2\times 10^{-4}$.
Based on prior experience~\cite{LeaciPrixDirectedFStatPRD}, these delays do not affect the metric estimation.
For the cross-correlation method's metric~\cite{ScoX1CrossCorr2015PRD}, the goal is to constrain the (pair-averaged) phase mismatch over $T_\mathrm{max}$ from offset $\delta \lambda_i$, which grows linearly proportionally to $T_\mathrm{max}$, so it is negligible from the phase mismatch over $(1+d\tau/dt)T_\mathrm{max}$.

Nor are average noise weightings affected much by resampling, because the normalization $N$ is a sum over $T_\mathrm{obs}$.
However, weightings are based on average noise per SFT.
To find the weights, we average noise for each $T_\mathrm{short}$ interval by interpolating with Equation~\ref{shannon-interp-formula}.
Terms $T_\mathrm{sft}$ in Section~\ref{crosscorr-stat} become replaced with $T_\mathrm{short}$.

The current implementation zero-pad gaps instead of skipping them.
These gaps contribute nothing to $\rho$, and, because the noise-weighted antenna functions $\hat a(t)$ and $\hat b (t)$ give gaps zero weight, they contribute nothing to $N$.

Compared to the non-resampling cross-correlation method~\cite{ScoX1CrossCorr2015PRD}, resampling yields two benefits.
First, $T_\mathrm{short}$ supercedes $T_\mathrm{sft}$, the latter being limited by modulation moving the signal out of bin.
Increasing $T_\mathrm{short}$ reduces the number of (new) pairs, $N_\mathrm{pairs} \approx N_\mathrm{det}^2 T_\mathrm{max} T_\mathrm{obs} T_\mathrm{short}^{-2}$ (replacing $T_\mathrm{sft}$ from Equation $3.27$ in~\cite{ScoX1CrossCorr2015PRD}).
Because sensitivity is, to zeroth order, proportional to $h_0^\mathrm{sens} \propto (N_\mathrm{det}^{2} T_\mathrm{obs} T_\mathrm{max})$, independent of $T_\mathrm{sft}$, but cost is linearly proportional to the number of templates times the number of pairs, it is optimal to minimize the number of pairs by maximizing $T_\mathrm{short}$.

Second, the number of frequency templates required is automatically supplied by an FFT.
An FFT over a time period $T_\mathrm{coh}$ is spaced at $1/T_\mathrm{coh} \propto T_\mathrm{max}^{-1}$.
This scaling comes from the metric element $g_{ff}$ for that lag-time, indepedent of $T_\mathrm{sft}$ and resampling.
Rather than needing to repeat this fine frequency grid for every SFT, resampling allows all the data to be gathered into one FFT with time $T_\mathrm{FFT} \geq T_\mathrm{coh}$.
(For finer sampling, the FFT can be zero-padded; for coarser, its output can be decimated).

\subsubsection{Pair selection for resampled statistic\label{pair-selection-for-resampling}}

Resampled $x'(\tau)$ as given by Equation~\ref{resampled-time-series-q-eq} must be divided into pairs to calculate the $\rho$ statistic.

The set of pairs $\mathcal{P}$ must be constructed.
Taking $Q$ detectors, they are indexed by $X$ for the first component of the cross-correlation method's pair and $Y$ for the second component.
These $X,Y$ indices range from $0$ to $Q-1$.
An option exists to exclude same-detector correlations, as in the stochastic radiometer.
Here, we allow same-detector correlations, except same-detector same-time correlations, that is, the auto-correlation.
We reuse indices $K$ and $L$ from previous sections but restrict the range of each to a single detector.
Indexing $T_\mathrm{short}$ intervals is marked by $K$ for detector $X$ and $L$ for detector $Y$.
Indices $K,L$ range from $0$ to $M = T_\mathrm{obs}/T_\mathrm{short}$, regardless of any gaps.
Approximating Equation~\ref{lag-time-constraint} in the detector frame, such that

\begin{eqnarray}
\{L|K\} &:& |K T_\mathrm{short} - L T_\mathrm{short}| \leq T_\mathrm{max},\\
    \implies && \{K - T_\mathrm{max}/T_\mathrm{short},\ldots,K + T_\mathrm{max}/T_\mathrm{short}\},\nonumber
\end{eqnarray}

\noindent which is straightforward when $T_\mathrm{max}$ is an integer multiple $R$ of $T_\mathrm{short}$.
(Performance is best in practice when $T_\mathrm{short} = T_\mathrm{max}$).
This set $\{L|K\}$ contains $M_L = 2R+1$ elements for cross-detector correlations and $R$ for same-detector correlations, to avoid double-counting.

Detector-time pairing is predictable, and it is acceptable because $K$ to $K+1$ differences are of order $d\tau/dt \approx 2\times 10^{-4}$.
Yet the resampled time series do not start at precisely the same source frame time.
Let $(\tau_X = \tau_K|K=0)$, $(\tau_Y = \tau_L|L=0)$.
They can differ by $(\vec r_X - \vec r_Y)\cdot \vec n/c$, which for ground-based detectors is of order $10$ ms at most.
This $\Delta \tau_{XY}$ is still a full cycle at $100$ Hz, and it must be accounted for, by timeshifting the resampled time series to the same starting epoch.
The correct factor is the physical frequency $f_0$.
Differences $\tau_K - \tau_L$ require a further timeshift at the heterodyned frequency, $f_0 - f_\mathrm{het}$, as they are internal to the resampled time series.

\subsubsection{Fourier transform size and phase shift}

The above definitions separate $\mathcal{P}$ pairs into intervals and detectors.
To construct $\rho$ from resampled data in these pairs using an FFT, we require the number of FFT samples, $N_\mathrm{FFT}$.
The metric resolution answers this question.
Then we will substitute the pair definition into $\rho$ to make an explicit quadruple sum.

The metric spacing $\delta \lambda_f$ will be achieved by an FFT of duration $1/(\delta \lambda_f)$.
For typical mismatch  $\mu_f$, Equation~\ref{metric-spacing-eq} and Equation 4.31a of~\cite{ScoX1CrossCorr2015PRD} yield $\delta \lambda_f < 1/T_\mathrm{coh}$.
Specifically, Equation 4.33~\cite{ScoX1CrossCorr2015PRD} becomes $(3/4)T_\mathrm{max}^2$ on the right-hand side in the case $T_\mathrm{max} = T_\mathrm{short}$,

\begin{equation}
\delta \lambda_f = \sqrt{\frac{6\mu_f}{\pi}}\frac{1}{T_\mathrm{coh}},
\end{equation}

\noindent which provides $\delta \lambda_f T_\mathrm{coh} < 1$ up to $\mu_f \approx 0.52$.
This is a high value of mismatch.
Any FFT with that mismatch or finer frequency spacing is automatically long enough to include all the data in $T_\mathrm{coh}$.
(For coarser mismatch, decimation by a ratio $\nu_D \equiv \mathrm{ceil}(\delta \lambda_f \times T_\mathrm{coh})$ after the FFT can select the frequencies of interest).
Conversely, if $\mu_f = 0.1$, $\delta \lambda_f$ implies FFT duration $\geq 2.3 T_\mathrm{coh}$.
Dirichlet frequency interpolation is replaced by zero-padding to the metric resolution.

The recovered fraction of spectral power is known from Equation 3.18~\cite{ScoX1CrossCorr2015PRD}, $\langle \Xi^2\rangle$ (to which $\rho$ is linearly proportional): for Dirichlet interpolation with $m$ bins,

\begin{equation}
\langle \Xi^2 \rangle = 2 \int_0^{m/2} \mathrm{sinc}^2 \kappa d \kappa.
\end{equation}

\noindent In that paper, $m = 2$ was recommended to capture $0.903$ of $\rho$.
The function $\delta_{T_\mathrm{sft}}(f-f')$ is a continuous function determined by data; only $\kappa_{Kk}$ are discrete.
Zero-padding from $T_\mathrm{coh}$ to $T_\mathrm{FFT}$ (and taking only $1$ bin of the FFT, so $m=1$) gives,

\begin{equation}
\langle \Xi^2 \rangle_\mathrm{resamp} = 2 \int_0^{1/2} \mathrm{sinc} \left(\frac{T_\mathrm{coh}}{T_\mathrm{FFT}} \kappa\right) \mathrm{sinc} \left(\frac{T_\mathrm{short}}{T_\mathrm{FFT}} \kappa\right) d \kappa.
\end{equation}

\noindent
Hence ($T_\mathrm{coh} = 3 T_\mathrm{short}$), $\langle \Xi^2\rangle \approx 0.861$ when $T_\mathrm{FFT} = T_\mathrm{coh}$, the minimal possible by design.
More typically, $\langle \Xi^2\rangle \approx 0.963$ when $T_\mathrm{FFT} = 2 T_\mathrm{coh}$, or $\approx 0.983$ when $T_\mathrm{FFT} = 3 T_\mathrm{coh}$.
This is sufficient to forego the cost of Dirichlet interpolation in the frequency domain.
Any desired improvement in $\langle \Xi^2 \rangle_\mathrm{resamp}$ can be obtained by requesting smaller $\mu_f$.

Practical considerations mean that FFT speed is most predictable when $N_\mathrm{FFT}$ is an integer power of $2$.
Our resampled time series has a fixed $\delta t'$, so the only way to increase the number of samples is to zero-pad further in time.
Starting with the required $N_\mathrm{FFT0}$,

\begin{eqnarray}
N_\mathrm{FFT0} 
 &=& \frac{\Delta f_\mathrm{load}}{\delta \lambda_f} \mathrm{ceil}(\delta \lambda_f \times T_\mathrm{coh}),\\
N_\mathrm{FFT} &=& 2^{\mathrm{ceil}\left( \log_2 N_\mathrm{FFT0} \right)}.
\label{n-fft-final}
\end{eqnarray}

\noindent
In time, $T_\mathrm{FFT} = \delta t' N_\mathrm{FFT}$.
The extension from $N_\mathrm{FFT0}$ to $N_\mathrm{FFT}$ causes over-sampling in the frequency domain.
From this we decimate by rounding down to the nearest bin with a real-valued ratio $\nu_R$,

\begin{eqnarray}
\nu_R = (\delta \lambda_f)(\delta t') N_\mathrm{FFT},
\end{eqnarray}

\noindent To maximize recovered power, we use bin-centered frequency.
Bin offset ($f_h \approx \bar f_h$ in Equation~\ref{heterodyne-shift-approx}) is solved with a shift $f_r^*$ to the nearest FFT bin:

\begin{eqnarray}
\mathrm{remainder}(a,b) &\equiv& a - \frac{a}{|a|}\mathrm{floor}\left(\frac{|a|}{|b|}\right),\\
f_r^* &=& \mathrm{remainder}\left( -f_\mathrm{band}/2 , T_\mathrm{FFT}^{-1} \right).
\end{eqnarray}

\noindent
We will multiply $a_r$ and $b_r$ each by $\exp{(-i2\pi f_r^* \tau)}$.
Preceeding time shifts using $f_h$ remain valid.
The smallest FFT frequency $f_\mathrm{FFT}$, at $k_0$, causes the smallest output frequency $f_\mathrm{min} = f_h - f_\mathrm{band}/2$ to be found at bin $k_0$:

\begin{eqnarray}
f_\mathrm{FFT} &=& f_h + f_r^* - \frac{1}{2} f_\mathrm{band} T_\mathrm{FFT},\\ 
k_0 &=& \mathrm{lround}\left(\frac{f_h - f_\mathrm{band}/2 - f_\mathrm{FFT} }{ T_\mathrm{FFT}^{-1} }\right),
\end{eqnarray}

\noindent 
where the $\mathrm{lround}$ function rounds to the integer less than its argument.

\subsubsection{Antenna function weighting~\label{antenna-function-weighting}}

Equation~\ref{resampled-time-series-q-eq} expresses a discrete time series $x_r' = x'(r\delta t')$ of $x'$ in $\tau = r \delta t'$.
Time series accounting for amplitude modulation by antenna functions $a$ and $b$ are returned.
The noise-weighted $\sqrt{2/(T_\mathrm{sft}S_h)} x_r'$  are multiplied by the noise-normalized $\hat a$,$\hat b$ antenna function time series.
This hat symbol equals multiplication by $\sqrt{2T_\mathrm{sft}/S_h}$.

In the following paragraphs, let us outline some practical considerations, because the implementation may otherwise be ambiguous.
When computed, elements $a_r$, $b_r$ should be normed to order unity for numerical stability~\cite{PrixFStatModel2011}.
A noise-normalization $\mathcal{S}_a = \sqrt{2T_\mathrm{sft}/S_h}\langle a\rangle$ should be used.
Multiplication by $\mathcal{S}_a$ can subsequently restore $\hat a$, $\hat b$ for the resampled time series.
An error-prone point is that we must use a factor of $\mathcal{S}_a \sqrt{T_\mathrm{short}/T_\mathrm{sft}}$ in the implementation of $\hat \Gamma^\mathrm{ave}_{KL}$ (because we have written new indices $K$,$L$ in terms of $T_\mathrm{short}$).
As the statistic contains factors of $a^2$, $b^2$, we track the ratio $T_\mathrm{short}/T_\mathrm{sft}$.
This choice preserves the correct normalization factor and ensures numerical stability at each stage.
(A clean-slate code implementation could be more straightforward).
The physically-meaningful values $a$, $b$ remain unchanged throughout.

The product of the normalizations equals $2/S_h$ (for $S_h$ approximated by the nearest SFT).
The kernel timestep is $\delta t'$ (in implementation, after the FFT).
Multiplication by the requisite frequency shift $f_r^*$ obtains $a_r$,$b_r$:

\begin{eqnarray}
a_r &\equiv& \frac{2 \delta t' }{S_h}  a(r \delta t') x'(r \delta t') e^{-\mathrm{i}2\pi f_r^* \tau}, \label{norm-a-eq}\\
b_r &\equiv& \frac{2 \delta t' }{S_h}  b(r \delta t') x'(r \delta t') e^{-\mathrm{i}2\pi f_r^* \tau}.
\end{eqnarray}

\noindent
Here $a(t)$ and $b(t)$ are real-valued amplitude modulations with period of one sidereal day.
They are not heterodyned.
(Their period is also greater than the maximum Roemer delay, giving $a(r \delta t') \approx a(t(r \delta t'))$, $b(r \delta t') \approx b(t(r \delta t'))$).
Antenna functions are effectively constant over $T_\mathrm{sft}$.
Multiplying $a(t)$ and $b(t)$ by $x(t)$ prepares the optimal filter for the $\mathcal{F}$-statistic~\cite{Jaranowski1998} as well as for our inner product.

\subsubsection{Phase shifts after Fourier transform}

Subsequent shifts are labeled $\Phi_\mathrm{out}$ and $\Phi_\mathrm{in}$.
$\Phi_{\mathrm{out}_K}$ is the shift at the physical frequency of bin $k$, $f = f_h - f_\mathrm{band}/2 + k(\delta \lambda_f)$, due to start time (epoch) for that detector's ($X$ for $K$, $Y$ for $L$) resampled time series.
$\Phi_{\mathrm{in}_K}$ is the shift at the heterodyned frequency of bin $k$, $[k_0 + \mathrm{floor}(\nu_R k)] T_\mathrm{FFT}^{-1}$, from different start times $K T_\mathrm{sft}$ within the resampled time series.

\begin{eqnarray} 
\Phi_{\mathrm{out}_K}(k) &=& 2\pi [f_h - f_\mathrm{band}/2 + k (\delta \lambda_f)] (\tau_X), \label{phi-out-eq}\\
\Phi_{\mathrm{in}_K}(k) &=& 2\pi [k_0 + \mathrm{floor}(\nu_R k)] T_\mathrm{FFT}^{-1} K T_\mathrm{short}.
\end{eqnarray}

Considerations include the antenna-weighted, phase-model corrected frequency-domain data $\hat a^K \zeta_K$.
The $\zeta_K$ term equals the product $\Xi_K z_K \exp{(-\mathrm{i} \Phi_K)}$.
This term is explored in Appendix~\ref{relationships-to-other-optimal-statistics}, Equation~\ref{from-here-resamp}; in contrast with Equation~\ref{fourier-transform-def}, $k$ refers to a frequency bin, instead of $m$.
In the Appendix, the index $m \equiv j - T_\mathrm{sft}/(2 \delta t)$ is introduced for a time-domain sample.

Let us now reconstruct $\hat a^K \zeta_K$ with resampling: $\delta t$ becomes $\delta t'$, $T_\mathrm{sft}$ becomes $T_\mathrm{short}$.
The index $m$ increases with $r$.
Precisely, $t_m = t_K + m\delta t$ is the overall time, analogous to $r \delta t'$.
So $m$ becomes $r - t_K/(\delta t')$.

We look at the time-domain limits of the data $\hat a^K \zeta_K$ as defined in Equation~\ref{from-here-resamp}.
The lower limit, $m = -T_\mathrm{sft}/(2\delta t')$, becomes $r = (t_K - T_\mathrm{short}/2)/(\delta t')$.
The upper limit becomes $r = (t_K + T_\mathrm{short}/2)/(\delta t')$.
We call them (non-integer) $r_{B,K}$ and $r_{U,K}$.
The discrete sum must round them.
No samples are missed when $r_{U,K} = r_{B,K+1}$.
As long as the ideal sample number, $N'_\mathrm{ideal} = T_\mathrm{short}/(\delta t')$, is $N'_\mathrm{ideal} \gg 1$, rounding is tolerable.
We will soon replace $r_{U,K}$ with the zero-padded $r_{B,K} + N_\mathrm{FFT}$.

The term $a^K x_K(t_m)$ contains $t_m = r \delta t'$.
Allowing $r_k \equiv \mathrm{round}(t_K/(\delta t'))$,
then $r = m + t_K/(\delta t')$ is simply $r = m + r_K$.
So $a^K x_K(t_M)$ translates to $a_{m + r_K}$. 
This is the $a_r$ weighted in Equation~\ref{norm-a-eq}.

Substitute the above into $\hat a^K \zeta_K$:

%
\begin{eqnarray}
\hat a^K \zeta_K  
   &=& \sum_{r=r_{B,K}}^{r=r_{U,K}} a_r e^{-\mathrm{i} 2\pi f_K r \delta t'},\label{fft-inklings}
\end{eqnarray}

\noindent observing that $\Phi_K = f_K t_K$ (source-frame frequency is constant).
Equation~\ref{fft-inklings} foretells a Fourier transform from $r$ into $k$.
Heterodyning has $f_K = f_0 - f_h$, discretely indexed as $k = f_K T_\mathrm{FFT}$.
Raise $r_{U,K}$ to $r_{B,K} + N_\mathrm{FFT}$.
Zero-padding (mathematically, using the Heaviside step function $H$) keeps the sum constant:

\begin{eqnarray}
\hat a^K \zeta_K 
   &=& \sum_{r=r_{B,K}}^{r=r_{B,K} + N_\mathrm{FFT}} \frac{H(r_{U,K} - r) a_r}{ \exp{(\mathrm{i} 2\pi k r / N_\mathrm{FFT})} },
\end{eqnarray}

\noindent
In practice, an FFT starts at $s = r - r_{B,K}$.
Re-indexing,

\begin{eqnarray}
\hat a^K \zeta_K &=& \sum_{s=k}^{N_\mathrm{FFT}} \frac{H(r_{U,K} - r_{B,K} - s) a_{s +r_{B,K}}}{ \exp{(\mathrm{i} 2\pi k [s + r_{B,K}] / N_\mathrm{FFT})} },
\end{eqnarray}

\noindent wherein $r_{B,K}$ factors in the kernel:

\begin{eqnarray}
\frac{2\pi k r_{B,K} }{ N_\mathrm{FFT} } 
   &=& 2\pi \frac{k_0 + (k-k_0)}{T_\mathrm{FFT}} K T_\mathrm{short},
\end{eqnarray}

\noindent expressing $k$ in terms of distance from a minimum $k_0$.
If we pick bins $\bar k$ above $k_0$ at a continuous decimation rate $\nu_R$,

\begin{eqnarray}
\frac{2\pi \bar k r_{B,K} }{N _\mathrm{FFT}}
   &=& \Phi_{\mathrm{in}_K}(\bar k).
\end{eqnarray}

Finally, 
as in Equation~\ref{phi-out-eq}, $\Phi_{\mathrm{out}_K}$ corrects an overall time shift in the resampling epoch, $\tau_X$.
When the heterodyning starts at epoch $\tau_X$ after reference time $\tau_0$,

\begin{eqnarray}
\frac{ 2\pi k [s + r_{B,K}]}{N_\mathrm{FFT}} &=& 2\pi \Phi(\tau-\tau_0),\\
\Phi(\tau-\tau_0) &=& f_0 \tau  H([\tau_0+\tau_X] - \tau) \\
  && + (f_0 - f_H) \tau H(\tau - [\tau_0+\tau_X]),\nonumber
\end{eqnarray}

\noindent expanding the first Heaviside function into a Boxcar $B$,

\begin{eqnarray}
H(\tau_0 + \tau_X - \tau) &=& H(\tau_0 - \tau) + B(\tau_0, \tau_0 + \tau_X),
\end{eqnarray}

\noindent so during $\tau_0$ to $\tau_X$, $f_0 \tau_X$ cycles are accumulated, justifying $\Phi_{\mathrm{out}_K}$.
(The second Heaviside function is null, because $r \delta t'$ starts at $\tau_X$).

\subsubsection{Frequencies returned from Fourier transform}

With a discrete Fourier transform (DFT) from time samples $s$ into frequencies $k$ being the operation $\mathcal{F}^s_k$,

\begin{eqnarray}
\mathcal{F}^s_k y_s &=& \sum_{s=k}^{N_\mathrm{FFT}} e^{-(\mathrm{i} 2\pi k s / N_\mathrm{FFT})} y_s,\\
(a^K \zeta_K)_k &=& e^{-\mathrm{i}2\pi[\Phi_{\mathrm{in}_K} + \Phi_{\mathrm{out}_K}](k) },\\
  && \times \mathcal{F}^s_k \left(H(r_{U,K} - r_{B,K} - s) a_{s +r_{B,K}} \right),\nonumber
\end{eqnarray}

\noindent DFTs return a frequency \textit{vector} indexed by $k$, rather than a scalar as in the previous demodulation search~\cite{ScoX1CrossCorr2015PRD}. 
We select the set of frequencies $\bar k$.
Mathematically, we represent this as a selection function $\delta_{\bar k}^k$ that reduces to the Dirichlet delta function when $\nu_R = 1$, so

\begin{equation}
(a^K \zeta_K)_{\bar k } = \delta_{\bar k}^k (a^K \zeta_K)_k.
\end{equation}

In the case of $\{L|K\}$, where $M_L$ multiple, often consecutive, $T_\mathrm{short}$ intervals are present at a single detector, we can do one Fourier transform, because $T_\mathrm{FFT}\geq T_\mathrm{coh}$.
Call the sum $S^{(L|K)}_{\bar k}$:

\begin{eqnarray}
S^{(L|K)}_{\bar k} 
  &=& \delta_{\bar k}^{k} e^{-\mathrm{i}2\pi[\Phi_{\mathrm{in}_{L_0}} + \Phi_{\mathrm{out}_{L_0}}](k) }\\
  && \times \mathcal{F}^s_k \left(H(r_{U,(L_0 + M_L)} - r_{B,L_0} - s) a_{s +r_{B,L_0}} \right),\nonumber
\end{eqnarray}

\noindent so the whole sum can be done in a single FFT.
This is because $\Phi_{\mathrm{out}_L}$ depends only on $L$ for its detector time epoch ($\tau_Y$), and $\Phi_{\mathrm{out}_L}$ is proportional to $L T_\mathrm{short}$, which is absorbed into the Fourier transform kernel.

If $S_{(L|K)}$ skips some term, \textit{e.g.,} auto-correlation where $L=K$ in the same detector, this is handled, both in theory (by subtracting a Boxcar function) and practice (by skipping that time and putting the next $T_\mathrm{short}$ at the following place in the zero-padded time series).
Segment $L$ depends implicitly on its detector $Y$.

\subsubsection{Statistic in resampled data and physical meaning}

Taking a look at $\rho$ from Equation~\ref{textbook-cc-rho},
we see it can be phrased in terms of $\hat a^K \zeta_{K}$ in the Appendix~\ref{relationships-to-other-optimal-statistics}, Equation~\ref{staged-to-turn-to-fafb}.
We will break it into explicit pairs over $Q$ detectors (first cross-correlation pair component indexed by $X$, second by $Y$), each of which has $M \equiv T_\mathrm{obs}/T_\mathrm{short}$ (zero-padded gapless) data segments, as in Section~\ref{pair-selection-for-resampling}.
A data segment index $K$ for the first component of a pair is matched by $M_L$ terms of the second component, starting from $L_0$.
Eliding $b$ terms,

\begin{eqnarray}
\rho &=& \frac{N}{5} \Re \sum_{X=0}^Q \sum_{K=0}^M \hat a^K \zeta_{K}^* \sum_{Y=0}^Q \sum_{L=L_0}^{M_L} \hat a^L \zeta_{L} + \ldots,
\end{eqnarray}

\noindent so we insert the Fourier transforms to get the vector $\rho_{\bar k}$,

\begin{eqnarray}
\rho_{\bar k} &=& \frac{N}{5} \Re \sum_{X=0}^Q \sum_{K=0}^M (\hat a^K \zeta_{K})_{\bar k}^* \sum_{Y=0}^Q S^{(L|K)}_{\bar k}+ \ldots,
\end{eqnarray}

A commonly-used projection of the in-phase data onto $a(t)$, the sub-interval integral $F_{a_I}$ (as in Appendix~\ref{relationships-to-other-optimal-statistics}, Equation~\ref{the-grand-id}) motivates us to name the key quantities:

\begin{eqnarray}
\bar F_{a_{K,\bar k}} &=& (\hat a^K \zeta_{K})_{\bar k},\\
\bar F_{a_{L,\bar k}} &=& S^{(L|K)}_{\bar k}.
\end{eqnarray}

\noindent
Note: unlike terms $F_a$ and $F_b$ in Appendix~\ref{relationships-to-other-optimal-statistics}, the above quantities include noise normalization.
Overall normalizations $\hat A^2_\mathcal{P}$, $\hat B^2_\mathcal{P}$, $\hat C^2_\mathcal{P}$ are the sums over pairs of $(\hat a^K \hat a^L)^2$, $(\hat b^K \hat b^L)^2$, and $(\hat a^K \hat b^L \hat a^L \hat b^K)$, respectively.
Then the resampled $\rho$ statistic parallels Equation~\ref{well-formatted-rho}:

\begin{eqnarray}
\rho_{\bar k} &=& \frac{\sqrt{2}
}{\sqrt{ \hat A_\mathcal{P}^2 + 2 \hat C^2_\mathcal{P}+ \hat B_\mathcal{P}^2 }} \times \ldots
\label{final-formatted-rho}
 \\
&&
\Re \sum_{X=0}^Q \sum_{K=0}^M \sum_{Y=0}^Q \left[ \bar F_{a_{K,\bar k}} \bar F_{a_{L,\bar k}} +  \bar F_{b_{K,\bar k}} \bar F_{b_{L,\bar k}} \right] \nonumber
.
\end{eqnarray}

Equation~\ref{final-formatted-rho} holds in any reference frame.
Dependence on detector and source motion has been absorbed by resampling, so the remaining formula is manifestly invariant.
This formula for $\rho$ is a semicoherent matched filter assuming a sinusoidal waveform.
In the (non-physical) case of zero Roemer delay, frequency is constant and no resampling is needed, so Equation~\ref{final-formatted-rho} exactly equals Equation~\ref{well-formatted-rho}.
Resampling is elegantly interpreted as a shift to a frame with zero Roemer delay, where the frequency is effectively constant (up to the accuracy of the resampling parameters and numerical precision).
It is unsurprising but reassuring that the result is independent of the original frame.

\subsubsection{Summary of resampling implementation}

Resampling has been ported from the $\mathcal{F}$-statistic computation into the cross-correlation method.
The implementation differs in that $\mathcal{F}$ needs no concept of $T_\mathrm{short}$: its coherence time is the FFT time, and because each segment is resampled individually without being subdivided into pairs, $\Phi_\mathrm{in} = 0$.
The $\mathcal{F}$-statistic includes auto-correlation, and there is no extra overlap.
(Some inefficiency in recalculating the same overlapping pairs in the cross-correlation method could be reduced by caching partial terms $F_a$, $F_b$ from Appendix~\ref{relationships-to-other-optimal-statistics}).

For the $\mathcal{F}$-statistic, resampling has already accelerated long $T_\mathrm{coh}$ searches.
Resampling should also speed-up the cross-correlation method.
Considering Equation~\ref{abstract-rho-matrix}, we have offloaded phase-correction from the $\textbf{W}$ matrix onto the $\textbf{z}$ vector, turning a quadratic operation into a linear one.
That the remaining matrix can be evaluated by an FFT is a further improvement.
In the next section, we measure computational speed and sensitivity.

\section{Computational cost and sensitivity\label{resamp_cost}}

\begin{figure*}
\includegraphics[width=0.9\paperwidth, trim={0 30 0 0}, clip]{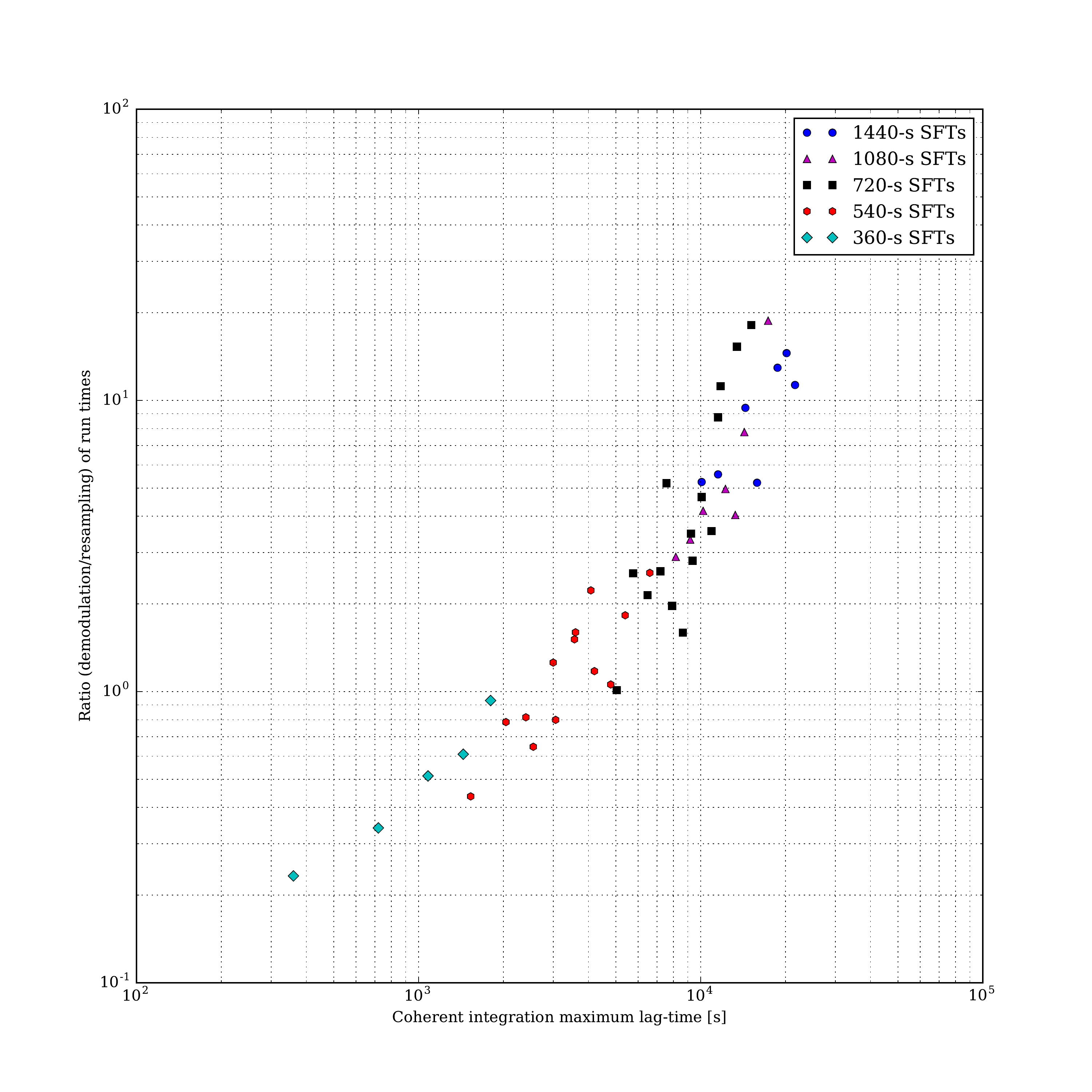}
\caption{
Ratio of demodulation run time to resampling run time \textit{vs} maximum lag-time $T_\mathrm{max}$ for representative $T_\mathrm{sft}$.
At longer $T_\mathrm{max}$, the relative advantage of resampling grows.
It is roughly 20 for the longest typical O1 set-ups, resulting from both the ability to pair independently of $T_\mathrm{sft}$ (using $T_\mathrm{short}$) and from the FFT yielding $\rho$ as a function of $f_0$ for a given set of binary orbital parameters.
Even longer $T_\mathrm{max}$ are attainable because of the asymptotic metric of orbital parameter space.
As in Figure~\ref{cost_per_template_figure}, Doppler wings large in proportion to $f_\mathrm{band}$ reduce resampling efficiency; these tests use $0.050$-Hz $f_\mathrm{band}$.
}
\label{ratio_of_runtimes_figure}
\end{figure*}

The computational speed and cost of resampling for the cross-correlation method is to be measured.
A first comparison (Figure~\ref{cost_per_template_figure}) takes overall run times of the demodulation and resampling techniques for a given number of templates.
The relative speed-up, in Figure~\ref{ratio_of_runtimes_figure}, governs how much can be re-invested in search depth.
Deeper understanding helps predict the computational cost in time required for conceivable use cases: the \textit{timing model}.

\subsection{Demodulation timing model}

First, define the timing model for the demodulation search.
Let each dimension have spacing $\delta \lambda$ determined by the metric, requiring $N_\lambda$ templates be searched in each dimension to cover a range $\Delta \lambda = N_\lambda \delta \lambda$.
Using a simple cubic lattice,

\begin{equation}
N_\mathrm{template} = \prod_{\lambda} \frac{\Delta \lambda}{\delta \lambda}.
\end{equation}

Take a test case for a single point in orbital parameter space.
With $n_\mathrm{bin} = 2$ Dirichlet interpolation bins, $N_\mathrm{template} = 55488$, $T_\mathrm{max} = 22800$~s, $T_\mathrm{obs} = 3.0\times 10^{6}$~s, $T_\mathrm{sft} = 1440$~s ($N_\mathrm{det} =2$) this case is measured to take a total time of $T_\mathrm{demod} = 159.80$~s. 
(Single-threaded without SIMD instructions on an Intel Core i7-4980HQ at 2.8 GHz).
Normalizing these parameters into a single timing constant, $\tau_\mathrm{demod}$ for two detectors, and with scalings taken from~\cite{ScoX1CrossCorr2015PRD}, we have a timing function,

\begin{eqnarray}
N_\mathrm{pairs} &\approx& \frac{N_\mathrm{det}(N_\mathrm{det}+1)}{2} T_\mathrm{max} T_\mathrm{obs} T_\mathrm{sft}^{-2},\\
T_\mathrm{demod} &=& \tau_\mathrm{demod} n_\mathrm{bin} N_\mathrm{template} N_\mathrm{pairs}.
\label{demod-timing-eq}
\end{eqnarray}

\noindent Using this measurement, $\tau_\mathrm{demod}$ is about $1.5 \times 10^{-8}$ s.

Note that this single measurement is based on gapless data.
In the presence of gaps, the demodulation search can easily skip to the next SFT (at present, resampling cannot skip gaps).

Template count $N_\mathrm{template}$ depends on every parameter's $\delta \lambda$.
Because $\delta \lambda$ depends on $T_\mathrm{max}$ for all four Doppler parameters, the computational cost increases with longer lag-time.
Each $\delta \lambda$ is proportional to the inverse square root of the corresponding metric element $g_{\lambda \lambda}$ as in Equation~\ref{metric-spacing-eq}.
Whelan \textit{et al}~\cite{ScoX1CrossCorr2015PRD} note that the metric element $g_{ff}$ increases with $T_\mathrm{max}^2$, while the orbital parameter elements also increase as $T_\mathrm{max}^2$ for $T_\mathrm{max} \ll P_\mathrm{orb}$ before asymptoting as $T_\mathrm{max}$ approaches the $P_\mathrm{orb}$.
Uncertainty in $P_\mathrm{orb}$ is low enough that a single template is enough to cover it for short $T_\mathrm{max}$, but not generally at high $T_\mathrm{max}$.
So the computational cost scaling for demodulation has $1+2+1 = 4$ powers of $T_\mathrm{max}$: it is $T_\mathrm{demod} \propto T_\mathrm{max}^4$ for short lag-time.
After the orbital period resolves and also asymptotes for long lag-time, the scaling is $\propto T_\mathrm{max}^2$, with a larger coefficient.
Contrast this case with resampling.

\subsection{Resampling timing model}

\begin{table*}[t]
\begin{tabular}{r | l l | l l}
Coefficient & Low $N_\mathrm{FFT}$ value [s] & Low $N_\mathrm{FFT}$ uncertainty [s] & High $N_\mathrm{FFT}$ value [s] & High $N_\mathrm{FFT} $ uncertainty [s] \\
\hline \\
$\tau_{0,\mathrm{CCbin}}$ & $1.01 \times 10^{-7}$ & $\pm1.10 \times 10^{-8}$ & $1.34 \times 10^{-7}$ & $\pm2.25 \times 10^{-8} $ \\
$\tau_{0,\mathrm{bary}}$ & $1.62 \times 10^{-7}$ & $\pm7.48 \times 10^{-10}$ & $1.62 \times 10^{-7}$ & $\pm3.20 \times 10^{-9}$ \\
$\tau_{0,\mathrm{FFT}}$ & $5.27 \times 10^{-10}$ & $\pm5.19 \times 10^{-11}$ & $1.40 \times 10^{-9}$ & $\pm6.00 \times 10^{-11}$
\end{tabular}
\caption{
Table of timing coefficients. 
Higher values indicate greater computational cost. 
Values obtained from fit to overall external total run time (see Figure~\ref{timing-constants-resamp}).
Timing coefficients are divided into low and high $N_\mathrm{FFT}$ values, with the threshold being $N_\mathrm{FFT} = 2^{18}$, above which $\tau_{0,FFT}$ is about $3\times$ higher. 
This difference is expected from the \textit{FFTW} library performance profile and may arise from cache sizes.
Uncertainty reported is $\pm 1 \sigma$ to the fit.
About 120 measurements each done for low and high $N_\mathrm{FFT}$ on the Atlas cluster, using a mix of Intel Xeon E3-1220v3 and E3-1231v3 processors.
Results accord with single-processor \textit{Callgrind} performance profiling, but future precise internal per-function measurements may be valuable.
}
\label{timing-coefficient-table}
\end{table*}

\begin{figure}
\includegraphics[width=1.0\columnwidth, trim={30 30 30 30}, clip]{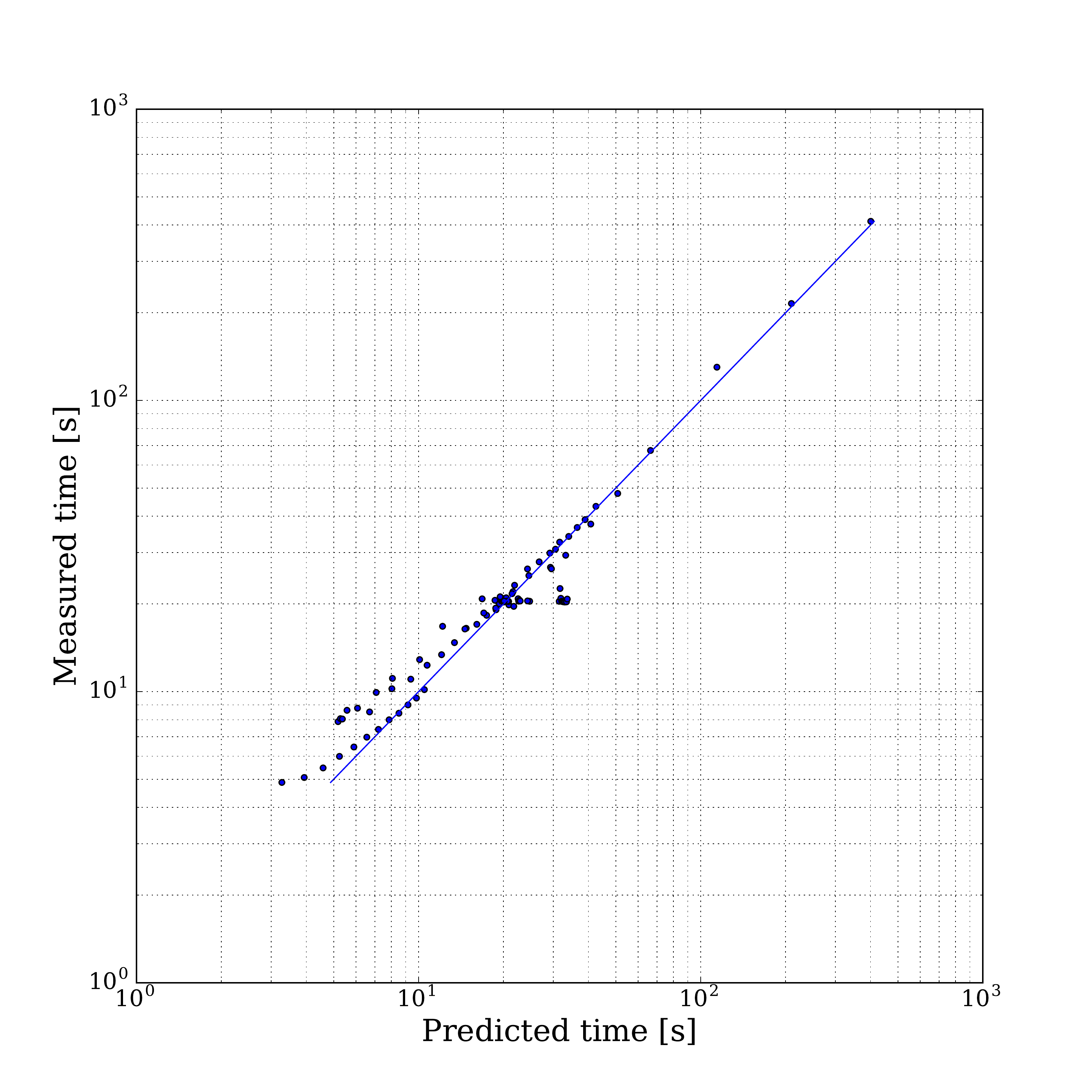}
\caption{
Resampling timing model.
Measured \textit{vs} predicted run time for (about 120) use cases, varying frequency band $f_\mathrm{band}$, observing time $T_\mathrm{obs}$, lag-time $T_\mathrm{max}$, number of observatories $N_\mathrm{det}$, starting frequency $f_\mathrm{min}$, projected semi-major axis $a_p$, allowed frequency mismatch $\mu_f$ in the statistic, and number of Dirichlet kernel terms $D$ for resampling. 
This set captures the \textit{low} $N_\mathrm{FFT}$ case, where the threshold for \textit{high} $N_\mathrm{FFT}$ is $2^{18}$ bins: for more bins, the cost per FFT bin $(\tau_\mathrm{(0,FFT)})$ is approximately $3$ times greater (see Table~\ref{timing-coefficient-table}).
The diagonal line marks an exact prediction of run time of the resampling code. 
A fit is made using the three resampling timing constants, $\tau_\mathrm{(0,CCbin)}$, $\tau_\mathrm{(0,bary)}$, and $\tau_\mathrm{(0,FFT)}$, of the timing model in Equation~\ref{resampTimingModelEq}.
Measurement done on Atlas cluster and may vary due to machines under realistic use conditions.
}
\label{timing-constants-resamp}
\end{figure}

Better \textit{scaling} is sought from the resampling timing function.
Longer lag-times are theoretically easier to achieve with resampling.
It is the measurements of the coefficients that determine whether the overall computational cost is affordable.

The resampling timing function is complicated: it involves three timing constants.
Table~\ref{timing-coefficient-table} lists these constants.
First is the timing constant $\tau_{0,\mathrm{CCbin}}$ for per-template (per-bin) operations, such as multiplying, adding, copying, and phase-shifting results to and from the FFT.
Second is the timing constant $\tau_{0,\mathrm{bary}}$: the cost for barycentering for each point in orbital parameter space.
Third and last is the timing $\tau_{0,\mathrm{FFT}}$: the cost of the FFT operation (using the \textit{FFTW} library) for each template.
This division into three parts is motivated by a pre-existing timing model for the $\mathcal{F}$-statistic~\cite{PrixTimingModel2017}.

The $\tau$ constants are measured using Atlas, the cluster at AEI Hannover, Germany.
A typical cluster node uses an Intel Xeon E3-1220v3 at 3.1 GHz; a smaller set of E3-1231v3 (3.4 GHz) and E5-1650v2 (3.5 GHz) CPUs are also in use.
Approximately 120 configurations, varying frequency band ($f_\mathrm{band}$), observing time ($T_\mathrm{obs}$), lag-time ($T_\mathrm{max}$), number of observatories ($N_\mathrm{det}$), starting frequency ($f_h - f_\mathrm{band}/2$), projected semi-major axis ($a_p$), allowed frequency mismatch ($\mu_f$) in the statistic, and number of Dirichlet kernel terms ($D$), are tested and fit to the three search parameters.
This fit minimizes the discrepancy between predicted and measured time, as shown in Figure~\ref{timing-constants-resamp}.

Time $T_\mathrm{resamp}$ is predicted as follows.
It is most efficient to take $T_\mathrm{max} = T_\mathrm{short}$.
We divide the analysis into bands of $\Delta f_\mathrm{load}$ (Equation~\ref{delta-f-load}).
We next separate $N_\mathrm{template} = N_\mathrm{orb} N_f$, into orbital $N_\mathrm{orb}$ and frequency $N_f$ template counts.
The FFT size is $N_\mathrm{FFT}$ (by Equation~\ref{n-fft-final}).
A `triangular' function accounts for detector pairings,

\begin{equation}
\mathrm{triang}(N) = 1 + \frac{N+1}{2}.
\end{equation}

\noindent
Taking a prefactor of $5$ for the FFT logarithmic term is based on~\cite{PrixTimingModel2017}, from which the basic scheme of our model is motivated.
Absorbing a typical number, $D=8$, into $\tau_{0,\mathrm{bary}}$, is efficient.
The total time is then $T_\mathrm{resamp}$:

\begin{eqnarray}
T_\mathrm{resamp} &=& N_\mathrm{orb} N_\mathrm{det} (T_\mathrm{obs} / T_\mathrm{max}) [\ldots \\
 && \tau_{0,\mathrm{CCbin}} N_f \mathrm{triang}(N_\mathrm{det}) +\ldots  \nonumber \\
 && \tau_{0,\mathrm{bary}} \left(2 \Delta f_\mathrm{load} \times T_\mathrm{max} \times (D/8) \right) + \ldots \nonumber \\
 && \tau_{0,\mathrm{FFT}} N_\mathrm{FFT} \times 5 \mathrm{log}_2(N_\mathrm{FFT}) \times \mathrm{triang}(N_\mathrm{det}) \nonumber ]
\label{resampTimingModelEq}
\end{eqnarray}

Observe that $N_\mathrm{FFT}$ is proportional, albeit through power-of-two steps, to $N_f$, and $N_f$ is proportional to $T_\mathrm{max}$ as before.
At low lag-time, $N_\mathrm{orb}=2$, so the resampling time scales $T_\mathrm{resamp} \propto T_\mathrm{max}^2 \mathrm{log} T_\mathrm{max}$.
At high lag-time, after the number of orbital templates has asymptoted and period dimension resolved, it is, with a larger coefficient,
$T_\mathrm{resamp} \propto \mathrm{log} T_\mathrm{max}$.
The improvement stems from two parts of the new code: the `SFT gain' by reducing the number of pairs saves a factor of $T_\mathrm{max}$, and the `FFT gain' by converting the $\mathbf{W}$ weights matrix into an FFT operator effects $T_\mathrm{max}^2 \rightarrow T_\mathrm{max} \log T_\mathrm{max}$. 

\textit{Caveats:} the \textit{FFTW} functions for FFTs alert us to a $3\times$ increase in $\tau_\mathrm{0,FFT}$ for $N_\mathrm{FFT}$ above about $2^{18}$.
This behavior is observed and is why Table~\ref{timing-coefficient-table} is divided into low and high $N_\mathrm{FFT}$ sections.
Our prediction for $T_\mathrm{resamp}$ applies a factor of $3$ multiplier when $N_\mathrm{FFT}$ is predicted to be in this slow regime.
A key \textit{caveat} is that the precise $N_\mathrm{FFT}$ is difficult to calculate \textit{a priori}.
(The \textit{post hoc} $N_\mathrm{FFT}$ is used to make $\tau$ estimates more accurate).
This difficulty comes from the metric calculation depending on the true phase derivatives instead of a simpler diagonal approximation (as explained in~\cite{ScoX1CrossCorr2015PRD}).
Slight misprediction in metric-derived spacing can be amplified by power-of-2 rounding in $N_\mathrm{FFT}$.
Future improvement in $T_\mathrm{resamp}$ estimation can be expected from reusing the exact code used for metric calculation in the timing predictor.

\subsection{Sensitivity of optimized set-up\label{sensitivity_gain}}

\begin{table*}[]
\begin{tabular}{r r | r r | r r }
$\min f_0$ [Hz] & $\max f_0$ [Hz] & $\max T_\mathrm{max}$ [s] & $\min T_\mathrm{max}$ [s] & $f_\mathrm{band} [Hz]$ & $T_\mathrm{sft}$ [s]\\
\hline \\
25   & 50   & 25920 & 10080 & 0.050 & 1440 \\
50   & 100  & 19380 &  8160 & 0.050 & 1080 \\
100  & 150  & 15120 &  6720 & 0.050 &  720 \\
150  & 200  & 11520 &  5040 & 0.050 &  720 \\
200  & 300  &  6600 &  2400 & 0.050 &  540 \\
300  & 400  &  4080 &  1530 & 0.050 &  540 \\
400  & 600  &  1800 &   360 & 0.050 &  360 \\
600  & 800  &   720 &   360 & 0.050 &  360 \\
800  & 1200 &   300 &   300 & 0.050 &  300 \\
1200 & 2000 &   240 &   240 & 0.050 &  240 \\
\end{tabular}
\caption{
O1 search set-up~\cite{ScoX1CrossCorr2017ApJO1}.
Set-up depends on the GW frequency, $f_0$, of a search band, as well as its position in orbital parameter space.
More likely regions in parameter space are allocated longer $T_\mathrm{max}$ to increase detection probability.
All search bands in O1 are $50$~mHz wide in $f_\mathrm{band}$.
The SFT duration $T_\mathrm{sft}$ varies with frequency to limit spectral leakage from orbital acceleration.
To find the set-up for a given point in parameter space, find the line bounding $f_0$ between $\min f_0$ and $\max f_0$ columns and consult~\cite{ScoX1CrossCorr2017ApJO1} to determine its placement in orbital parameters.
This set-up offers significant potential for re-optimization~\cite{MingSetup2015}.
}
\label{search-setup-table}
\end{table*}

Sensitivity depth $D^C$ for the semi-coherent cross-correlation method search scales $T_\mathrm{max}^{1/4}$~\cite{ScoX1CrossCorr2015PRD}, up to an uncertain time where spin-wandering makes longer integration incoherent.
The demodulation technique gives an effective scaling of $D^C \propto (T_\mathrm{demod})^{1/16}$ for low lag-time $T_\mathrm{max}$, compared to $P_\mathrm{orb}$, or $\propto (T_\mathrm{demod})^{1/8}$ for high lag-time.
Resampling, dropping the logarithmic term, offers $D^C \propto T_\mathrm{resamp}^{1/8}$ for low lag-time or $D^C \approx \mathrm{(constant)}$ for high.

Once the computational cost reaches the orbital parameter metric plateau and asymptotes, additional sensitivity is nearly cost-free with resampling.
Surprisingly, in the frequency dimension, the number of templates continues to increase $\propto T_\mathrm{max}$, but because $T_\mathrm{short} = T_\mathrm{max}$, the number of semicoherent segments decreases linearly as $T_\mathrm{max}$ increases, so there are longer but fewer FFTs to do.
Small cost increases do continue, in the logarithmic FFT term.
Two caveats: the number of period templates still depends on $T_\mathrm{obs}$, and power-law scalings assume a large number of semicoherent segments.
The conceivable case of $T_\mathrm{max} = 10$~days, $T_\mathrm{obs} = 3$~months may be close to the limit where this approximation holds, and excluding the auto-correlation means that the ratios of $T_\mathrm{obs}/T_\mathrm{max} < 5$ (approximately) may exclude some data.
(The latter is partly-solvable by decoupling $T_\mathrm{short}$ from $T_\mathrm{max}$).
Nevertheless, the ease of high $T_\mathrm{max}$ with resampling helps both future searches and follow-ups.

Gains in search sensitivity depend on the measured timing constants.
We iteratively estimate the maximum $T_\mathrm{max}$ possible with the resampling code for the same computing resources made available, in a given band, as to the demodulation O1 search~\cite{ScoX1CrossCorr2017ApJO1}.
For future searches, the distribution across bands can be re-allocated to maximize detection probability.

\begin{figure*}
\includegraphics[width=0.41\paperwidth, trim={0 30 0 0}, clip]{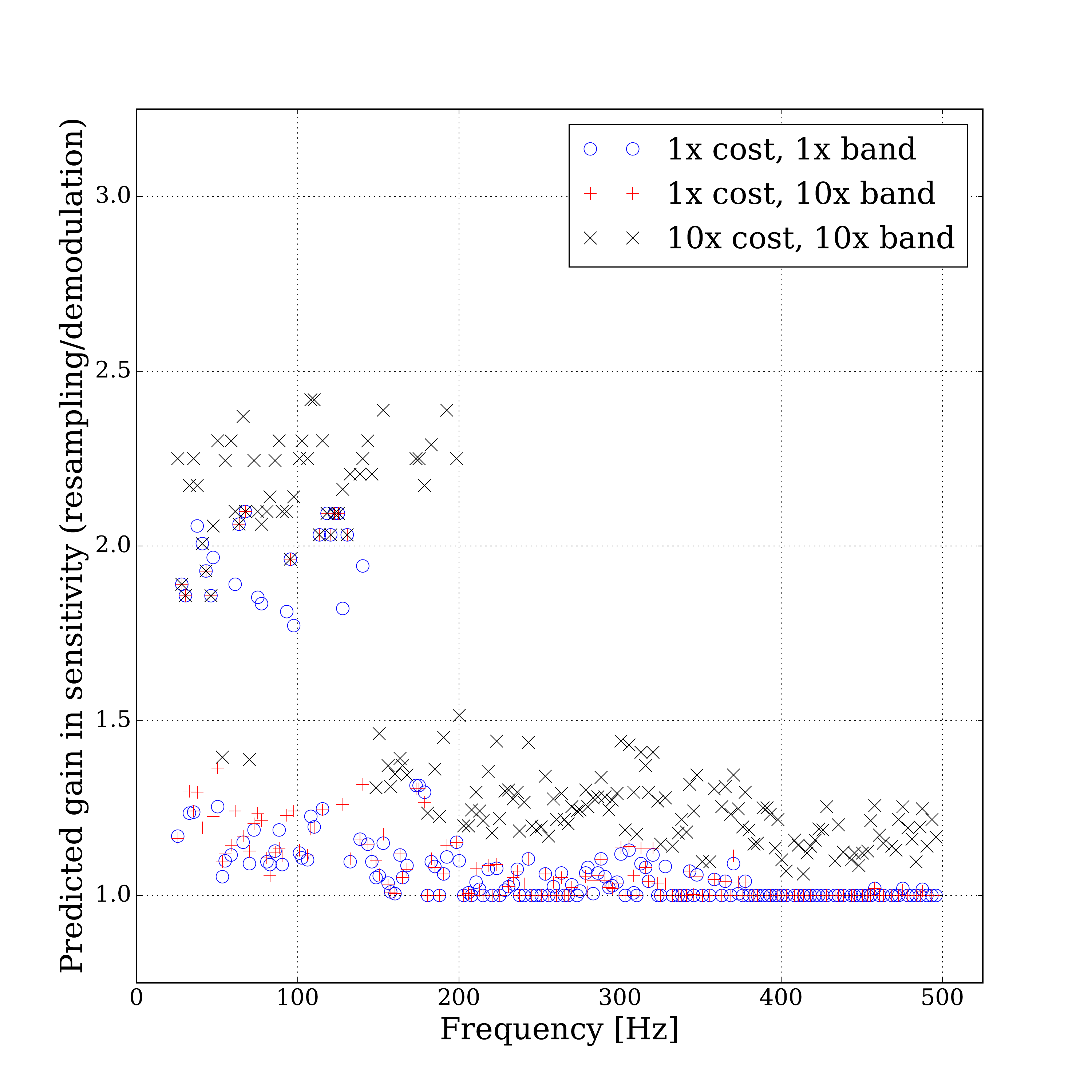}
\includegraphics[width=0.41\paperwidth, trim={0 30 0 0}, clip]{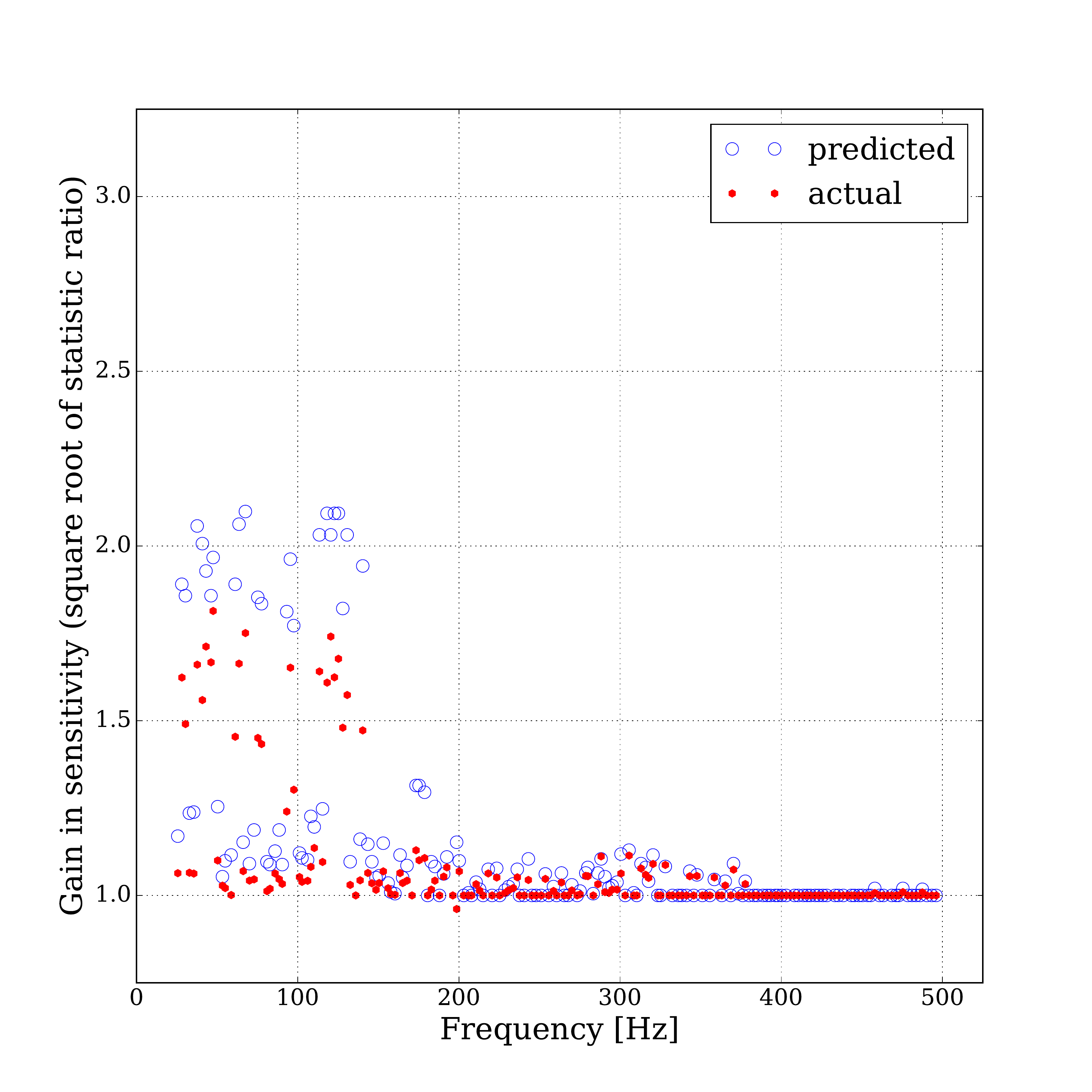}
\caption{
\textit{(Left)}
Predicted gain in sensitivity for resampling over demodulation, \textit{vs} frequency, based on $h_0$ upper limits being proportional to $T_\mathrm{max}^{-1/4}$.
$T_\mathrm{max}$ is capped at $3$ days in this Figure; compare Figure~\ref{10-days-spin-wander-gains}.
An observation time of $T_\mathrm{obs} = 1.5\times 10^{7}$~s is assumed.
The timing model (Equation~\ref{resampTimingModelEq}) estimates cost for incrementally-longer $T_\mathrm{max}$ until constraints reached.
Our cost allowance is predicated on the  measured ratio of resampling to demodulation times on the Atlas cluster.
Whenever resampling is slower, the result defaults to $1$.
Symbols on the figure indicate the following constraints: $(\circ)$ equal computational cost, $(+)$ $f_\mathrm{band}$ $10\times$ wider, $(\times)$ $f_\mathrm{band}$ $10\times$ wider and given $10\times$ computing time.
No other re-optimization of set-up is done.
Fluctation in the results occurs, because benefit scales non-linearly with increased $T_\mathrm{max}$; less-probable regions of orbital parameter space are allocated lower $T_\mathrm{max}$ and see less benefit at fixed cost.
A distinct high-gain population is seen where benefits are limited by spin wandering, computing cost having asymptoted in the binary metric.
The $(+)$ set is worse for lower frequencies because some bands move into high $N_\mathrm{fft}$, but mediun frequencies benefit from reduced Doppler wings.
The $(\times)$ set shows improvement up to about $500$~Hz, with more relative gain because Table~\ref{search-setup-table} set-ups allocate shorter lag-time to those frequencies, so affording greater room for improvement.
For equal cost $(\circ)$, median gain from $20$~Hz to $125$~Hz is $51\%$ and from $20$~Hz to $250$~Hz is $11\%$.
\textit{(Right)}
Empirical results of simulation for $3$-day limit, equal-cost, equal-band.
\textit{(Red hexagons)} show the square-root of the ratio of $\rho$ (resampling) divided by $\rho$ (demodulation).
Improvement exists but is less than predicted, possibly because $h_0 \propto \rho^2 \propto T_\mathrm{max}^4$ scaling laws are imprecise.
}
\label{03-days-spin-wander-gains}
\end{figure*}

\begin{figure}
\includegraphics[width=0.41\paperwidth, trim={0 30 0 0}, clip]{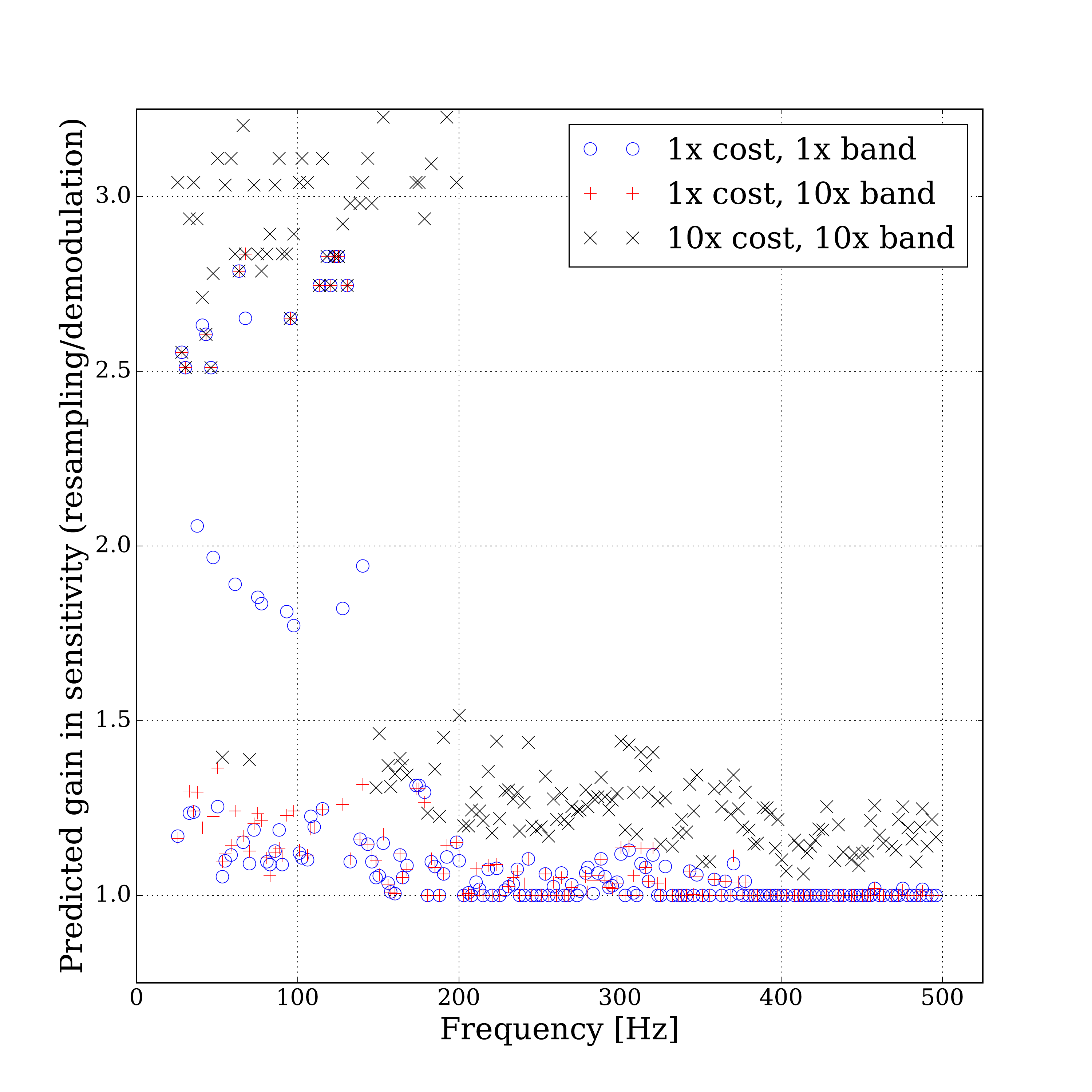}
\caption{
Predicted gain in sensitivity for resampling over demodulation, \textit{vs} frequency.
$T_\mathrm{max}$ capped at $10$ days; compare Figure~\ref{03-days-spin-wander-gains}.
Bands limited by computational cost rather than spin-wandering see no change.
In this optimistic scenario, the median gain from the $(\times)$ $10\times$ cost, $10\times$ $f_\mathrm{band}$ case, from $20$~Hz to $125$~Hz, is $2.83\times$, and from $20$~Hz to $250$~Hz it is $2.75\times$.
This is the best improvement that we consider.
As in Figure~\ref{03-days-spin-wander-gains}, actual gains may be less.
}
\label{10-days-spin-wander-gains}
\end{figure}

\begin{figure*}
\includegraphics[width=0.9\paperwidth, trim={0 30 0 0}, clip]{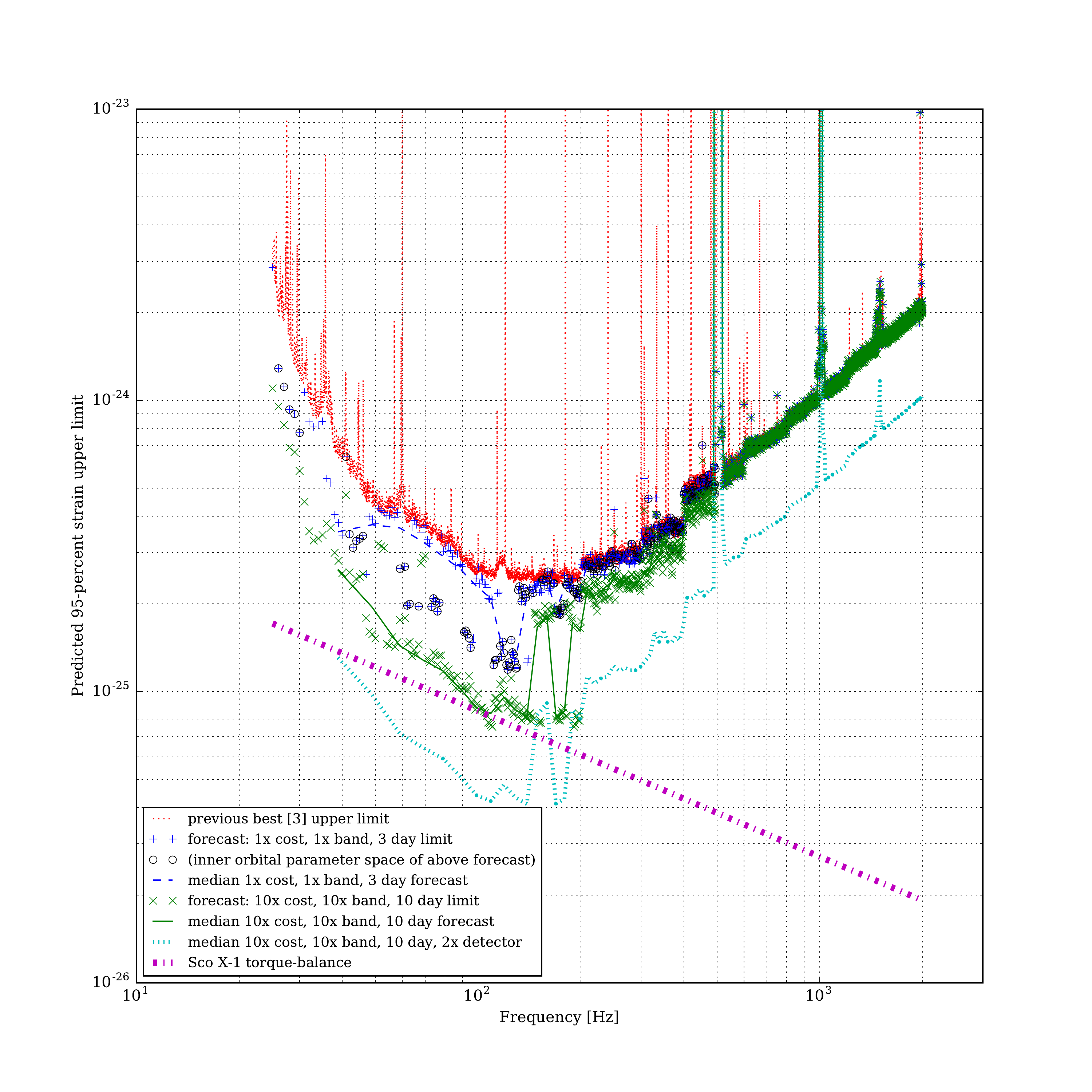}
\caption{
Theoretically-forecast (hypothetical) upper limits extrapolated from the O1 cross-correlation method's $95\%$ marginalized Sco X-1 search, without $\cos \iota$ information~\cite{ScoX1CrossCorr2017ApJO1}
Extrapolation based on gains in Figures~\ref{03-days-spin-wander-gains} and~\ref{10-days-spin-wander-gains}.
As in the former, actual gains may be less.
O1 limit shown in \textit{(red dots)}.
Extrapolation based on equal cost, 3-day spin-wandering limit in \textit{(blue dashed line)}, and based on $10\times$ cost, $10\times f_\mathrm{band}$, 10-day spin-wandering in \textit{(green solid line)}.
Respective $(\textit{blue } + )$ and $(\textit{(green } \times )$ indicate every 1-Hz interval (original upper limit used $50$-mHz intervals).
Lines trace (non-running) median of $10$-Hz bins.
Fluctuations seen in lines because some bands limited by spin-wandering, others not.
The $(+)$ bands are circled $(\circ)$ if they are in a long $T_\mathrm{max}$ part of orbital parameter space, defined as $a_p \leq 2.1663$~s, $1131415225 < T_\mathrm{asc} \leq 1131415583 $ based on O1 setup~\cite{ScoX1CrossCorr2017ApJO1}.
Longer $T_\mathrm{max}$ times benefit more from resampling, as noted in Figure~\ref{03-days-spin-wander-gains}.
Caution: correct upper limits would require estimation of detection efficiency, not done here.
Present extrapolation suggests torque-balance might be attained in the best $( \textit{green } \times)$ case at 100~Hz, or \textit{(small cyan dashes)}, with a $2\times$ improved detector noise floor ($T_\mathrm{obs}$ same as O1), from 40~to 140~Hz.
Compare to Figure~\ref{projected-sens-depth}.
}
\label{projected-ul}
\end{figure*}

\begin{figure*}
\includegraphics[width=0.9\paperwidth, trim={0 30 0 0}, clip]{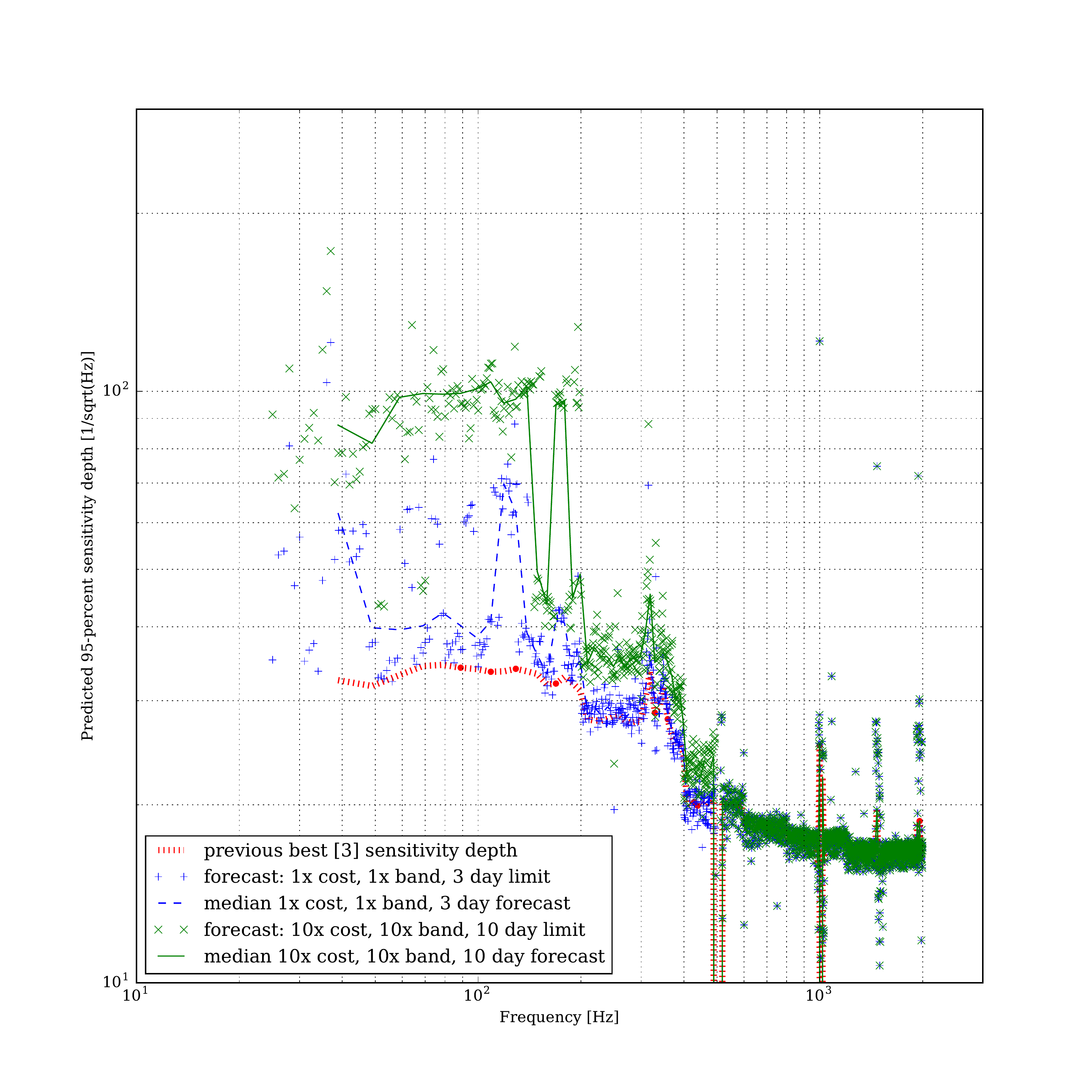}
\caption{
Projected sensitivity depth \textit{vs} frequency.
Compare to Figure~\ref{projected-ul}, from which this graph is derived by dividing by the harmonic mean of the Hanford (H1) and Livingston (L1) detector ASD noise floors.
Additional uncertainty is introduced by contaminating lines in the detectors.
The equal-cost case $(\textit{blue } +)$ shows bands up to about $200$~Hz fluctuating between the mid-$30$ and mid-$60$~Hz$^{-1/2}$ range; as noted in Figure~\ref{03-days-spin-wander-gains}, gain is less at fixed cost for less-probable parts of orbital parameter space, because they are allocated shorter $T_\mathrm{max}$ in O1.
For the best, $10\times$ cost, $10\times f_\mathrm{band}$, 10-day spin-wandering case $(\textit{green } \times)$, most parts of orbital parameter space reach approximately $100$~Hz$^{-1/2}$.
Caution: results are hypothetical and theoretically extrapolated from the timing model, as noted in Figure~\ref{projected-ul}; also, eccentricity is assumed negligible.
Results may improve depending on computational cost re-optimization (confer~\cite{MingSetup2015}).
}
\label{projected-sens-depth}
\end{figure*}

For now, we consult Figures~\ref{03-days-spin-wander-gains} and~\ref{10-days-spin-wander-gains}.
These figures show, using the $D^C \propto T_\mathrm{max}^{1/4}$ assumption, the forecast sensitivity gain from resampling's speed-up relative to demodulation.
The exact same test set-ups are run for both resampling and demodulation and the run time is measured.
Then Equation~\ref{resampTimingModelEq} is used to predict the run time of resampling with longer $T_\mathrm{max}$, iteratively increasing $T_\mathrm{max}$ by $1\%$ until the original demodulation cost is predicted to be reached.
The quarter root of the final $T_\mathrm{max}$ is taken as the forecast gain.
(If resampling already takes as least as long as demodulation for a given set-up, this gain defaults to unity).
Gains depend on the test bands' set-ups (Table~\ref{search-setup-table}).
Figure~\ref{03-days-spin-wander-gains}'s right side contrasts empirical sensitivity gains with predictions.
The actual relative gain, given by the square root of the ratio of the $\rho$ statistic with given $T_\mathrm{max}$, tends to be less than the power-law prediction.
Sensitivity forecasts in Figures~\ref{projected-ul} and~\ref{projected-sens-depth} should thus be read as cautiously optimistic.

Long lag-times benefit the most from resampling.
Figure~\ref{ratio_of_runtimes_figure} illustrates that resampling is only faster than demodulation for bands of $T_\mathrm{max} \gtrsim 2000$ to $5000$~s, which Table~\ref{search-setup-table} shows to be in frequency bands less than roughly $200$ to $400$~Hz in the O1 setup.
These $T_\mathrm{max}$ allocations~\cite{ScoX1CrossCorr2017ApJO1} were designed to maximize detection probability by investing integration time in high-probability regions of orbital parameter space and frequencies near the torque balance level.
Where $T_\mathrm{max}$ is already large, resampling offers more acceleration, thus more computing to be reinvested, and Figures~\ref{03-days-spin-wander-gains} and~\ref{10-days-spin-wander-gains} show bigger gains.
In principle, the cost allocation is a global problem: we want to maximize the detection probability of the entire search, not one band.
This problem has been addressed not only in~\cite{ScoX1CrossCorr2017ApJO1} but also~\cite{MingSetup2015}.
In the future, these methods can be turned to the complicated task of re-optimizing the resampling cost allocation to maximize detection probability.
For this paper, forecasts are based on the O1 allocation.
Also note that we assume that the sensitivity gains $\propto T_\mathrm{max}$ will uniformly scale the detection efficiency curves that set upper limits.
Taking this product of averages is only approximate: the true sensitivity is an average constructed from the products of gains in each band. 
As the $\rho$ statistic ratio from long $T_\mathrm{max}$ is less than predicted, a systematic study is needed about the sensitivity gain from computational reinvestment.
In the future, we expect our assumptions to be tested by a second Mock Data Challenge (following~\cite{ScoX1MDC2015PRD}).

At present, results are suggestive.
Figure~\ref{projected-ul} shows the projected upper limits that are forecast based on O1 results~\cite{ScoX1CrossCorr2017ApJO1}, divided by the sensitivity gain estimated for each band.
Figure~\ref{projected-sens-depth} shows these upper limits divided by the noise ASD of the detector, to show sensitivity depth $D^C$, which is easier to compare with other methods.
Both figures refer to results marginalized over $\cos \iota$, as the inclination angle of Sco X-1 is unknown.
Long $T_\mathrm{max}$ bands at low frequencies can potentially double to triple in sensitivity.
Given equal cost allowance and the assumption of $T_\mathrm{max}$ limited to $3$~days by spin-wandering, the gain is limited: from $20$~to $125$~Hz, the median gain is $51\%$, and from $20$~to $250$~Hz, it is only $11\%$, with minimal benefit at higher frequencies.
The sensitivity depth varies between the mid-$30$s and mid-$60$s~Hz$^{-1/2}$, depending on position in orbital parameter space.
Given tenfold resources and the assumption of $T_\mathrm{max}$ limited to $10$~days, the gains are respectively $2.83\times$ and $2.75\times$ over O1.
This sensitivity depth is approximately $100$~Hz$^{-1/2}$.
Given O1 noise, the latter scenario would just touch the torque-balance level at $100$~Hz.
Given twofold detector improvement, the upper limits would scale linearly, and resampling could potentially reach below torque balance from approximately $40$~to $140$~Hz.
Longer observing runs should improve sensitivity with the usual $T_\mathrm{obs}^{1/4}$ scaling~\cite{ScoX1CrossCorr2015PRD}.

Future computational enhancements in the cross-correlation method, such as GPU acceleration for the barycentering and FFT operations, may make the tenfold gain in cost allowance realistic, as may access to larger computing resources.
For example, one \textit{Einstein@home} Month (EM) of computing power assumes 12 thousand cores~\cite{MingSetup2015}, or roughly $8.64$~million CPU hours.
Depending on CPU performance compared to the Atlas cluster, multi-EM allocations could extend the cross-correlation method's depth.
It may be possible to use Bessel functions, as in~\cite{SidebandBessel2017}, or a loosely-coherent approach~\cite{PowerFluxMethod2010}, to accelerate moving through the orbital parameter space: the phase modulation can in principle be `resampled' in the frequency domain as well as our time-domain approach, and some fusion of the two may be faster.
Even now, resampling can accelerate longer lag-time follow-ups (progressive $4\times$ increases in $T_\mathrm{max}$ for search candidates~\cite{ScoX1CrossCorr2017ApJO1}) and improve the low-frequency search.

The `CrossCorr' cross-correlation method is not the only method that may reach such performance.
The `Sideband' method~\cite{SidebandMarkovModelSuvorova2016} is under active development, and a binary-oriented, resampled $\mathcal{F}$-statistic code~\cite{LeaciPrixDirectedFStatPRD} has offered even greater sensitivity depth.
The latter predicts that torque-balance could be reach up to $160-200$~Hz for conservative assumptions about eccentricity or $500-600$~Hz if eccentricity is assumed to be well-constrained.
(By assuming the eccentricity to be circular, our result of $140$~Hz is comparable to the latter case).
Predictions are highly sensitive to the timing function and cost allowances of the final code, as well as to assumptions about spin-wandering.
Here we have presented our estimates based on working search code and extrapolations from the finished O1 search using the cross-correlation method.

\section{Conclusions\label{conclusion}}

Resampling accelerates the deepest current search for Sco X-1 and similar LMXBs, the cross-correlation method~\cite{ScoX1CrossCorr2017ApJO1}.
By calculating the cross-correlation method's $\rho$ statistic using barycentric interpolation to the source frame, followed by an FFT, speed-up is possible for long coherent integration lag-times.
Because of the plateauing of the binary orbital parameter space, this acceleration can drive the cross-correlation method's forecast sensitivity to the torque-balance level in conceivable scenarios.
In the most optimistic case with O1-like data, it may graze this level at 100~Hz; with a detector twice as sensitive (closer to Advanced LIGO design sensitivity), this range may extend from 40~to 140~Hz.
Re-optimization of the computational cost distribution across parameter space~\cite{MingSetup2015} can focus resources where detection is most probable.
Reaching torque-balance might then be possible without large increases in computing power.
Future improvement may allow it to compete up to higher frequencies, as might other proposed methods~\cite{LeaciPrixDirectedFStatPRD}.
The cross-correlation method with resampling works already.
This success is possible thanks to the deep similarity between the $\mathcal{F}$-statistic and $\rho$-statistic and the shared codebase of the \textit{LIGO Applications Library}, which allowed the importation of large portions of the resampling algorithm, once the mathematics were understood.
Future improvements to any of this family of methods might be transplanted to benefit all.

Many unknowns remain in Sco X-1.
The depicted torque-balance level assumes a $10$-km radius and $1.4$-solar mass for a NS that itself has not been confirmed in the system; the level varies with the object's moment of inertia.
Expectation has held that Sco X-1's luminosity makes it a promising target.
Other systems may prove promising alternative targets, particularly if they have a known spin frequency.
Known frequency, or much more precise orbital parameters, could reduce the cost of the cross-correlation method and similar semicoherent searches by many orders of magnitude.
Then a sensitivity limited only by spin-wandering might be easily reached, regardless of location on the spectrum.
Until then, computational optimizations will play a pivotal role in broadband searches.
We see potential in applying this proven method to Advanced LIGO searches -- gravitational waves from Sco X-1 have never been closer to detection.

\begin{acknowledgments}
This work was partly funded by the Max-Planck-Institut. 
JTW and YZ were supported by NSF grants No.~PHY-1207010 and
No.~PHY-1505629.
JTW acknowledges the hospitality of the Max Planck Institute for
Gravitational Physics (Albert Einstein Institute) in Hannover.
These investigations use data and computing resources from the LIGO Scientific Collaboration.
Further thanks to the Albert-Einstein-Institut Hannover and the Leibniz Universit\"{a}t Hannover for support of the Atlas cluster, on which most of the computing for this project was done.
Many people offered helpful comments, especially R. Prix for extensive knowledge on the resampling code implementation, K. Wette for familiarity with the LIGO Applications Library, along with V. Dergachev, A. Mukherjee, K. Riles, S. Walsh, S. Zhu, E. Goetz, M. Cabero-M\"{u}ller, C. Messenger, C. Aulbert, H. Fehrmann, C. Beer, O. Bock, H.-B. Eggenstein and B. Maschenschalk, L. Sun, E. Thrane, A. Melatos, B. Allen, B. Schutz, and all members of the AEI and LIGO Scientific Collaboration-Virgo continuous waves (CW) groups.
We also thank our referee for helpful reading and comments.
This document bears LIGO Document Number DCC-P1600327.
\end{acknowledgments}

\appendix

\section{Relationships to other optimal statistics\label{relationships-to-other-optimal-statistics}}

Terms called $F_a$ and $F_b$~\cite{Dhurandhar2008} relate $\rho$ to the $\mathcal{F}$-statistic, already amenable to resampling~\cite{Patel:2009qe}.
These $F_a$ and $F_b$ are the components of the statistic that are respectively projections of data along the $a$ and $b$ time series.
To investigate these components, we will look at the phase-model corrected frequency-domain data, $\zeta_K$.
(Precisely, $\zeta_K = \Xi_K z_K \exp{(-i\Phi_K)}$ for $z_K$, $\Xi_K$ from Equation~\ref{total-data-eq}).
We can arrange the data $z_{Kk}$, indexed by frequency bin $k$, to include phase shift $\exp{(-\mathrm{i}\Phi_K)}$,

\begin{equation}
\zeta_{K} \equiv \sum_{k\in\mathcal{K}_K}(\mathrm{i})^{2k} \mathrm{sinc}(\kappa_{Kk}) z_{Kk} e^{-\mathrm{i}\Phi_K},
\end{equation}

\noindent and likewise $\zeta_{L}$, substituting (real-valued) Equation~\ref{geometric-filter} into $\rho$ ($\Re$ denoting the real part)
and grouping terms: 

\begin{eqnarray}
\rho &=& \frac{N}{5} \Re \sum_{KL\in\mathcal{P}} \left[(\hat a^K \zeta_{K})^* \hat a^L \zeta_{L} + (\hat b^K \zeta_{K})^* \hat b^L \zeta_{L} \right],
\label{staged-to-turn-to-fafb}
\end{eqnarray}

\noindent which merits inspection of $\hat a^K \zeta_K$.
Insert $\hat a^K$ and Equation~\ref{fourier-transform-def},
noting 
$(-1)^{k-l} = (-1)^{l-k}$, 
$(\mathrm{i})^{2k} = \exp(\mathrm{i} \pi k)$:

\begin{eqnarray}
\hat a^K \zeta_K 
= &&\frac{2}{S_K} a^K \sum_{k\in\mathcal{K}_K} \mathrm{sinc}(\kappa_{Kk}) \\
  &\times& \sum_{j=0}^{N-1} x_K (t_K - T_\mathrm{sft}/2 + j \delta t) \nonumber \\
  &\times& e^{-\mathrm{i} 2\pi k (j \delta t - T_\mathrm{sft}/2) / T_\mathrm{sft}} \delta t  e^{-\mathrm{i}\Phi_K},\nonumber
\end{eqnarray}

\noindent Whereas $a(t)$, $b(t)$ amplitude modulations have a period on the order of a sideral day, as in Section~\ref{antenna-function-weighting}, the $a^K$, $S_K$ terms vary much more slowly than $f_K$,
So we 
take $m \equiv j - T_\mathrm{sft}/(2\delta t)$, $t_j \equiv t_K - T_\mathrm{sft}/2 + j \delta t$,
$t_m \equiv t_K + m \delta t$,
moving the antenna functions inside the sum over $m$,

\begin{eqnarray}
\hat a^K \zeta_K = && \sum_{m=-T_\mathrm{sft}/(2\delta t)}^{m=T_\mathrm{sft}/(2\delta t)-1} \delta t \frac{2}{ S_K} a^K x_K (t_m) \label{slowly-varying-a-zeta}\\
 &\times& \sum_{k\in\mathcal{K}_K}\mathrm{sinc}(\kappa_{Kk}) e^{-\mathrm{i} (2\pi k m \delta t / T_\mathrm{sft} + \Phi_K)}.\nonumber
\end{eqnarray}

\noindent 
Instead of including all frequency bins $k$ for $z_K$ of Equation~\ref{total-data-eq}, the SFT signal-resolution can be zero-padded (see Equation~9.3 of~\cite{Allen2002}).
Zero-padding $k$ brings the nearest bin closer to $f_K$, approaching $\kappa_{Kk}\approx 0$,

\begin{eqnarray}
\hat a^K \zeta_K \approx \sum_{m=-T_\mathrm{sft}/(2\delta t)}^{m=T_\mathrm{sft}/(2\delta t)-1} && \delta t \frac{2}{S_K} a^K x_K (t_m) \label{from-here-resamp}\\
  &\times&  e^{-\mathrm{i} (2\pi (f_K T_\mathrm{sft}) m \delta t / T_\mathrm{sft} + \Phi_K)}.\nonumber
\end{eqnarray}

\subsection{Statistic in conventional quantities}

Proceeding to continuous time, Equation 5.10 of~\cite{Dhurandhar2008} has sub-interval integral $F_{a_I}$,

\begin{eqnarray}
F_{a_I} &=& \int_{T_I - \Delta T/2}^{T_I + \Delta T/2} a(t) x(t) e^{-\mathrm{i} \varphi (t)} dt,\\
\label{crosscorr-fa-i}
F_{a} &=& \sum_I F_{a_I}.
\label{f-stat-fa}
\end{eqnarray}


Treating sums as integrals, $\delta t \rightarrow dt$, $T_\mathrm{sft} \rightarrow \Delta T$, $a_K \rightarrow a(t)$, $x_K (t_m) \rightarrow x(t)$, and $\exp{(-\mathrm{i} [2\pi (f_K T_\mathrm{sft}) m \delta t / T_\mathrm{sft} + \Phi_K] )} \rightarrow \exp{(-\mathrm{i} \varphi(t))}$. 
Observe that with $2 \pi f_K = (d\Phi_K / dt | t= t^K)$, $t = m\delta t$, $\varphi$ is a Taylor approximation of $\Phi$:

\begin{eqnarray}
\hat a^K \zeta_K &=& \frac{2}{S_K} F_{a_K}.\label{the-grand-id}
\end{eqnarray}

Taking $T_0 \rightarrow T_\mathrm{sft}$ in Equation 42 of~\cite{Jaranowski1998}, we write an inner product,

\begin{equation}
(x || y) \equiv \frac{2}{T_0}\int_{-T_0 / 2}^{T_0/2} x(t) y(t) dt,
\label{jks-double-inner-product}
\end{equation}

\noindent so $F_{a_K} = [T_\mathrm{sft} / 2] ( a\cdot x || \exp{(-\mathrm{i}\varphi)} )$ can be viewed as a projection of the amplitude-modulated data onto the phase-model basis.
A reader may wonder whether this is not a Fourier transform.
Not quite: $\varphi(t)$ is phase-modulated and does not increase linearly in evenly-sampled detector time $m \delta t$.
Before addressing this problem with resampling (Section~\ref{resampling}), we connect $\rho$ to related statistics.

Using $F_{a_K}$ in Equation~\ref{staged-to-turn-to-fafb},

\begin{eqnarray}
\rho &=& \frac{N}{5} 4\Re \sum_{KL\in\mathcal{P}} \left[ \frac{F_{a_K}^*}{S_K} \frac{F_{a_L}}{S_L} + \frac{F_{b_K}^*}{S_K} \frac{F_{b_L}}{S_L} \right].
\end{eqnarray}

\noindent
In the bin-centered limit (Equation 3.18 in~\cite{ScoX1CrossCorr2015PRD}), $\langle \Xi^2 \rangle \approx 1$. 
Establishing $N$ without $\Xi$ but with,

\begin{eqnarray}
\hat A^2_\mathcal{P} &\equiv& \sum_{KL\in\mathcal{P}} (\hat a^K \hat a^L)^2, \\
\hat B^2_\mathcal{P} &\equiv& \sum_{KL\in\mathcal{P}} (\hat b^K \hat b^L)^2, \\
\hat C^2_\mathcal{P} &\equiv& \sum_{KL\in\mathcal{P}} (\hat a^K \hat b^L \hat a^L \hat b^K),
\end{eqnarray}

\noindent we obtain,
\begin{eqnarray}
N &=& \frac{10}{\sqrt{2}} \left[ \hat A_\mathcal{P}^2 + 2 \hat C_\mathcal{P}^2+ \hat B_\mathcal{P}^2\right]^{-1/2},
\end{eqnarray}

\begin{eqnarray}
\rho &=& \frac{4\sqrt{2}\Re \sum_{KL\in\mathcal{P}} \left[ \frac{F_{a_K}^*}{S_K} \frac{F_{a_L}}{S_L} + \frac{F_{b_K}^*}{S_L} \frac{F_{b_L}}{S_L} \right] }{\sqrt{ \hat A_\mathcal{P}^2 + 2 \hat C^2_\mathcal{P}+ \hat B_\mathcal{P}^2 }} 
.
\label{well-formatted-rho}
\end{eqnarray}

Compare $\rho$ to the $\mathcal{F}$-statistic in a specific case.
Take $Q$ detectors indexed by $X$, $Y$, each with $M\equiv T_\mathrm{obs}/T_\mathrm{sft}$ SFTs.
Assume a frequency-dependent, stationary noise PSD $S_h(f)$.
Allow all pairs $\mathcal{P}$, so the sum expands into a double sum of a double sum seen in Equation~\ref{f-stat-fa},

\begin{eqnarray}
\rho = && 4\sqrt{2} \left( 
S_h^2(f) \sqrt{\hat A_\mathcal{P}^2 + 2 \hat C^2_\mathcal{P}+ \hat B_\mathcal{P}^2 }
\right)^{-1} \label{explicit-detector-rho-comparison}\\
  &\times&  \Re \left[ \sum_X^Q \sum^M_{K(X)} F_{a_{K(X)}}^* \sum_Y^Q \sum^M_{L(X)} F_{a_{L(Y)}} \ldots \right. \nonumber \\
  && \left. + \sum_X^Q \sum^M_{K(X)} F_{b_{K(X)}}^* \sum_Y^Q \sum^M_{L(Y)} F_{b_{L(Y)}} \right]
,\nonumber
\end{eqnarray}

\noindent
As the index $I$ in Equation~\ref{f-stat-fa} is detector independent,

\begin{eqnarray}
\sum_X^Q \sum^M_{K(X)} F_{a_{K(X)}}^* 
  &=& F_a^*, 
\end{eqnarray}

\noindent so too for the $L$ index and $b$ terms, allowing (self-) \textit{auto-correlations.}
Normally, the cross-correlation method does not allow auto-correlations~\cite{ScoX1CrossCorr2015PRD}, but it can~\cite{Dhurandhar2008}, such that,

\begin{eqnarray}
\rho &=& 4\sqrt{2}\frac{|F_a|^2 + |F_b|^2 }{S_h^2(f) \sqrt{\hat A_\mathcal{P}^2 + 2 \hat C^2_\mathcal{P}+ \hat B_\mathcal{P}^2 }} 
,
\end{eqnarray}

Simplifying the denominator,

\begin{eqnarray}
\hat A^2_\mathcal{P} 
  &=& \left(\sum_I^{QM} (\hat a^I)^2 \right)^2,
\\
\hat A_\mathcal{P} &=& \frac{1}{T_\mathrm{sft}} \sum_I^{QM} \hat a^I \hat a^I T_\mathrm{sft},
\end{eqnarray}

\noindent which Riemann integrates for $a(t)$, $b(t)$ that vary slowly compared to $T_\mathrm{sft}$ (faster than $T_\mathrm{obs}$, so an overall shift is negligible and $T_0 \rightarrow T_\mathrm{obs}$ in Equation~\ref{jks-double-inner-product}),

\begin{eqnarray}
\hat A_\mathcal{P} &\approx & 
\frac{Q T_\mathrm{obs}}{S_h (f)} (a || a) 
\end{eqnarray}

\noindent Forming norms $A\equiv (a||a)$, $B\equiv (b||b)$, $C\equiv(a||b)$~\cite{Jaranowski1998}:

\begin{eqnarray}
\rho = 4 \sqrt{2} \frac{|F_a|^2 + |F_b|^2}{S_h(f) Q T_\mathrm{obs} \sqrt{A^2 + 2C^2 + B^2}}.
\end{eqnarray}

\subsection{Comparison to the $\mathcal{F}$-statistic}

The $\mathcal{F}$-statistic is a maximum-likelihood (ML) estimator.
Values of $\mathcal{A}^\mu$ are chosen where the likelihood ratio is a maximum, $\Lambda_\mathrm{ML}$.
Composing frequency-integrated projections $x_\mu$ onto the basis $h^\mu$ in Equation~\ref{decomposition-projection-f}, with $\mathcal{M}^{\mu\nu}$ the ML projections of $h^\mu$ onto $h^\nu$~\cite{Jaranowski1998,CutlerMulti2005,BStatPrix2009,Patel:2009qe,WhelanNewAmplitude2014CQG}:

\begin{eqnarray}
\Lambda_\mathrm{ML} &=& e^\mathcal{F},\label{abstract-fstat}\\ 
\mathcal{F} &\equiv& \frac{1}{2} x_\mu \mathcal{M}^{\mu\nu} x_\nu,\\
 &=& \frac{4}{S_h(f) T_0}\frac{B|F_a|^2 + A|F_b|^2 - 2C\Re(F_a F_b^*)}{A\cdot B - C^2}. \nonumber
\label{the-fstat}
\end{eqnarray}

\noindent
Both $\rho$ and $\mathcal{F}$ are dimensionless.
As after Equation~5.15 in~\cite{Dhurandhar2008}, $\rho$ and $\mathcal{F}$ are proportional when $A\approx B$, $C \ll A,B$:

\begin{eqnarray}
\rho &\approx& 4\sqrt{2} \frac{|F_a|^2 + |F_b|^2}{S_h(f) Q T_\mathrm{obs} \sqrt{2 A^2}},
\end{eqnarray}

\noindent
equating $\mathcal{F}$ with $T_0 = Q T_\mathrm{obs}$,

\begin{eqnarray}
\mathcal{F} &\approx& \frac{4}{S_h(f) T_0}\frac{|F_a|^2 + |F_b|^2}{A}.
\end{eqnarray}

\noindent Even for multiple detectors, (all-pairs) $\rho$ can converge to the (fully-coherent) $\mathcal{F}$-statistic.
Illustrating the crossover is now possible.

Dhurandhar \textit{et al}~\cite{Dhurandhar2008} introduce the cross-correlation method starting from two data streams, like the stochastic radiometer~\cite{Ballmer2006CQG}, instead of the multi-detector $\mathcal{F}$-statistic~\cite{CutlerMulti2005}. 
The weight matrix $\textbf{W}$ of Whelan \textit{et al}~\cite{ScoX1CrossCorr2015PRD} can merge these viewpoints.
Any SFT, from any detector, is a dimension in $\textbf{z}$ (`flattening' SFTs over the Greek indices also represented as boldface in~\cite{CutlerMulti2005} to represent different detectors).
Cutler \& Schutz Equation~3.8~\cite{CutlerMulti2005} has $2\mathcal{F} = \sum_{a,d}(\Gamma^{-1})^{ad} (\mathbf{x}|\mathbf{h}_\mathbf{a})(\mathbf{x}|\mathbf{h}_\mathbf{d})$: $a,d$ are the waveform components $\mu,\nu$ in our Equation~\ref{abstract-fstat}.
Their inner products of $\mathbf{x}$ with the waveforms $\mathbf{h}_{\mathbf{a}},\mathbf{h}_\mathbf{d}$ are scalar-valued vectors indexed by $a$ and $d$, equivalent to summing $F_a$ or $F_b$ from multiple detectors.
Only then is $\mathcal{F}$ computed.
The sum of fully-coherent single-detector $\mathcal{F}$ does not equal the \textit{fully}-coherent multiple-detector $\mathcal{F}$, which takes into account the cross-detector terms and converges with the ideal cross-correlation method.

Divergence can occur with \textit{semi}-coherent methods~\cite{CutlerSemi2005,PrixShaltev2012}.
Semicoherent calculations with $T_\mathrm{coh} < T_\mathrm{obs}$ are more efficient, having higher sensitivity at fixed computational cost, than fully-coherent methods~\cite{HierarchicalBrady2000,CutlerSemi2005,PrixShaltev2012}.
The sum of $\mathcal{F}$-statistics over $T_\mathrm{obs}/T_\mathrm{coh}$ segments of $\mathcal{F}$-statistics is computed, albeit with reduced sensitivity compared to the much more expensive fully-coherent search.
Joint- and single-detector $\mathcal{F}$ can both be computed for each $T_\mathrm{coh}$.
(Comparison between joint and single is the basis of the $\mathcal{F}$-statistic consistency veto~\cite{EinsteinHomeS52013}).
The main difference between the cross-correlation method and the semicoherent $\mathcal{F}$-statistic is that the former, distinguishing $K$ and $L$, helps to exclude auto-correlations.

Examine the optimal amplitude parameters in $\mathcal{M}^{\mu\nu}$ and weights $\textbf{W}$.
Despite Equation~\ref{abstract-rho-matrix}'s resemblance to Equation~\ref{the-fstat}, $\textbf{W}$ and $\mathcal{M}^{\mu\nu}$ are matrices over different spaces.
$\textbf{W}$ (implicit indices) is of SFTs, whereas $\mathcal{M}^{\mu\nu}$ (explicit indices) is of four amplitude parameters.
The amplitude-parameter space metric is $\mathcal{M}^{\mu\nu}$, so $2\mathcal{F} = x^\mu x_\nu$~\cite{BStatPrix2009}.
In principle, $\rho_\mathrm{ideal}$ might not use $\hat \Gamma^\mathrm{ave}_{KL}$ (chosen to avoid specifying $\cos \iota$ and $\psi$~\cite{ScoX1CrossCorr2015PRD}), but instead $\Gamma$ based on maximization or marginalization~\cite{BStatPrix2009} of $\mathcal{A}^\mu$. 
A start would be projections, $\textbf{z}_\mu$, of $\textbf{z}$ onto the $h_\mu$ basis.
Each $\textbf{z}$ (a data vector, implicitly indexed, \textit{e.g.}, by SFTs) can be projected to extract the components along the $4$ amplitude-parameter space dimensions, producing the $N_\mathrm{sft}\times 4$ matrix, $\textbf{z}_\mu$.
Schematically,

\begin{equation}
\rho_\mathrm{ideal} = \frac{1}{2} \mathcal{M}^{\mu\nu}\textbf{z}_\mu^\dag \textbf{W} \textbf{z}_\nu.
\end{equation}

\noindent
Hence $\textbf{z}_\mu$ absorb $\Phi$ and can be thought of as Fourier transforms of source-frame data.
The matrix $\mathcal{M}$ absorbs $\Gamma$ from $\textbf{W}$, leaving $\textbf{W}$ a binary-valued index of which SFTs to pair.
Such a statistic would echo the likelihood ratio mentioned in Section V of~\cite{Dhurandhar2008}.
Recall, $\rho$ and $\mathcal{F}$ converge when $A\approx B$ and $C \ll A,B$, \textit{i.e.,} when $\mathcal{M}$ is proportional to the identity matrix.
So, when amplitude space is flat and auto-correlations are included, $\rho \approx \mathcal{F}$.

One impetus for cross-detector correlation is that only signal should be coherent.
This ground underlies stochastic searches: GW strains between detectors are related, but the noise is statistically independent (see Section III of~\cite{Allen1999}).
Appendix~\ref{glitch-merits-of-crosscorr} will revisit the robustness and merits of pairing choices for $\textbf{W}$.

\subsection{Comparison to the stochastic search}

The stochastic search~\cite{Allen1999,Ballmer2006CQG,ThraneStochastic2009} is built on same-time cross-correlation of multiple detectors~\cite{ChristensenStochastic1992}, whereby sensitivity depends on an overlap reduction function~\cite{FlanaganStochastic1993}.
In stochastic literature, $S$ often indicates detector strains and $P$ indicates noise PSDs.
To keep consistency in this paper, $S$ will be replaced by $X$ for strain and $P$ by $S$ for noise.
For the stochastic radiometer $Y$-statistic~\cite{FlanaganStochastic1993,Ballmer2006CQG},

\begin{equation}
Y = \int_{-\infty}^{+\infty} d f \int_{-\infty}^{+\infty} df \delta_T (f - f') X_1^* (f) Q(f') X_2 (f'),
\end{equation}

\noindent
where $\delta_T$ is a finite-time Dirac delta function approximation, $X_1$, $X_2$ are Fourier-transformed detector strains, and $Q$ is an optimal filter.
For sky direction $\hat \Omega ' = \hat \Omega$,

\begin{equation}
Y_{\hat \Omega '} = (\lambda T) \int_{-\infty}^{+\infty} df \frac{\gamma^*_{\hat \Omega '} H}{S_1 S_2} X_1^* X_2.
\label{radiometer-directed-y}
\end{equation}

\noindent
Expanding, with normalization factor $\lambda$, measurement duration $T$, and $S_1$ and $S_2$ the noise PSD, as well as $H(f)$ the strain power of the stochastic background, with overlap reduction function $\gamma_{\hat \Omega '}$ and polarizations $A\in \{+,\times\}$ and detector separation vector $\Delta \vec x$:

\begin{eqnarray}
\gamma_{\hat \Omega '} &=& \frac{1}{2} \sum_A e^{\mathrm{i} 2\pi f \hat \Omega \cdot \frac{\Delta \vec x}{c}} F_1^A (\hat \Omega)F_2^A (\hat \Omega ).
\end{eqnarray}

\noindent This $Y$ is effectively a case of the simultaneous cross-correlation method's $\rho$ restricted to different detectors~\cite{ScoX1CrossCorr2015PRD}.
Notationally, $\hat \Omega = \vec n$.
Radiometer $\Delta x$ is $c$ times detector arrival time difference $\Delta d_{KL} \equiv (\vec r_K(t) - \vec r_L(t)) \cdot \vec n /c $, stemming from Equation~\ref{t-ssb-equation}.
Then, the radiometer phase difference $2\pi f \hat \Omega \cdot (\Delta \vec x /c)$ equals $\Delta \Phi_{KL}$ in Equation~\ref{textbook-cc-rho} and is $2\pi f_0 \Delta d_{KL}$.
Because $10 \Gamma^\mathrm{ave}_{KL} = F_+^K F_+^L + F_\times^K F_\times^L$~\cite{ScoX1CrossCorr2015PRD},

\begin{equation}
\gamma_{\hat \Omega'} = 5 \frac{\sqrt{S_K S_L}}{2 T_\mathrm{sft}} e^{i \Delta \Phi_{KL}} \hat \Gamma_{KL}^\mathrm{ave}.
\end{equation}

To be exact~\cite{MitraRadiometer2008}, where $\tilde Q(\hat \Omega, t, f; H) = Q(f')$, $\Delta t = T_\mathrm{sft}$ is time segment length and $\gamma^*(\Omega, t, f) = \gamma_{\hat \Omega'}$,

\begin{eqnarray}
\tilde{Q}(\hat \Omega, t, f; H) &=& \lambda(\Omega, t) \frac{H(f) \gamma^* (\Omega, t,f)}{S_1(t; |f|) S_2(t;|f|)},
\end{eqnarray}

\noindent and $\lambda(\Omega,t) = \lambda T$, absorbing $\lambda$ in Equation~\ref{radiometer-directed-y}.

Absent a $\Phi$ model, radiometer must equally sum frequency bin contributions over $\Delta f \geq \Delta f_\mathrm{obs}$ (Equation~\ref{delta-f-obs}).
This width means $\Xi \approx 1$ and $X_1 = \sum_K \sum_k \tilde{x}_{Kk}$, $X_2 = \sum_L \sum_l \tilde{x}_{Ll}$ (referring to Equation~\ref{normalized-z-bin}; this is imprecise when radiometer uses overlapping, windowed bins~\cite{ThraneStochastic2009} and the cross-correlation method uses non-overlapping rectangular bins).
Moreover, $S_1 = S_K$, $S_2 = S_L$, $T = \Delta t$.

Compare to looking for an isolated point with no other sources and refer to the discussion following Equation 3.36 of~\cite{MitraRadiometer2008}.
If the stochastic background is taken as constant in frequency, $H^2(f)=1$, $\lambda$ simplifies (integrating over frequency and substituting Equation 3.34 of~\cite{MitraRadiometer2008} as directed for network power $P_{NW}^2$),

\begin{eqnarray}
\lambda(t) &\approx& [\Delta t P^2_{NW}(t)]^{-1},
\end{eqnarray}
\begin{eqnarray}
[\lambda(\Omega, t)\Delta t]^{-1}
&\approx&\frac{5}{T_\mathrm{sft}} \frac{1}{\sqrt{S_K S_L}} \hat{\Gamma}^\mathrm{ave}_{KL},
\\
\lambda(\Omega, t) &=& \frac{\sqrt{S_K S_L}}{5}\frac{1}{\hat \Gamma^\mathrm{ave}_{KL}},\\
N  
  &=& \frac{5}{\sqrt{2}} \bar \lambda(\Omega),
\end{eqnarray}

\noindent where $N$ is the cross-correlation method's normalization and $\bar \lambda (\Omega)$ is harmonic root mean square radiometer normalization.

In that case,
after all substitutions and considering the cross-correlation method's $\rho$ evaluated over all bins and only between the same SFT pairs as radiometer,

\begin{eqnarray}
\rho 
 &\approx& 4\sqrt{2} \Re \left(Y_{\Omega '}\right),
\end{eqnarray}

\noindent by taking sums over cross-correlation method indices $K$ and $L$ to produce radiometer $S_1$ and $S_2$.
Exact equality results for a single pair, such as the fully-coherent, cross-detector-only $\rho$.
This conclusion bolsters Whelan \textit{et al}~\cite{ScoX1CrossCorr2015PRD} (notably Section III.D), stating that the cross-correlation method is similar to the radiometer with a phase model to allow different-time correlations.

The cross-correlation method, the radiometer, and the $\mathcal{F}$-statistic, which all are described as near-\textit{optimal} under different conditions, do converge in certain limits.
Understanding the cross-correlation method's intersections aids theory and practice.

In theory, viewing $\mathcal{F}$ as approximating the Bayesian $\mathcal{B}$-statistic~\cite{BStatPrix2009} informs $\rho$ as an approximate function of the likelihood ratio~\cite{Dhurandhar2008}.
This perspective might facilitate Bayesian model selection for vetoes using alternative line hypotheses to compare against the signal hypothesis~\cite{KeitelRobust2014}.
It may also link search set-up optimization for detection probability to rigorous statements about posterior probability~\cite{MingSetup2015}.
Radiometric techniques might generate a deconvolved sky map of future detections~\cite{MitraRadiometer2008,ThraneStochastic2009}.

This paper should resolve confusion about the cross-correlation method.
It does not use cross-detector data as its template.
The cross-correlation method is a matched-filter-based semicoherent search, with the template corresponding to the signal model with chosen amplitude parameters, searched over the Doppler parameters.
It differs in which filtered data are conjugated in the real-valued statistic.

In practice, at present, ties between the statistics help bring resampling from the $\mathcal{F}$-statistic into the cross-correlation method.
Resampling solves the problem that $\Phi(t)$, particularly the time-varying $\phi(t)$, is not increasing at uniform frequency $f_0$, because of the time-varying Doppler shifts.
If Doppler modulation were constant, then a Fourier transform could supply $F_a$ (or the cross-correlation method's components $F_{a_K}$ and $F_{a_L}$) and $F_b$, providing an entire frequency band.
The data must be moved into the \textit{source frame}, in which velocity with respect to the source is constant.

\section{Downsampling and heterodyning\label{downsampling-and-heterodyning}}

Section~\ref{resampling} is done with downsampled data, heterodyned downwards in frequency by $f_h$.
Consider a bandpass-limited sample (subscript $p$) of Short Fourier Transform data for SFT $K$, equivalent to a rectangular frequency-domain window with starting bin $k_a$ and ending bin $k_b$.
Gaps in the set of SFTs are zero-padded to yield $M = T_\mathrm{obs}/T_\mathrm{sft}$ .
Equation~\ref{inverse-fft-eq} says that the time index with respect to SFT start time is $j$, and with respect to the observation run is $q_K$.

The $q_K$ are non-overlapping integers from $0$ to $MN -1$, whereas $j$ (implicitly depending on $K$) range from $0$ to $N-1$.
Start times and SFT durations are integer multiples of the sampling time, $t_K - T_\mathrm{sft}/2 \equiv K T_\mathrm{sft}$, and
$j \delta t = q_K \delta t - t_K  + T_\mathrm{sft}/2$.

The ideal \textit{bandpassed} data $x_{K,p}$ from an inverse Fourier transform of the whole $T_\mathrm{obs}$ would be,

\begin{eqnarray}
x_{K,p}(q_K\delta t) &\equiv& \sum_{k=k_a}^{k = k_b} e^{\mathrm{i} 2\pi q_K \delta t \frac{k}{T_\mathrm{sft} }  }\frac{z_{Kk}}{T_\mathrm{sft}},\\
     & = & \sum_{k=k_a}^{k = k_b} e^{\mathrm{i} 2\pi j \delta t \frac{k}{T_\mathrm{sft} }  }\frac{z_{Kk}}{T_\mathrm{sft}},\nonumber
\end{eqnarray}

\noindent
because $t_K - T_\mathrm{sft}/2$ is an integer.
When $k_a = 0$, $k_b = N-1$, $x_{K,p}$ is equivalent to $x_K$ in Equation~\ref{inverse-fft-eq}. 
Yet we want not simply bandpassed data, but downsampled, heterodyned data.
Since $M>1$, we handle the sum over $K$.
The difficulty is keeping phase coherence between inverse Fourier transforms.

\textit{Heterodyne frequency} $f_h$ is in the center of the band, near central bin $k_h \equiv (k_a + k_b)/2$.
For discrete bins, the nearest frequency $\bar f_h \equiv k_h T_\mathrm{sft}^{-1}$.
The frequency $\bar f_h = f_h - f_r$ is of the nearest integer bin to the ideal heterodyne $f_h$, where $f_r$ is the remainder.

Let $l \equiv k - k_h$, so $k = l + k_h$:

\begin{eqnarray}
x_{K,p}(q_K\delta t)
  &=& \sum_{(l+k_h)=k_a}^{(l+k_h) = k_b} e^{\mathrm{i} 2\pi j \delta t \frac{(l+k_h)}{T_\mathrm{sft} }  }\frac{z_{K(l+k_h)}}{T_\mathrm{sft}} \label{bandpassed-data-eq}\\
  &=& e^{\mathrm{i} 2\pi j \delta t \frac{k_h}{T_\mathrm{sft} }  } \sum_{l= k_a - k_h}^{l = k_b - k_h} e^{\mathrm{i} 2\pi j \delta t \frac{l}{T_\mathrm{sft} }  }\frac{z_{K(l+k_h)}}{T_\mathrm{sft}}\nonumber.
\end{eqnarray}

\noindent
The sum contains all information on $[k_a,k_b]$. 
Call it $x^h_K$:


\begin{eqnarray}
x^h_K (q_K \delta t) &\equiv&
\sum_{l= k _a - k_h}^{l = k_b - k_h } e^{\mathrm{i} 2\pi j \delta t \frac{l}{T_\mathrm{sft} }  }\frac{z_{K(l+k_h)}}{T_\mathrm{sft}}, \label{freq-shifted-eq}\\
x_{K,p}(q_K\delta t) &=& e^{i 2\pi j \delta t \frac{k_h}{T_\mathrm{sft}}} x^h_K(q_K\delta t),\label{up-to-phase-het-corr}
\end{eqnarray}

\noindent expressing bandpassed $x_{K,p}$ in terms of the desired, frequency-shifted $x^h_K$.
In continuous time, Equation~\ref{up-to-phase-het-corr} is the expression $x_{K,p}(t) = \exp{(i2\pi f_h t)} x^h_K(t)$, where $x_{K,p}$ is the bandpassed data (frequency content at $f$) and $x^h_K$ is the heterodyned data (frequency content at $f-f_h$).
Many derivations stop here, but we need the phase corrections for heterodyning a set of SFTs.

To represent \textit{complex, downsampled} data in a frequency band $f_\mathrm{band}$ without aliasing, we need a total bandwidth of $\Delta f_\mathrm{load}$.
Note that $\Delta f_\mathrm{load}$ must cover not only all frequencies of interest but also frequency modulation's Doppler wings, $\Delta f_\mathrm{drift}$, with additional bins to account for spectral leakage, including $D$ `Dirichlet terms'.
The total width $\Delta f_\mathrm{load}$ is~\cite{PrixTimingModel2017},

\begin{equation}
\Delta f_\mathrm{load} = \left(1+\frac{4}{2 D +1} \right)\left(f_\mathrm{band} + \Delta f_\mathrm{drift} + \frac{16}{T_\mathrm{sft}}\right).
\label{delta-f-load}
\end{equation}

\noindent
Then we find the new sampling time interval is not $\delta t$ but rather $\delta t' = 1/\Delta f_\mathrm{load}$.
The old number of samples in an SFT is $N = T_\mathrm{sft}/(\delta t)$ and the new number is $N' = \Delta f_\mathrm{load} T_\mathrm{sft}$; $N' \leq N$.
($N'$ can be rounded up to ensure it is an integer).
Create the new coordinate $q_K'$, so $t = q_K' \delta t'$.
The Fourier transform kernel must contain an integer, and
$q_K' \approx (t_K - T_\mathrm{sft}/2 + j \delta t)/(\delta t')$ is not generally an integer.
Additional phase corrections thus arise.
Note,

\begin{eqnarray}
q_K' &=& \Delta f_\mathrm{load} q_K \delta t,\\
q_K' k / N' &=& (q_K \delta t) k / T_\mathrm{sft},
\end{eqnarray}

Meanwhile we can choose $k_a$ and $k_b$ with a difference $k_b - k_a = \Delta f_\mathrm{load} T_\mathrm{sft}$, ergo $k_b - k_a = N'$:

\begin{eqnarray}
k_a &=& \left(f_h - \frac{1}{2} \Delta f_\mathrm{load}\right) T_\mathrm{sft},\\
k_b &=& \left(f_h + \frac{1}{2} \Delta f_\mathrm{load}\right) T_\mathrm{sft}.
\end{eqnarray}

In practice, we will use the minimum frequency of interest, $f_a = f_\mathrm{min}$, to choose a heterodyne frequency $f_h = f_\mathrm{min} + \frac{1}{2} f_\mathrm{band}$.

As $t = q_K \delta t$, we can substitute $q_K' \delta t'$ into the argument of $x_{K,p}(q_K \delta t)$ as defined in Equation~\ref{bandpassed-data-eq}.
Using $N'$,
%

\begin{eqnarray}
x_{K,p}(q_K' \delta t')  &=& \sum_{l=k_a - k_h}^{l=k_b - k_h} e^{\mathrm{i} 2\pi q_K' \frac{l+k_h}{2 \Delta f_\mathrm{load} T_\mathrm{sft} }  }\frac{z_{K(l + k_h)}}{T_\mathrm{sft}}, \label{first-shift-approx}\\
  &=& e^{\mathrm{i} 2\pi q_K' k_h / N' } \sum_{l=k_a - k_h}^{l=k_b - k_h} e^{\mathrm{i} 2\pi q_K' l / N' }  \frac{z_{K(l+k_h)}}{T_\mathrm{sft}}, \nonumber
\end{eqnarray}

\noindent
We need to break apart $q_K'$ in the sum:

\begin{eqnarray}
q_K' /N' 
 &=& \frac{t_K - T_\mathrm{sft}/2}{ T_\mathrm{sft}} + \frac{j \delta t}{ T_\mathrm{sft} },
\end{eqnarray}

\noindent where again, because $(t_K - T_\mathrm{sft}/2)$ is always an integer multiplied by integer $l$, the first term evaluates to unity in the sum exponent.
In $q_K' k_h/N'$, however, though $k_h$ is also integer, we leave the term so we can see the effect of approximating $\bar f_h$.
We find,

\begin{eqnarray}
x_{K,p}(q_K' \delta t') 
 &=& e^{\mathrm{i} 2\pi q_K \delta t \frac{k_h}{T_\mathrm{sft} }} x_K^h(q_K \delta t).
 \label{eq-with-broken-sum}
\end{eqnarray}

\noindent This result concords with Equation~\ref{up-to-phase-het-corr}.
Considering $\bar f_h$,

\begin{eqnarray}
x_{K,p}(q_K' \delta t') 
 &=& e^{\mathrm{i} 2\pi q_K \delta t (f_h - f_r)} x_K^h(q_K \delta t),
\end{eqnarray}

\noindent
where an approximation is used for this Appendix,

\begin{eqnarray}
x_{K,p}(q_K' \delta t')
 &\approx& e^{\mathrm{i} 2\pi q_K \delta t f_h } x_K^h(q_K \delta t).
\label{heterodyne-shift-approx}
\end{eqnarray}

\noindent
Generally the code will have access to $f_h$ but not $k_h$; the remainder $f_r$ is fixed by later by rounding to the nearest bin (in the paper body, $f_r^*$).

Next, we seek $x_K^h$ in downsampled time.
Our goal is $x^h$ covering all the observing time, but we must go through $x_{K,p}$ to preserve the phase shifts between SFTs.
For the single-SFT case, we could just substitute $q_K' \delta t'$ into the argument for $x_K^h$ and be done.

Notice that $k_a - k_h = -N'/2$, $k_b - k_h = +N'/2 - 1$ (for an even number of samples including 0).
For any point in time, comparison with Equation~\ref{freq-shifted-eq} shows,

\begin{eqnarray} 
x_K^h (q_K \delta t) 
 &=& \sum_{l=-N'/2}^{l=N'/2 - 1} e^{\mathrm{i} 2\pi j (\delta t/\delta t') \frac{l}{N'} }  \frac{z_{K(l+k_h)}}{T_\mathrm{sft}},
\label{sum-for-heterodyne}
\end{eqnarray}

\noindent we could define the generally non-integer $j' = j (\delta t/\delta t')$; fortunately, $j'/N' = j/N$.
The exponent is then $\exp{(\mathrm{i} 2 \pi j' l /N)}$.

For a detour, note that $x_K^h$ is almost fit for a Fourier transform, but it requires an index shift.
Periodicity in the Fourier transform means that any substitution $j k \rightarrow j k +  Q N$ for a transform with time steps $j$, frequency steps $k$, and number of samples $N$, by integer $Q$, leaves the result invariant.
For half-integer $Q$, the substitution moves positive frequencies into negative frequencies (increasing in the same direction as before) and vice versa. 
Choose new index $m \equiv l + N'/2$, so $l = m - N'/2$:

\begin{eqnarray}
x_K^h (q_K \delta t)
&=& (-1)^{-j'}  \sum_{m=0}^{m=N - 1} B(0,N') \nonumber \\
&& \times e^{\mathrm{i} 2\pi j m / N } \frac{z_{K(m+k_h-N'/2)}}{T_\mathrm{sft}},\label{shifted-for-fft}
\end{eqnarray}

\noindent where, for illustration, $B(0, N')$ is the Boxcar function, acting as a bandpass.
This $x^h_K$ is at the full sampling rate and is only theoretical.
The sum term is a straightforward inverse Fourier transform, from $m$ to $j$, of the $z_K$ data from frequency bins $k_h - N'/2$ to $k_h + N'/2 -1$.
In practice, the $(-1)^{-j'}$ factor (the move from positive to negative frequencies) depends into the conventions of Fast Fourier Transform programs.
Care is required to ensure the right convention.
For us, the interface with the \textit{FFTW} library absorbs this factor.
We will use this Fourier transform after constructing the time series.


To construct the full time-series for the entire observing run, use the time-shift Equation~\ref{heterodyne-shift-approx} for $x_{K,p}$ and Equation~\ref{sum-for-heterodyne} for $x_K^h$, noting that $\bar f_h \approx f_h$:

\begin{eqnarray}
\exp{(i2\pi f_h q_K \delta t)} 
&\approx& e^{i2\pi f_h [t_K - T_\mathrm{sft}/2 + j' k_h/N']},
\end{eqnarray}

\noindent
whereby the \textit{frequency-shifted, heterodyned, downsampled} $x_K^h(q_K' \delta t')$ has the SFT start time phase shift with respect to $x_K^h(q_K \delta t)$:

\begin{eqnarray}
x_K^h(q_K' \delta t') = e^{-i2\pi f_h [t_K - \frac{T_\mathrm{sft}}{2} + \frac{j' k_h}{N'}]} x_{K,p}(q'_K \delta t').
\end{eqnarray}

The $j' k_h/N'$ can be absorbed into the bandpassing by a change of index, providing a quantity amenable to an FFT.
Returning to Equation~\ref{bandpassed-data-eq} for $x_{K,p}(q_K \delta t)$, which equals $x_{K,p}(q_K' \delta t')$ at equal times $t$:

\begin{eqnarray}
x_K^h(q_K \delta t) i
 &=& e^{-i2\pi f_h [t_K - \frac{T_\mathrm{sft}}{2}]} \\
 &&\times\sum_{l=-N'/2}^{N'/2+1}  e^{\mathrm{i} 2\pi \left(\frac{j' (l+k_h)}{N'} - \frac{j' k_h}{N'}\right)} \frac{z_{K(l+k_h)}}{T_\mathrm{sft}}.\nonumber
\end{eqnarray}

\noindent Invoking Equation~\ref{shifted-for-fft},

\begin{eqnarray}
x_K^h(q_K' \delta t') &=& e^{-i2\pi f_h [t_K - \frac{T_\mathrm{sft}}{2}]} (-1)^{-j'} \\
&& \times \sum_{m=0}^{N'-1} e^{\mathrm{i} 2\pi j' m / N'} \frac{z_{K(m+k_h - N'/2)} }{T_\mathrm{sft}}  \nonumber
\end{eqnarray}

While for arbitrary $q_K$, $j'$ is not an integer, the downsampled time-series $q_K'$ is specifically chosen for times where it is.
Then the sum is indeed an inverse discrete Fourier transform of bandpassed data (which by itself is $x_{K,p}$), but also shifted by $k_h$.
Including the negative-frequency sign convention with $(-1)^{-j'}$, call this $x_{K,s}$:

\begin{eqnarray}
x_{K,s}(q'_K \delta t') &\equiv& (-1)^{-j'} \\
 &&  \times \sum_{m=0}^{N'-1} e^{\mathrm{i} 2\pi j' m / N'} \frac{z_{K(m+k_h - N'/2)} }{T_\mathrm{sft}} \nonumber\\
x_K^h(q_K' \delta t') &=& e^{-i2\pi f_h [t_K - \frac{T_\mathrm{sft}}{2}]} x_{K,s}(q_K' \delta t').
\end{eqnarray}

\noindent This result for the exponent depends on the heterodyne frequency $f_h$ and SFT mid-time $t_K$ but not the index $q_K$.
In comparison with Equation~\ref{heterodyne-shift-approx}, the index $j \delta t$ has been absorbed.
So it is generally true of any time $t = q_K \delta t$, including $t = q_K' \delta t'$.
\noindent Carefully note, however, that $x_K^h$ is \textit{still} heterodyned in the sense that a Fourier transform will yield the spectrum shifted by $f_h$.
All the correction has done is shift the phase so that different SFTs are in-phase.
We now use this alignment to construct the complete time series from the SFTs.

Since $q_K$ is distinct for the entire time series, that series of $x_{q'}^h\equiv x^h(q'\delta t')$ is the sum (neglecting windowing),

\begin{eqnarray}
x^h(q' \delta t') 
  &=& \sum_{K=0}^M e^{-\mathrm{i}2\pi f_h [t_K - T_\mathrm{sft}/2]} x_{K,s}(q_K' \delta t').
\label{long-time-series-het-corrected}
\end{eqnarray}

\noindent In practice, the quantity $x_{K,s}(q_K' \delta t')$ is computed from the inverse Fourier transform of a band of data centered around $f_h$ with bandwidth $f_\mathrm{band}$, so Equation~\ref{long-time-series-het-corrected} is the simplest construction of the complete downsampled time series.
Again, any signal at frequency $f_0$ in $x$ is at $f_0 - f_h$ in $x'$.
Downsampling also reduces the computation cost of interpolating into the BB frame.

\section{Interpretation and degeneracies of statistic\label{stat-interp-rho}}

Several properties of the $\rho$ statistic should be noted that do not neatly fit into the main text.
In the fully-coherent limit, just as the $\mathcal{F}$-statistic is proportional to the log-likelihood ratio of a sinusoidal waveform hypothesis compared to Gaussian noise~\cite{Jaranowski1998}, so too should the $\rho$ statistic be interpreted.
In this limit, the set of output $\rho(f_0,\lambda)$ from a search constitutes a sampling of the likelihood surface.
This likelihood surface is amenable to composite hypothesis testing, as well as Bayesian interpretation~\cite{BStatPrix2009}.

Locally, the `likelihood surface' of $\rho$ is well-described by the metric approximation~\cite{ScoX1CrossCorr2015PRD}.
Globally, long-range degeneracies appear.
Degeneracies step mainly from surfaces of $d\Phi = 0$ in the phase model, Equation~\ref{phase-model-eq}.
In the $(f_0, a_p, T_\mathrm{asc})$ space, these degeneracies form a cone along the $f_0$ axis, with the vertex at the maximum $\rho$.
The surface of the cone arises from the largest component of the set of sidebands from residual phase modulation when $(a_p, T_\mathrm{asc})$ are offset from their true values.
This surface has been noted elsewhere in cross-section as a 2-dimensional $X$ shape, for example in the $(f_0, a_p)$ plane~\cite{SidebandMarkovModelSuvorova2016,MeadorsDirectedMethods2016}.
Because this extended surface correlates neighboring templates, na\"{i}ve division by a trials factor equal to the number of templates (Bonferroni correction) may yield an overly-conservative $p$-value.
The metric may also be too conservative for high values of mismatch~\cite{WettePRD2016}.

Semicoherent statistics such as $\rho$ grow proportionally to $(T_\mathrm{obs} T_\mathrm{max})^{1/4}$, and they also grow proportionally to $h_0^2$.
This is in contrast to fully-coherent statistics, which take $T_\mathrm{max} = T_\mathrm{obs}$ and therefore grow proportionally to $T_\mathrm{obs}^{1/2}$.
However, another class of power-based statistics, such as the `TwoSpect' method~\cite{GoetzTwoSpectMethods2011}, also grows as $T_\mathrm{obs}^{1/4}$ but, differently from the semicoherent case, as $h_0^4$.
GW phase coherence is not used over timescales longer than one SFT in these power-based statistics, and the final statistic depends on the power of a second FFT, over the orbital cycle.

The cross-correlation method's code must calculate $\rho$ as efficiently as possible in a sample of the likelihood surface that does not miss its peak.
Viewed as semi-coherent choice of the weights matrix $\textbf{W}$, the goal is the calculate the largest number of elements of the weights matrix for the lowest cost.
Skipping the auto-correlation in our code comes at the cost of the statistic contribution from that element.
Avoidance of auto-correlation is natural from the standpoint of the \textit{Radiometer}, which only permits same-time correlations and has no signal model.
For the radiometer, auto-correlation would contaminate the search with the noise of the detector.
From the standpoint of the $\mathcal{F}$-statistic, it is conversely natural to include the auto-correlation, because it fits in the middle of an FFT.
Capturing the adjacent elements of the weights matrix from the cross-correlation method with an FFT requires additional overlap of a factor of $T_\mathrm{coh}/T_\mathrm{short} \geq 3$.
It should be determined whether the cost of this overlap is worth the exclusion of noise (and signal) contributions from the auto-correlation.

\section{Merits of the cross-correlation method in noisy data\label{glitch-merits-of-crosscorr}}

The cross-correlation method, unlike the $\mathcal{F}$-statistic but like the \textit{Radiometer} method, avoids auto-correlation by default.
Consider the presence of some sine-Gaussian glitch in the data that might justify this avoidance:

\begin{equation}
g(t) = A e^{-(t-t_0)^2/(2\sigma^2)} \sin{\omega t - \phi_0}.
\end{equation}

\noindent In the Fourier domain in which the cross-correlation method computes its statistic, the Fourier transform of $g(t)$, $\tilde{g}(f)$, is the convolution of the Fourier transforms of the Gaussian and sinusoidal terms, which are respectively also Gaussian and a Dirac delta function.
The glitch does contribute noise in a Gaussian frequency distribution around the frequency $\omega$, with amplitude proportional to $A$.
By removing the auto-correlation, such glitches will never correlate with themselves.
Assuming that $\omega$ and $t_0$ are randomly-distributed, they will be unlikely to correlate with other glitches at different times.
Therefore, the noise background of the cross-correlation method could conceivably be lower.

Empirically, values of $\rho$ and $\mathcal{F}$ appear similar for comparable noise and signal strength.
Whether the theoretically lower background of the cross-correlation method holds in real data is an important test.
If the two statistics recover signals comparably well for the same coherent integration time, then whichever calculates a given coherence time most efficiently is best.
This paper has established a path between the two methods.

\bibliography{bibliography.bib}

\begin{thebibliography}{77}%
\makeatletter
\providecommand \@ifxundefined [1]{%
 \@ifx{#1\undefined}
}%
\providecommand \@ifnum [1]{%
 \ifnum #1\expandafter \@firstoftwo
 \else \expandafter \@secondoftwo
 \fi
}%
\providecommand \@ifx [1]{%
 \ifx #1\expandafter \@firstoftwo
 \else \expandafter \@secondoftwo
 \fi
}%
\providecommand \natexlab [1]{#1}%
\providecommand \enquote  [1]{``#1''}%
\providecommand \bibnamefont  [1]{#1}%
\providecommand \bibfnamefont [1]{#1}%
\providecommand \citenamefont [1]{#1}%
\providecommand \href@noop [0]{\@secondoftwo}%
\providecommand \href [0]{\begingroup \@sanitize@url \@href}%
\providecommand \@href[1]{\@@startlink{#1}\@@href}%
\providecommand \@@href[1]{\endgroup#1\@@endlink}%
\providecommand \@sanitize@url [0]{\catcode `\\12\catcode `\$12\catcode
  `\&12\catcode `\#12\catcode `\^12\catcode `\_12\catcode `\%12\relax}%
\providecommand \@@startlink[1]{}%
\providecommand \@@endlink[0]{}%
\providecommand \url  [0]{\begingroup\@sanitize@url \@url }%
\providecommand \@url [1]{\endgroup\@href {#1}{\urlprefix }}%
\providecommand \urlprefix  [0]{URL }%
\providecommand \Eprint [0]{\href }%
\providecommand \doibase [0]{http://dx.doi.org/}%
\providecommand \selectlanguage [0]{\@gobble}%
\providecommand \bibinfo  [0]{\@secondoftwo}%
\providecommand \bibfield  [0]{\@secondoftwo}%
\providecommand \translation [1]{[#1]}%
\providecommand \BibitemOpen [0]{}%
\providecommand \bibitemStop [0]{}%
\providecommand \bibitemNoStop [0]{.\EOS\space}%
\providecommand \EOS [0]{\spacefactor3000\relax}%
\providecommand \BibitemShut  [1]{\csname bibitem#1\endcsname}%
\let\auto@bib@innerbib\@empty
\bibitem [{\citenamefont {Abbott}\ \emph {et~al.}(2016)\citenamefont {Abbott}
  \emph {et~al.}}]{GW150914LIGO}%
  \BibitemOpen
  \bibfield  {author} {\bibinfo {author} {\bibfnamefont {B.P}\ \bibnamefont
  {Abbott}} \emph {et~al.} (\bibinfo {collaboration} {LIGO Scientific
  Collaboration and Virgo Collaboration}),\ }\href {\doibase
  10.1103/PhysRevLett.116.061102} {\bibfield  {journal} {\bibinfo  {journal}
  {Phys. Rev. Lett.}\ }\textbf {\bibinfo {volume} {116}},\ \bibinfo {pages}
  {061102} (\bibinfo {year} {2016})}\BibitemShut {NoStop}%
\bibitem [{\citenamefont {Brady}\ \emph {et~al.}(1998)\citenamefont {Brady},
  \citenamefont {Creighton}, \citenamefont {Cutler},\ and\ \citenamefont
  {Schutz}}]{Brady1998}%
  \BibitemOpen
  \bibfield  {author} {\bibinfo {author} {\bibfnamefont {P.R.}\ \bibnamefont
  {Brady}}, \bibinfo {author} {\bibfnamefont {T.}~\bibnamefont {Creighton}},
  \bibinfo {author} {\bibfnamefont {C.}~\bibnamefont {Cutler}}, \ and\ \bibinfo
  {author} {\bibfnamefont {B.F.}\ \bibnamefont {Schutz}},\ }\bibfield  {title}
  {\enquote {\bibinfo {title} {Searching for periodic sources with {LIGO}},}\
  }\href@noop {} {\bibfield  {journal} {\bibinfo  {journal} {Phys. {R}ev. {D}}\
  }\textbf {\bibinfo {volume} {57}},\ \bibinfo {pages} {2101} (\bibinfo {year}
  {1998})}\BibitemShut {NoStop}%
\bibitem [{\citenamefont {Abbott}\ \emph
  {et~al.}(2017{\natexlab{a}})\citenamefont {Abbott} \emph
  {et~al.}}]{ScoX1CrossCorr2017ApJO1}%
  \BibitemOpen
  \bibfield  {author} {\bibinfo {author} {\bibfnamefont {B.P}\ \bibnamefont
  {Abbott}} \emph {et~al.},\ }\bibfield  {title} {\enquote {\bibinfo {title}
  {Upper limits on gravitational waves from {S}corpius {X-1} from a model-based
  cross-correlation search in {A}dvanced {LIGO} data},}\ }\href {\doibase
  10.3847/1538-4357} {\bibfield  {journal} {\bibinfo  {journal} {Astrophys
  {J}}\ }\textbf {\bibinfo {volume} {847}},\ \bibinfo {pages} {47} (\bibinfo
  {year} {2017}{\natexlab{a}})}\BibitemShut {NoStop}%
\bibitem [{\citenamefont {Papaloizou}\ and\ \citenamefont
  {Pringle}(1978)}]{PapaloizouPringle1978}%
  \BibitemOpen
  \bibfield  {author} {\bibinfo {author} {\bibfnamefont {J.}~\bibnamefont
  {Papaloizou}}\ and\ \bibinfo {author} {\bibfnamefont {J.E.}\ \bibnamefont
  {Pringle}},\ }\bibfield  {title} {\enquote {\bibinfo {title} {Gravitational
  radiation and the stability of rotating stars},}\ }\href@noop {} {\bibfield
  {journal} {\bibinfo  {journal} {{MNRAS}}\ }\textbf {\bibinfo {volume}
  {184}},\ \bibinfo {pages} {501} (\bibinfo {year} {1978})}\BibitemShut
  {NoStop}%
\bibitem [{\citenamefont {Wagoner}(1984)}]{Wagoner1984}%
  \BibitemOpen
  \bibfield  {author} {\bibinfo {author} {\bibfnamefont {R.V.}\ \bibnamefont
  {Wagoner}},\ }\bibfield  {title} {\enquote {\bibinfo {title} {Gravitational
  radiation from accreting neutron stars},}\ }\href@noop {} {\bibfield
  {journal} {\bibinfo  {journal} {{A}p. {J}.}\ }\textbf {\bibinfo {volume}
  {278}},\ \bibinfo {pages} {345} (\bibinfo {year} {1984})}\BibitemShut
  {NoStop}%
\bibitem [{\citenamefont {Bildsten}(1998)}]{Bildsten1998}%
  \BibitemOpen
  \bibfield  {author} {\bibinfo {author} {\bibfnamefont {L.}~\bibnamefont
  {Bildsten}},\ }\bibfield  {title} {\enquote {\bibinfo {title} {Gravitational
  radiation and rotation of accreting neutron stars},}\ }\href {\doibase
  10.1086/311440} {\bibfield  {journal} {\bibinfo  {journal} {Astrophys. {J}.
  {L}ett.}\ }\textbf {\bibinfo {volume} {501}},\ \bibinfo {pages} {L89}
  (\bibinfo {year} {1998})}\BibitemShut {NoStop}%
\bibitem [{\citenamefont {Giacconi}\ \emph {et~al.}(1962)\citenamefont
  {Giacconi}, \citenamefont {Gursky}, \citenamefont {Paolini},\ and\
  \citenamefont {Rossi}}]{Giacconi1962}%
  \BibitemOpen
  \bibfield  {author} {\bibinfo {author} {\bibfnamefont {R.}~\bibnamefont
  {Giacconi}}, \bibinfo {author} {\bibfnamefont {H.}~\bibnamefont {Gursky}},
  \bibinfo {author} {\bibfnamefont {F.R.}\ \bibnamefont {Paolini}}, \ and\
  \bibinfo {author} {\bibfnamefont {B.B.}\ \bibnamefont {Rossi}},\ }\bibfield
  {title} {\enquote {\bibinfo {title} {Evidence for {X} rays from sources
  outside the solar system},}\ }\href@noop {} {\bibfield  {journal} {\bibinfo
  {journal} {Phys. {R}ev. {L}ett.}\ }\textbf {\bibinfo {volume} {9}},\ \bibinfo
  {pages} {439} (\bibinfo {year} {1962})}\BibitemShut {NoStop}%
\bibitem [{\citenamefont {Shawhan}(2010)}]{Shawhan2010}%
  \BibitemOpen
  \bibfield  {author} {\bibinfo {author} {\bibfnamefont {P.}~\bibnamefont
  {Shawhan}},\ }\bibfield  {title} {\enquote {\bibinfo {title}
  {Gravitational-wave astronomy: observational results and their impact},}\
  }\href@noop {} {\bibfield  {journal} {\bibinfo  {journal} {Class. {Q}uant.
  {G}rav.}\ }\textbf {\bibinfo {volume} {27}},\ \bibinfo {pages} {084017}
  (\bibinfo {year} {2010})}\BibitemShut {NoStop}%
\bibitem [{\citenamefont {Owen}(2010)}]{Owen2010}%
  \BibitemOpen
  \bibfield  {author} {\bibinfo {author} {\bibfnamefont {B.J.}\ \bibnamefont
  {Owen}},\ }\bibfield  {title} {\enquote {\bibinfo {title} {How to adapt
  broad-band gravitational-wave searches for $r$-modes},}\ }\href@noop {}
  {\bibfield  {journal} {\bibinfo  {journal} {Phys. {R}ev. {D}}\ }\textbf
  {\bibinfo {volume} {82}},\ \bibinfo {pages} {104002} (\bibinfo {year}
  {2010})}\BibitemShut {NoStop}%
\bibitem [{\citenamefont {Andersson}(1998)}]{Andersson1998}%
  \BibitemOpen
  \bibfield  {author} {\bibinfo {author} {\bibfnamefont {N.}~\bibnamefont
  {Andersson}},\ }\bibfield  {title} {\enquote {\bibinfo {title} {A new class
  of unstable modes of rotating relativistic stars},}\ }\href@noop {}
  {\bibfield  {journal} {\bibinfo  {journal} {APJ}\ }\textbf {\bibinfo {volume}
  {502}},\ \bibinfo {pages} {708} (\bibinfo {year} {1998})}\BibitemShut
  {NoStop}%
\bibitem [{\citenamefont {Friedman}\ and\ \citenamefont
  {Morsink}(1998)}]{Friedman1998}%
  \BibitemOpen
  \bibfield  {author} {\bibinfo {author} {\bibfnamefont {J.L}\ \bibnamefont
  {Friedman}}\ and\ \bibinfo {author} {\bibfnamefont {S.M.}\ \bibnamefont
  {Morsink}},\ }\bibfield  {title} {\enquote {\bibinfo {title} {Axial
  instability of rotating relativistic stars},}\ }\href@noop {} {\bibfield
  {journal} {\bibinfo  {journal} {APJ}\ }\textbf {\bibinfo {volume} {502}},\
  \bibinfo {pages} {714} (\bibinfo {year} {1998})}\BibitemShut {NoStop}%
\bibitem [{\citenamefont {Owen}\ \emph {et~al.}(1998)\citenamefont {Owen},
  \citenamefont {Lindblom}, \citenamefont {Cutler}, \citenamefont {Schutz},
  \citenamefont {Vecchio},\ and\ \citenamefont {Andersson}}]{Owen1998}%
  \BibitemOpen
  \bibfield  {author} {\bibinfo {author} {\bibfnamefont {B.J.}\ \bibnamefont
  {Owen}}, \bibinfo {author} {\bibfnamefont {L.}~\bibnamefont {Lindblom}},
  \bibinfo {author} {\bibfnamefont {C.}~\bibnamefont {Cutler}}, \bibinfo
  {author} {\bibfnamefont {B.F.}\ \bibnamefont {Schutz}}, \bibinfo {author}
  {\bibfnamefont {A.}~\bibnamefont {Vecchio}}, \ and\ \bibinfo {author}
  {\bibfnamefont {N.}~\bibnamefont {Andersson}},\ }\bibfield  {title} {\enquote
  {\bibinfo {title} {Gravitational waves from hot young rapidly rotating
  neutron stars},}\ }\href@noop {} {\bibfield  {journal} {\bibinfo  {journal}
  {Phys. {R}ev. {D}}\ }\textbf {\bibinfo {volume} {58}},\ \bibinfo {pages}
  {084020} (\bibinfo {year} {1998})}\BibitemShut {NoStop}%
\bibitem [{\citenamefont {Jaranowski}\ \emph {et~al.}(1998)\citenamefont
  {Jaranowski}, \citenamefont {Kr\'{o}lak},\ and\ \citenamefont
  {Schutz}}]{Jaranowski1998}%
  \BibitemOpen
  \bibfield  {author} {\bibinfo {author} {\bibfnamefont {P.}~\bibnamefont
  {Jaranowski}}, \bibinfo {author} {\bibfnamefont {A.}~\bibnamefont
  {Kr\'{o}lak}}, \ and\ \bibinfo {author} {\bibfnamefont {B.F.}\ \bibnamefont
  {Schutz}},\ }\bibfield  {title} {\enquote {\bibinfo {title} {Data analysis of
  gravitational-wave signals from spinning neutron stars: the signal and its
  detection},}\ }\href@noop {} {\bibfield  {journal} {\bibinfo  {journal}
  {Phys. {R}ev. {D}}\ }\textbf {\bibinfo {volume} {58}},\ \bibinfo {pages}
  {063001} (\bibinfo {year} {1998})}\BibitemShut {NoStop}%
\bibitem [{\citenamefont {Brady}\ and\ \citenamefont
  {Creighton}(2000)}]{HierarchicalBrady2000}%
  \BibitemOpen
  \bibfield  {author} {\bibinfo {author} {\bibfnamefont {P.R.}\ \bibnamefont
  {Brady}}\ and\ \bibinfo {author} {\bibfnamefont {T.}~\bibnamefont
  {Creighton}},\ }\bibfield  {title} {\enquote {\bibinfo {title} {Searching for
  periodic sources with {LIGO.} {II.} {H}ierarchical searches},}\ }\href@noop
  {} {\bibfield  {journal} {\bibinfo  {journal} {{P}hys. {R}ev. {D}}\ }\textbf
  {\bibinfo {volume} {61}},\ \bibinfo {pages} {082001} (\bibinfo {year}
  {2000})}\BibitemShut {NoStop}%
\bibitem [{\citenamefont {Patel}\ \emph {et~al.}(2010)\citenamefont {Patel},
  \citenamefont {Siemens}, \citenamefont {Dupuis},\ and\ \citenamefont
  {Betzwieser}}]{Patel:2009qe}%
  \BibitemOpen
  \bibfield  {author} {\bibinfo {author} {\bibfnamefont {Pinkesh}\ \bibnamefont
  {Patel}}, \bibinfo {author} {\bibfnamefont {Xavier}\ \bibnamefont {Siemens}},
  \bibinfo {author} {\bibfnamefont {Rejean}\ \bibnamefont {Dupuis}}, \ and\
  \bibinfo {author} {\bibfnamefont {Joseph}\ \bibnamefont {Betzwieser}},\
  }\bibfield  {title} {\enquote {\bibinfo {title} {{Implementation of
  barycentric resampling for continuous wave searches in gravitational wave
  data}},}\ }\href {\doibase 10.1103/PhysRevD.81.084032} {\bibfield  {journal}
  {\bibinfo  {journal} {{P}hys. {R}ev. {D}}\ }\textbf {\bibinfo {volume}
  {81}},\ \bibinfo {pages} {084032} (\bibinfo {year} {2010})},\ \Eprint
  {http://arxiv.org/abs/0912.4255} {arXiv:0912.4255 [gr-qc]} \BibitemShut
  {NoStop}%
\bibitem [{\citenamefont {Dhurandhar}\ \emph {et~al.}(2008)\citenamefont
  {Dhurandhar}, \citenamefont {Krishnan}, \citenamefont {Mukhopadhyay},\ and\
  \citenamefont {Whelan}}]{Dhurandhar2008}%
  \BibitemOpen
  \bibfield  {author} {\bibinfo {author} {\bibfnamefont {Sanjeev}\ \bibnamefont
  {Dhurandhar}}, \bibinfo {author} {\bibfnamefont {Badri}\ \bibnamefont
  {Krishnan}}, \bibinfo {author} {\bibfnamefont {Himan}\ \bibnamefont
  {Mukhopadhyay}}, \ and\ \bibinfo {author} {\bibfnamefont {John~T.}\
  \bibnamefont {Whelan}},\ }\bibfield  {title} {\enquote {\bibinfo {title}
  {Cross-correlation search for periodic gravitational waves},}\ }\href
  {\doibase 10.1103/PhysRevD.77.082001} {\bibfield  {journal} {\bibinfo
  {journal} {Phys. Rev. D}\ }\textbf {\bibinfo {volume} {77}},\ \bibinfo
  {pages} {082001} (\bibinfo {year} {2008})},\ \Eprint
  {http://arxiv.org/abs/0712.1578} {arXiv:0712.1578} \BibitemShut {NoStop}%
\bibitem [{\citenamefont {Chung}\ \emph {et~al.}(2011)\citenamefont {Chung},
  \citenamefont {Melatos}, \citenamefont {Krishnan},\ and\ \citenamefont
  {Whelan}}]{Chung2011}%
  \BibitemOpen
  \bibfield  {author} {\bibinfo {author} {\bibfnamefont {C.T.Y.}\ \bibnamefont
  {Chung}}, \bibinfo {author} {\bibfnamefont {A.}~\bibnamefont {Melatos}},
  \bibinfo {author} {\bibfnamefont {B.}~\bibnamefont {Krishnan}}, \ and\
  \bibinfo {author} {\bibfnamefont {J.T.}\ \bibnamefont {Whelan}},\ }\bibfield
  {title} {\enquote {\bibinfo {title} {Designing a cross-correlation search for
  continuous-wave gravitational radiation from a neutron star in the supernova
  remnant {SNR} 1987{A}},}\ }\href {\doibase 10.1111/j.1365-2966.2011.18585.x}
  {\bibfield  {journal} {\bibinfo  {journal} {MNRAS}\ }\textbf {\bibinfo
  {volume} {414}},\ \bibinfo {pages} {2650} (\bibinfo {year}
  {2011})}\BibitemShut {NoStop}%
\bibitem [{\citenamefont {Whelan}\ \emph {et~al.}(2015)\citenamefont {Whelan},
  \citenamefont {Sundaresan}, \citenamefont {Zhang},\ and\ \citenamefont
  {Peiris}}]{ScoX1CrossCorr2015PRD}%
  \BibitemOpen
  \bibfield  {author} {\bibinfo {author} {\bibfnamefont {John~T.}\ \bibnamefont
  {Whelan}}, \bibinfo {author} {\bibfnamefont {Santosh}\ \bibnamefont
  {Sundaresan}}, \bibinfo {author} {\bibfnamefont {Yuanhao}\ \bibnamefont
  {Zhang}}, \ and\ \bibinfo {author} {\bibfnamefont {Prabath}\ \bibnamefont
  {Peiris}},\ }\bibfield  {title} {\enquote {\bibinfo {title} {Model-based
  cross-correlation search for gravitational waves from {S}corpius {X-1}},}\
  }\href {\doibase 10.1103/PhysRevD.91.102005} {\bibfield  {journal} {\bibinfo
  {journal} {{P}hys. {R}ev. {D}}\ }\textbf {\bibinfo {volume} {91}},\ \bibinfo
  {pages} {102005} (\bibinfo {year} {2015})}\BibitemShut {NoStop}%
\bibitem [{\citenamefont {Messenger}\ \emph {et~al.}(2015)\citenamefont
  {Messenger}, \citenamefont {Bulten}, \citenamefont {Crowder}, \citenamefont
  {Dergachev}, \citenamefont {Galloway}, \citenamefont {Goetz}, \citenamefont
  {Jonker}, \citenamefont {Lasky}, \citenamefont {Meadors}, \citenamefont
  {Melatos}, \citenamefont {Premachandra}, \citenamefont {Riles}, \citenamefont
  {Sammut}, \citenamefont {Thrane}, \citenamefont {Whelan},\ and\ \citenamefont
  {Zhang}}]{ScoX1MDC2015PRD}%
  \BibitemOpen
  \bibfield  {author} {\bibinfo {author} {\bibfnamefont {C.}~\bibnamefont
  {Messenger}}, \bibinfo {author} {\bibfnamefont {H.~J.}\ \bibnamefont
  {Bulten}}, \bibinfo {author} {\bibfnamefont {S.~G.}\ \bibnamefont {Crowder}},
  \bibinfo {author} {\bibfnamefont {V.}~\bibnamefont {Dergachev}}, \bibinfo
  {author} {\bibfnamefont {D.~K.}\ \bibnamefont {Galloway}}, \bibinfo {author}
  {\bibfnamefont {E.}~\bibnamefont {Goetz}}, \bibinfo {author} {\bibfnamefont
  {R.~J.~G.}\ \bibnamefont {Jonker}}, \bibinfo {author} {\bibfnamefont {P.~D.}\
  \bibnamefont {Lasky}}, \bibinfo {author} {\bibfnamefont {G.~D.}\ \bibnamefont
  {Meadors}}, \bibinfo {author} {\bibfnamefont {A.}~\bibnamefont {Melatos}},
  \bibinfo {author} {\bibfnamefont {S.}~\bibnamefont {Premachandra}}, \bibinfo
  {author} {\bibfnamefont {K.}~\bibnamefont {Riles}}, \bibinfo {author}
  {\bibfnamefont {L.}~\bibnamefont {Sammut}}, \bibinfo {author} {\bibfnamefont
  {E.~H.}\ \bibnamefont {Thrane}}, \bibinfo {author} {\bibfnamefont {J.~T.}\
  \bibnamefont {Whelan}}, \ and\ \bibinfo {author} {\bibfnamefont
  {Y.}~\bibnamefont {Zhang}},\ }\bibfield  {title} {\enquote {\bibinfo {title}
  {Gravitational waves from {S}corpius {X-1}: A comparison of search methods
  and prospects for detection with advanced detectors},}\ }\href {\doibase
  10.1103/PhysRevD.92.023006} {\bibfield  {journal} {\bibinfo  {journal} {Phys.
  Rev. D}\ }\textbf {\bibinfo {volume} {92}},\ \bibinfo {pages} {023006}
  (\bibinfo {year} {2015})}\BibitemShut {NoStop}%
\bibitem [{\citenamefont {Abbott}\ \emph {et~al.}(2007)\citenamefont {Abbott}
  \emph {et~al.}}]{AbbottScoX12007}%
  \BibitemOpen
  \bibfield  {author} {\bibinfo {author} {\bibfnamefont {B.}~\bibnamefont
  {Abbott}} \emph {et~al.},\ }\bibfield  {title} {\enquote {\bibinfo {title}
  {Searches for periodic gravitational waves from unknown isolated sources and
  {Scorpius X-1}: results from the second {LIGO} science run},}\ }\href@noop {}
  {\bibfield  {journal} {\bibinfo  {journal} {Phys. {R}ev. {D}}\ }\textbf
  {\bibinfo {volume} {76}},\ \bibinfo {pages} {082001} (\bibinfo {year}
  {2007})}\BibitemShut {NoStop}%
\bibitem [{\citenamefont {Abadie}\ \emph {et~al.}(2011)\citenamefont {Abadie}
  \emph {et~al.}}]{AbadieStoch2011}%
  \BibitemOpen
  \bibfield  {author} {\bibinfo {author} {\bibfnamefont {J.}~\bibnamefont
  {Abadie}} \emph {et~al.},\ }\bibfield  {title} {\enquote {\bibinfo {title}
  {Directional limits on persistent gravitational waves using {LIGO S5} science
  data},}\ }\href@noop {} {\bibfield  {journal} {\bibinfo  {journal} {Phys.
  {R}ev. {L}ett.}\ }\textbf {\bibinfo {volume} {107}},\ \bibinfo {pages}
  {271102} (\bibinfo {year} {2011})}\BibitemShut {NoStop}%
\bibitem [{\citenamefont {Aasi}\ \emph
  {et~al.}(2014{\natexlab{a}})\citenamefont {Aasi} \emph
  {et~al.}}]{GoetzTwoSpectResults2014}%
  \BibitemOpen
  \bibfield  {author} {\bibinfo {author} {\bibfnamefont {J.}~\bibnamefont
  {Aasi}} \emph {et~al.},\ }\bibfield  {title} {\enquote {\bibinfo {title}
  {First all-sky search for continuous gravitational waves from unknown sources
  in binary systems},}\ }\href@noop {} {\bibfield  {journal} {\bibinfo
  {journal} {Phys. {R}ev. {D}}\ }\textbf {\bibinfo {volume} {90}},\ \bibinfo
  {pages} {062010} (\bibinfo {year} {2014}{\natexlab{a}})}\BibitemShut
  {NoStop}%
\bibitem [{\citenamefont {Sammut}\ \emph {et~al.}(2014)\citenamefont {Sammut},
  \citenamefont {Messenger}, \citenamefont {Melatos},\ and\ \citenamefont
  {Owen}}]{Sammut2014PRD}%
  \BibitemOpen
  \bibfield  {author} {\bibinfo {author} {\bibfnamefont {L.}~\bibnamefont
  {Sammut}}, \bibinfo {author} {\bibfnamefont {C.}~\bibnamefont {Messenger}},
  \bibinfo {author} {\bibfnamefont {A.}~\bibnamefont {Melatos}}, \ and\
  \bibinfo {author} {\bibfnamefont {{B.J.}}\ \bibnamefont {Owen}},\ }\bibfield
  {title} {\enquote {\bibinfo {title} {Implementation of the
  frequency-modulated sideband search method for gravitational waves from low
  mass x-ray binaries},}\ }\href {\doibase 10.1103/PhysRevD.89.043001}
  {\bibfield  {journal} {\bibinfo  {journal} {{P}hys. {R}ev. {D}}\ }\textbf
  {\bibinfo {volume} {89}},\ \bibinfo {pages} {043001} (\bibinfo {year}
  {2014})}\BibitemShut {NoStop}%
\bibitem [{\citenamefont {Aasi}\ \emph
  {et~al.}(2015{\natexlab{a}})\citenamefont {Aasi} \emph
  {et~al.}}]{Sideband2015}%
  \BibitemOpen
  \bibfield  {author} {\bibinfo {author} {\bibfnamefont {J.}~\bibnamefont
  {Aasi}} \emph {et~al.},\ }\bibfield  {title} {\enquote {\bibinfo {title}
  {Directed search for gravitational waves from {S}corpius {X-1} with initial
  {LIGO} data},}\ }\href {\doibase 10.1103/PhysRevD.91.062008} {\bibfield
  {journal} {\bibinfo  {journal} {{P}hys. {R}ev. {D}}\ }\textbf {\bibinfo
  {volume} {91}},\ \bibinfo {pages} {062008} (\bibinfo {year}
  {2015}{\natexlab{a}})}\BibitemShut {NoStop}%
\bibitem [{\citenamefont {Meadors}\ \emph {et~al.}(2017)\citenamefont
  {Meadors}, \citenamefont {Goetz}, \citenamefont {Riles}, \citenamefont
  {Creighton},\ and\ \citenamefont {Robinet}}]{MeadorsS6ScoX1PRD2017}%
  \BibitemOpen
  \bibfield  {author} {\bibinfo {author} {\bibfnamefont {G.D.}\ \bibnamefont
  {Meadors}}, \bibinfo {author} {\bibfnamefont {E.}~\bibnamefont {Goetz}},
  \bibinfo {author} {\bibfnamefont {K.}~\bibnamefont {Riles}}, \bibinfo
  {author} {\bibfnamefont {T.}~\bibnamefont {Creighton}}, \ and\ \bibinfo
  {author} {\bibfnamefont {F.}~\bibnamefont {Robinet}},\ }\bibfield  {title}
  {\enquote {\bibinfo {title} {Searches for continuous gravitational waves from
  {S}corpius {X-1} and {XTE} {J1705-305} in {LIGO}'s sixth science run},}\
  }\href {\doibase 10.1103/PhysRevD.95.042005} {\bibfield  {journal} {\bibinfo
  {journal} {{P}hys {R}ev {D}}\ }\textbf {\bibinfo {volume} {95}},\ \bibinfo
  {pages} {042005} (\bibinfo {year} {2017})}\BibitemShut {NoStop}%
\bibitem [{\citenamefont {Abbott}\ \emph
  {et~al.}(2017{\natexlab{b}})\citenamefont {Abbott} \emph
  {et~al.}}]{O1Radiometer2017}%
  \BibitemOpen
  \bibfield  {author} {\bibinfo {author} {\bibfnamefont {B.P.}\ \bibnamefont
  {Abbott}} \emph {et~al.},\ }\bibfield  {title} {\enquote {\bibinfo {title}
  {Directional limits on persistent gravitational waves from {A}dvanced
  {LIGO}'s first observing run},}\ }\href {\doibase
  10.1103/PhysRevLett.118.121102} {\bibfield  {journal} {\bibinfo  {journal}
  {{P}hys {R}ev {Lett}}\ }\textbf {\bibinfo {volume} {118}},\ \bibinfo {pages}
  {121102} (\bibinfo {year} {2017}{\natexlab{b}})}\BibitemShut {NoStop}%
\bibitem [{\citenamefont {Abbott}\ \emph
  {et~al.}(2017{\natexlab{c}})\citenamefont {Abbott} \emph
  {et~al.}}]{O1Sideband2017}%
  \BibitemOpen
  \bibfield  {author} {\bibinfo {author} {\bibfnamefont {B.P.}\ \bibnamefont
  {Abbott}} \emph {et~al.},\ }\bibfield  {title} {\enquote {\bibinfo {title}
  {Search for gravitational waves from {S}corpius {X-1} in the first {A}dvanced
  {LIGO} observing run with a hidden {M}arkov model},}\ }\href {\doibase
  10.1103/PhysRevD.95.122003} {\bibfield  {journal} {\bibinfo  {journal}
  {{P}hys {R}ev {D}}\ }\textbf {\bibinfo {volume} {95}},\ \bibinfo {pages}
  {122003} (\bibinfo {year} {2017}{\natexlab{c}})}\BibitemShut {NoStop}%
\bibitem [{\citenamefont {Chakrabarty}\ \emph {et~al.}(2003)\citenamefont
  {Chakrabarty} \emph {et~al.}}]{Chakrabarty2003}%
  \BibitemOpen
  \bibfield  {author} {\bibinfo {author} {\bibfnamefont {D.}~\bibnamefont
  {Chakrabarty}} \emph {et~al.},\ }\bibfield  {title} {\enquote {\bibinfo
  {title} {Nuclear-powered millisecond pulsars and the maximum spin frequency
  of neutron stars},}\ }\href@noop {} {\bibfield  {journal} {\bibinfo
  {journal} {Nature}\ }\textbf {\bibinfo {volume} {424}},\ \bibinfo {pages}
  {42} (\bibinfo {year} {2003})}\BibitemShut {NoStop}%
\bibitem [{\citenamefont {Watts}\ \emph {et~al.}(2008)\citenamefont {Watts},
  \citenamefont {Krishnan}, \citenamefont {Bildsten},\ and\ \citenamefont
  {Schutz}}]{Watts2008}%
  \BibitemOpen
  \bibfield  {author} {\bibinfo {author} {\bibfnamefont {A.L.}\ \bibnamefont
  {Watts}}, \bibinfo {author} {\bibfnamefont {B.}~\bibnamefont {Krishnan}},
  \bibinfo {author} {\bibfnamefont {L.}~\bibnamefont {Bildsten}}, \ and\
  \bibinfo {author} {\bibfnamefont {B.F.}\ \bibnamefont {Schutz}},\ }\bibfield
  {title} {\enquote {\bibinfo {title} {Detecting gravitational wave emission
  from the known accreting neutron stars},}\ }\href@noop {} {\bibfield
  {journal} {\bibinfo  {journal} {MNRAS}\ }\textbf {\bibinfo {volume} {389}},\
  \bibinfo {pages} {839} (\bibinfo {year} {2008})}\BibitemShut {NoStop}%
\bibitem [{\citenamefont {Aasi}\ \emph
  {et~al.}(2015{\natexlab{b}})\citenamefont {Aasi} \emph
  {et~al.}}]{ALIGOStandardRef}%
  \BibitemOpen
  \bibfield  {author} {\bibinfo {author} {\bibfnamefont {J.}~\bibnamefont
  {Aasi}} \emph {et~al.} (\bibinfo {collaboration} {LIGO Scientific}),\
  }\bibfield  {title} {\enquote {\bibinfo {title} {{Advanced LIGO}},}\ }\href
  {\doibase 10.1088/0264-9381/32/7/074001} {\bibfield  {journal} {\bibinfo
  {journal} {Class. Quant. Grav.}\ }\textbf {\bibinfo {volume} {32}},\ \bibinfo
  {pages} {074001} (\bibinfo {year} {2015}{\natexlab{b}})},\ \Eprint
  {http://arxiv.org/abs/1411.4547} {arXiv:1411.4547 [gr-qc]} \BibitemShut
  {NoStop}%
\bibitem [{\citenamefont {Acernese}\ \emph {et~al.}(2015)\citenamefont
  {Acernese} \emph {et~al.}}]{AVirgoStandardRef}%
  \BibitemOpen
  \bibfield  {author} {\bibinfo {author} {\bibfnamefont {F.}~\bibnamefont
  {Acernese}} \emph {et~al.} (\bibinfo {collaboration} {VIRGO}),\ }\bibfield
  {title} {\enquote {\bibinfo {title} {{Advanced Virgo: a second-generation
  interferometric gravitational wave detector}},}\ }\href {\doibase
  10.1088/0264-9381/32/2/024001} {\bibfield  {journal} {\bibinfo  {journal}
  {Class. Quant. Grav.}\ }\textbf {\bibinfo {volume} {32}},\ \bibinfo {pages}
  {024001} (\bibinfo {year} {2015})},\ \Eprint {http://arxiv.org/abs/1408.3978}
  {arXiv:1408.3978 [gr-qc]} \BibitemShut {NoStop}%
\bibitem [{\citenamefont {Aso}\ \emph {et~al.}(2013)\citenamefont {Aso},
  \citenamefont {Michimura}, \citenamefont {Somiya}, \citenamefont {Ando},
  \citenamefont {Miyakawa}, \citenamefont {Sekiguchi}, \citenamefont
  {Tatsumi},\ and\ \citenamefont {Yamamoto}}]{KAGRAStandardRef}%
  \BibitemOpen
  \bibfield  {author} {\bibinfo {author} {\bibfnamefont {Yoichi}\ \bibnamefont
  {Aso}}, \bibinfo {author} {\bibfnamefont {Yuta}\ \bibnamefont {Michimura}},
  \bibinfo {author} {\bibfnamefont {Kentaro}\ \bibnamefont {Somiya}}, \bibinfo
  {author} {\bibfnamefont {Masaki}\ \bibnamefont {Ando}}, \bibinfo {author}
  {\bibfnamefont {Osamu}\ \bibnamefont {Miyakawa}}, \bibinfo {author}
  {\bibfnamefont {Takanori}\ \bibnamefont {Sekiguchi}}, \bibinfo {author}
  {\bibfnamefont {Daisuke}\ \bibnamefont {Tatsumi}}, \ and\ \bibinfo {author}
  {\bibfnamefont {Hiroaki}\ \bibnamefont {Yamamoto}} (\bibinfo {collaboration}
  {The KAGRA Collaboration}),\ }\bibfield  {title} {\enquote {\bibinfo {title}
  {Interferometer design of the {KAGRA} gravitational wave detector},}\ }\href
  {\doibase 10.1103/PhysRevD.88.043007} {\bibfield  {journal} {\bibinfo
  {journal} {Phys. Rev. D}\ }\textbf {\bibinfo {volume} {88}},\ \bibinfo
  {pages} {043007} (\bibinfo {year} {2013})}\BibitemShut {NoStop}%
\bibitem [{\citenamefont {Behnke}\ \emph {et~al.}(2015)\citenamefont {Behnke},
  \citenamefont {Papa},\ and\ \citenamefont {Prix}}]{BehnkeGalacticCenter2015}%
  \BibitemOpen
  \bibfield  {author} {\bibinfo {author} {\bibfnamefont {B.}~\bibnamefont
  {Behnke}}, \bibinfo {author} {\bibfnamefont {M.A.}\ \bibnamefont {Papa}}, \
  and\ \bibinfo {author} {\bibfnamefont {R.}~\bibnamefont {Prix}},\ }\bibfield
  {title} {\enquote {\bibinfo {title} {Postprocessing methods used in the
  search for continuous gravitational-wave signals from the {G}alactic
  {C}enter},}\ }\href@noop {} {\bibfield  {journal} {\bibinfo  {journal}
  {{PRD}}\ }\textbf {\bibinfo {volume} {91}},\ \bibinfo {pages} {064007}
  (\bibinfo {year} {2015})}\BibitemShut {NoStop}%
\bibitem [{\citenamefont {Leaci}\ and\ \citenamefont
  {Prix}(2015)}]{LeaciPrixDirectedFStatPRD}%
  \BibitemOpen
  \bibfield  {author} {\bibinfo {author} {\bibfnamefont {Paola}\ \bibnamefont
  {Leaci}}\ and\ \bibinfo {author} {\bibfnamefont {Reinhard}\ \bibnamefont
  {Prix}},\ }\bibfield  {title} {\enquote {\bibinfo {title} {Directed searches
  for continuous gravitational waves from binary systems: Parameter-space
  metrics and optimal {S}corpius {X-1} sensitivity},}\ }\href {\doibase
  10.1103/PhysRevD.91.102003} {\bibfield  {journal} {\bibinfo  {journal} {Phys.
  Rev. D}\ }\textbf {\bibinfo {volume} {91}},\ \bibinfo {pages} {102003}
  (\bibinfo {year} {2015})}\BibitemShut {NoStop}%
\bibitem [{\citenamefont {Riles}(2013)}]{Riles2013}%
  \BibitemOpen
  \bibfield  {author} {\bibinfo {author} {\bibfnamefont {K.}~\bibnamefont
  {Riles}},\ }\bibfield  {title} {\enquote {\bibinfo {title} {Gravitational
  waves: sources, detectors and searches},}\ }\href@noop {} {\bibfield
  {journal} {\bibinfo  {journal} {Prog. in {P}article \& {N}ucl. {P}hys.}\
  }\textbf {\bibinfo {volume} {68}},\ \bibinfo {pages} {1} (\bibinfo {year}
  {2013})}\BibitemShut {NoStop}%
\bibitem [{\citenamefont {Krishnan}\ \emph {et~al.}(2004)\citenamefont
  {Krishnan}, \citenamefont {Sintes}, \citenamefont {Papa}, \citenamefont
  {Schutz}, \citenamefont {Frasca},\ and\ \citenamefont
  {Palomba}}]{HoughTransformKrishnan2004}%
  \BibitemOpen
  \bibfield  {author} {\bibinfo {author} {\bibfnamefont {Badri}\ \bibnamefont
  {Krishnan}}, \bibinfo {author} {\bibfnamefont {Alicia~M.}\ \bibnamefont
  {Sintes}}, \bibinfo {author} {\bibfnamefont {Maria~Alessandra}\ \bibnamefont
  {Papa}}, \bibinfo {author} {\bibfnamefont {Bernard~F.}\ \bibnamefont
  {Schutz}}, \bibinfo {author} {\bibfnamefont {Sergio}\ \bibnamefont {Frasca}},
  \ and\ \bibinfo {author} {\bibfnamefont {Cristiano}\ \bibnamefont
  {Palomba}},\ }\bibfield  {title} {\enquote {\bibinfo {title} {Hough transform
  search for continuous gravitational waves},}\ }\href {\doibase
  10.1103/PhysRevD.70.082001} {\bibfield  {journal} {\bibinfo  {journal} {Phys.
  {R}ev. {D}}\ }\textbf {\bibinfo {volume} {70}},\ \bibinfo {pages} {082001}
  (\bibinfo {year} {2004})}\BibitemShut {NoStop}%
\bibitem [{\citenamefont {Abbott}(2008)}]{LSCPulsarS4}%
  \BibitemOpen
  \bibfield  {author} {\bibinfo {author} {\bibfnamefont {B.}~\bibnamefont
  {Abbott}},\ }\href@noop {} {\bibfield  {journal} {\bibinfo  {journal} {Phys.
  {R}ev. {D}}\ }\textbf {\bibinfo {volume} {77}},\ \bibinfo {pages} {022001}
  (\bibinfo {year} {2008})}\BibitemShut {NoStop}%
\bibitem [{\citenamefont {Abbott}\ \emph {et~al.}(2009)\citenamefont {Abbott}
  \emph {et~al.}}]{LSCPowerFlux2009}%
  \BibitemOpen
  \bibfield  {author} {\bibinfo {author} {\bibfnamefont {B.}~\bibnamefont
  {Abbott}} \emph {et~al.},\ }\bibfield  {title} {\enquote {\bibinfo {title}
  {All-sky {LIGO} search for periodic gravitational waves in the early
  fifth-science-run data},}\ }\href {\doibase 10.1103/PhysRevLett.102.111102}
  {\bibfield  {journal} {\bibinfo  {journal} {Phys. {R}ev. {L}ett}\ }\textbf
  {\bibinfo {volume} {102}},\ \bibinfo {pages} {111102} (\bibinfo {year}
  {2009})}\BibitemShut {NoStop}%
\bibitem [{\citenamefont {Dergachev}(2010)}]{PowerFluxMethod2010}%
  \BibitemOpen
  \bibfield  {author} {\bibinfo {author} {\bibfnamefont {V.}~\bibnamefont
  {Dergachev}},\ }\bibfield  {title} {\enquote {\bibinfo {title} {On blind
  searches for noise dominated signals: a loosely coherent approach},}\ }\href
  {\doibase 10.1088/0264-9381/27/20/205017} {\bibfield  {journal} {\bibinfo
  {journal} {Class. {Q}uant. {G}rav.}\ }\textbf {\bibinfo {volume} {27}},\
  \bibinfo {pages} {205017} (\bibinfo {year} {2010})}\BibitemShut {NoStop}%
\bibitem [{\citenamefont {Abadie}\ \emph {et~al.}(2012)\citenamefont {Abadie}
  \emph {et~al.}}]{PowerFluxAllSky2012}%
  \BibitemOpen
  \bibfield  {author} {\bibinfo {author} {\bibfnamefont {J.}~\bibnamefont
  {Abadie}} \emph {et~al.},\ }\bibfield  {title} {\enquote {\bibinfo {title}
  {All-sky search for periodic gravitational waves in the full {S5} {LIGO}
  data},}\ }\href {\doibase 10.1103/PhysRevD.85.022001} {\bibfield  {journal}
  {\bibinfo  {journal} {Phys. {R}ev. {D}}\ }\textbf {\bibinfo {volume} {85}},\
  \bibinfo {pages} {022001} (\bibinfo {year} {2012})}\BibitemShut {NoStop}%
\bibitem [{\citenamefont {Dupuis}\ and\ \citenamefont
  {Woan}(2005)}]{DupuisWoan2005}%
  \BibitemOpen
  \bibfield  {author} {\bibinfo {author} {\bibfnamefont {R.J.}\ \bibnamefont
  {Dupuis}}\ and\ \bibinfo {author} {\bibfnamefont {G.}~\bibnamefont {Woan}},\
  }\bibfield  {title} {\enquote {\bibinfo {title} {Bayesian estimation of
  pulsar parameters from gravitational wave data},}\ }\href {\doibase
  10.1103/PhysRevD.72.102002} {\bibfield  {journal} {\bibinfo  {journal} {Phys.
  {R}ev. {D}}\ }\textbf {\bibinfo {volume} {72}},\ \bibinfo {pages} {102002}
  (\bibinfo {year} {2005})}\BibitemShut {NoStop}%
\bibitem [{\citenamefont {Aasi}\ \emph
  {et~al.}(2014{\natexlab{b}})\citenamefont {Aasi} \emph
  {et~al.}}]{AasiPulsarInitialResults2014}%
  \BibitemOpen
  \bibfield  {author} {\bibinfo {author} {\bibfnamefont {J.}~\bibnamefont
  {Aasi}} \emph {et~al.},\ }\bibfield  {title} {\enquote {\bibinfo {title}
  {Gravitational-waves from known pulsars: results from the initial detector
  era},}\ }\href@noop {} {\bibfield  {journal} {\bibinfo  {journal} {Astrophys.
  {J}}\ }\textbf {\bibinfo {volume} {785}},\ \bibinfo {pages} {119} (\bibinfo
  {year} {2014}{\natexlab{b}})}\BibitemShut {NoStop}%
\bibitem [{\citenamefont {Messenger}\ and\ \citenamefont
  {Woan}(2007)}]{Messenger2007CQG}%
  \BibitemOpen
  \bibfield  {author} {\bibinfo {author} {\bibfnamefont {C.}~\bibnamefont
  {Messenger}}\ and\ \bibinfo {author} {\bibfnamefont {G.}~\bibnamefont
  {Woan}},\ }\bibfield  {title} {\enquote {\bibinfo {title} {A fast search
  strategy for gravitational waves from low-mass x-ray binaries},}\ }\href
  {\doibase 10.1088/0264-9381/24/19/S10} {\bibfield  {journal} {\bibinfo
  {journal} {{C}lassical and {Q}uantum {G}ravity}\ }\textbf {\bibinfo {volume}
  {24}},\ \bibinfo {pages} {S469} (\bibinfo {year} {2007})}\BibitemShut
  {NoStop}%
\bibitem [{\citenamefont {Suvorova}\ \emph {et~al.}(2016)\citenamefont
  {Suvorova}, \citenamefont {Sun}, \citenamefont {Melatos}, \citenamefont
  {Moran},\ and\ \citenamefont {Evans}}]{SidebandMarkovModelSuvorova2016}%
  \BibitemOpen
  \bibfield  {author} {\bibinfo {author} {\bibfnamefont {S.}~\bibnamefont
  {Suvorova}}, \bibinfo {author} {\bibfnamefont {L.}~\bibnamefont {Sun}},
  \bibinfo {author} {\bibfnamefont {A.}~\bibnamefont {Melatos}}, \bibinfo
  {author} {\bibfnamefont {W.}~\bibnamefont {Moran}}, \ and\ \bibinfo {author}
  {\bibfnamefont {R.J.}\ \bibnamefont {Evans}},\ }\bibfield  {title} {\enquote
  {\bibinfo {title} {Hidden {M}arkov model tracking of continuous gravitational
  waves from a neutron star with wandering spin},}\ }\href {\doibase
  10.1103/PhysRevD.93.123009} {\bibfield  {journal} {\bibinfo  {journal}
  {{P}hys {R}ev {D}}\ }\textbf {\bibinfo {volume} {93}},\ \bibinfo {pages}
  {123009} (\bibinfo {year} {2016})}\BibitemShut {NoStop}%
\bibitem [{\citenamefont {Goetz}\ and\ \citenamefont
  {Riles}(2011)}]{GoetzTwoSpectMethods2011}%
  \BibitemOpen
  \bibfield  {author} {\bibinfo {author} {\bibfnamefont {E.}~\bibnamefont
  {Goetz}}\ and\ \bibinfo {author} {\bibfnamefont {K.}~\bibnamefont {Riles}},\
  }\bibfield  {title} {\enquote {\bibinfo {title} {An all-sky search algorithm
  for continuous gravitational waves from spinning neutron stars in binary
  systems},}\ }\href@noop {} {\bibfield  {journal} {\bibinfo  {journal}
  {{C}lass. {Q}uant. {G}rav.}\ }\textbf {\bibinfo {volume} {28}},\ \bibinfo
  {pages} {215006} (\bibinfo {year} {2011})}\BibitemShut {NoStop}%
\bibitem [{\citenamefont {Meadors}\ \emph {et~al.}(2016)\citenamefont
  {Meadors}, \citenamefont {Goetz},\ and\ \citenamefont
  {Riles}}]{MeadorsDirectedMethods2016}%
  \BibitemOpen
  \bibfield  {author} {\bibinfo {author} {\bibfnamefont {G.D.}\ \bibnamefont
  {Meadors}}, \bibinfo {author} {\bibfnamefont {E.}~\bibnamefont {Goetz}}, \
  and\ \bibinfo {author} {\bibfnamefont {K.}~\bibnamefont {Riles}},\ }\bibfield
   {title} {\enquote {\bibinfo {title} {Tuning into {S}corpius {X-1}: adapting
  a continuous gravitational-wave search for a known binary system},}\ }\href
  {\doibase {10.1088/0264-9381/33/10/105017}} {\bibfield  {journal} {\bibinfo
  {journal} {Class. {Q}uant. {G}rav.}\ }\textbf {\bibinfo {volume} {33}},\
  \bibinfo {pages} {105017} (\bibinfo {year} {2016})}\BibitemShut {NoStop}%
\bibitem [{\citenamefont {Ballmer}(2006)}]{Ballmer2006CQG}%
  \BibitemOpen
  \bibfield  {author} {\bibinfo {author} {\bibfnamefont {Stefan~W}\
  \bibnamefont {Ballmer}},\ }\bibfield  {title} {\enquote {\bibinfo {title} {A
  radiometer for stochastic gravitational waves},}\ }\href {\doibase
  10.1088/0264-9381/23/8/S23} {\bibfield  {journal} {\bibinfo  {journal}
  {Classical and Quantum Gravity}\ }\textbf {\bibinfo {volume} {23}},\ \bibinfo
  {pages} {S179} (\bibinfo {year} {2006})}\BibitemShut {NoStop}%
\bibitem [{\citenamefont {{van~der~Putten}}\ \emph {et~al.}(2010)\citenamefont
  {{van~der~Putten}}, \citenamefont {{Bulten}}, \citenamefont
  {{van~den~Brand}},\ and\ \citenamefont {{Holtrop}}}]{2010JPhCS.228a2005V}%
  \BibitemOpen
  \bibfield  {author} {\bibinfo {author} {\bibfnamefont {S.}~\bibnamefont
  {{van~der~Putten}}}, \bibinfo {author} {\bibfnamefont {H.~J.}\ \bibnamefont
  {{Bulten}}}, \bibinfo {author} {\bibfnamefont {J.~F.~J.}\ \bibnamefont
  {{van~den~Brand}}}, \ and\ \bibinfo {author} {\bibfnamefont {M.}~\bibnamefont
  {{Holtrop}}},\ }\bibfield  {title} {\enquote {\bibinfo {title} {{Searching
  for gravitational waves from pulsars in binary systems: An all-sky
  search}},}\ }\href {\doibase 10.1088/1742-6596/228/1/012005} {\bibfield
  {journal} {\bibinfo  {journal} {Journal of Physics Conference Series}\
  }\textbf {\bibinfo {volume} {228}},\ \bibinfo {eid} {012005} (\bibinfo {year}
  {2010})}\BibitemShut {NoStop}%
\bibitem [{\citenamefont {{Skrutskie}}\ \emph {et~al.}(2006)\citenamefont
  {{Skrutskie}} \emph {et~al.}}]{2mass06}%
  \BibitemOpen
  \bibfield  {author} {\bibinfo {author} {\bibfnamefont {M.~F.}\ \bibnamefont
  {{Skrutskie}}} \emph {et~al.},\ }\bibfield  {title} {\enquote {\bibinfo
  {title} {{The Two Micron All Sky Survey (2MASS)}},}\ }\href {\doibase
  10.1086/498708} {\bibfield  {journal} {\bibinfo  {journal} {The Astronomical
  Journal}\ }\textbf {\bibinfo {volume} {131}},\ \bibinfo {pages} {1163--1183}
  (\bibinfo {year} {2006})}\BibitemShut {NoStop}%
\bibitem [{\citenamefont {Bradshaw}\ \emph {et~al.}(1999)\citenamefont
  {Bradshaw}, \citenamefont {Fomalont},\ and\ \citenamefont
  {Geldzahler}}]{Bradshaw1999}%
  \BibitemOpen
  \bibfield  {author} {\bibinfo {author} {\bibfnamefont {C.F.}\ \bibnamefont
  {Bradshaw}}, \bibinfo {author} {\bibfnamefont {E.B.}\ \bibnamefont
  {Fomalont}}, \ and\ \bibinfo {author} {\bibfnamefont {B.J.}\ \bibnamefont
  {Geldzahler}},\ }\bibfield  {title} {\enquote {\bibinfo {title}
  {High-resolution parallax measurements of {S}corpius {X-1}},}\ }\href@noop {}
  {\bibfield  {journal} {\bibinfo  {journal} {ApJ}\ }\textbf {\bibinfo {volume}
  {512}},\ \bibinfo {pages} {L121} (\bibinfo {year} {1999})}\BibitemShut
  {NoStop}%
\bibitem [{\citenamefont {{Galloway}}\ \emph {et~al.}(2014)\citenamefont
  {{Galloway}}, \citenamefont {{Premachandra}}, \citenamefont {{Steeghs}},
  \citenamefont {{Marsh}}, \citenamefont {{Casares}},\ and\ \citenamefont
  {{Cornelisse}}}]{Galloway2014}%
  \BibitemOpen
  \bibfield  {author} {\bibinfo {author} {\bibfnamefont {D.~K.}\ \bibnamefont
  {{Galloway}}}, \bibinfo {author} {\bibfnamefont {S.}~\bibnamefont
  {{Premachandra}}}, \bibinfo {author} {\bibfnamefont {D.}~\bibnamefont
  {{Steeghs}}}, \bibinfo {author} {\bibfnamefont {T.}~\bibnamefont {{Marsh}}},
  \bibinfo {author} {\bibfnamefont {J.}~\bibnamefont {{Casares}}}, \ and\
  \bibinfo {author} {\bibfnamefont {R.}~\bibnamefont {{Cornelisse}}},\
  }\bibfield  {title} {\enquote {\bibinfo {title} {{Precision Ephemerides for
  Gravitational-wave Searches. I. Sco X-1}},}\ }\href {\doibase
  10.1088/0004-637X/781/1/14} {\bibfield  {journal} {\bibinfo  {journal} {Ap
  J}\ }\textbf {\bibinfo {volume} {781}},\ \bibinfo {eid} {14} (\bibinfo {year}
  {2014})},\ \Eprint {http://arxiv.org/abs/1311.6246} {arXiv:1311.6246
  [astro-ph.HE]} \BibitemShut {NoStop}%
\bibitem [{\citenamefont {Wang}\ \emph {et~al.}(2016)\citenamefont {Wang},
  \citenamefont {Steeghs},\ and\ \citenamefont
  {Galloway}}]{WangSteeghsGalloway2016}%
  \BibitemOpen
  \bibfield  {author} {\bibinfo {author} {\bibfnamefont {L.}~\bibnamefont
  {Wang}}, \bibinfo {author} {\bibfnamefont {D.}~\bibnamefont {Steeghs}}, \
  and\ \bibinfo {author} {\bibfnamefont {D.}~\bibnamefont {Galloway}},\
  }\href@noop {} {\enquote {\bibinfo {title} {Sco {X}-1 parameters summary},}\
  }\bibinfo {howpublished} {private communication} (\bibinfo {year}
  {2016})\BibitemShut {NoStop}%
\bibitem [{\citenamefont {{Steeghs}}\ and\ \citenamefont
  {{Casares}}(2002)}]{2002ApJ...568..273S}%
  \BibitemOpen
  \bibfield  {author} {\bibinfo {author} {\bibfnamefont {D.}~\bibnamefont
  {{Steeghs}}}\ and\ \bibinfo {author} {\bibfnamefont {J.}~\bibnamefont
  {{Casares}}},\ }\bibfield  {title} {\enquote {\bibinfo {title} {{The Mass
  Donor of Scorpius X-1 Revealed}},}\ }\href {\doibase 10.1086/339224}
  {\bibfield  {journal} {\bibinfo  {journal} {Astrophys. J.}\ }\textbf
  {\bibinfo {volume} {568}},\ \bibinfo {pages} {273--278} (\bibinfo {year}
  {2002})},\ \Eprint {http://arxiv.org/abs/astro-ph/0107343} {astro-ph/0107343}
  \BibitemShut {NoStop}%
\bibitem [{\citenamefont {Allen}\ and\ \citenamefont
  {Romano}(1999)}]{Allen1999}%
  \BibitemOpen
  \bibfield  {author} {\bibinfo {author} {\bibfnamefont {B.}~\bibnamefont
  {Allen}}\ and\ \bibinfo {author} {\bibfnamefont {{J.D.}}\ \bibnamefont
  {Romano}},\ }\bibfield  {title} {\enquote {\bibinfo {title} {Detecting a
  stochastic background of gravitational radiation: signal processing
  strategies and sensitivities},}\ }\href@noop {} {\bibfield  {journal}
  {\bibinfo  {journal} {Phys. Rev. D}\ }\textbf {\bibinfo {volume} {59}},\
  \bibinfo {pages} {102001} (\bibinfo {year} {1999})}\BibitemShut {NoStop}%
\bibitem [{\citenamefont {Cutler}\ and\ \citenamefont
  {Schutz}(2005)}]{CutlerMulti2005}%
  \BibitemOpen
  \bibfield  {author} {\bibinfo {author} {\bibfnamefont {C.}~\bibnamefont
  {Cutler}}\ and\ \bibinfo {author} {\bibfnamefont {B.F.}\ \bibnamefont
  {Schutz}},\ }\bibfield  {title} {\enquote {\bibinfo {title} {Generalized
  {F}-statistic: {M}ultiple detectors and multiple gravitational wave
  pulsars},}\ }\href@noop {} {\bibfield  {journal} {\bibinfo  {journal} {{P}hys
  {R}ev {D}}\ }\textbf {\bibinfo {volume} {72}},\ \bibinfo {pages} {063006}
  (\bibinfo {year} {2005})}\BibitemShut {NoStop}%
\bibitem [{\citenamefont {Prix}\ and\ \citenamefont
  {Krishnan}(2009)}]{BStatPrix2009}%
  \BibitemOpen
  \bibfield  {author} {\bibinfo {author} {\bibfnamefont {R.}~\bibnamefont
  {Prix}}\ and\ \bibinfo {author} {\bibfnamefont {B.}~\bibnamefont
  {Krishnan}},\ }\bibfield  {title} {\enquote {\bibinfo {title} {Targeted
  search for continuous gravitational waves: {B}ayesian versus
  maximum-likelihood statistics},}\ }\href {\doibase
  10.1088/0264-9381/26/20/204013} {\bibfield  {journal} {\bibinfo  {journal}
  {{C}lass {Q}uant {G}rav}\ }\textbf {\bibinfo {volume} {26}},\ \bibinfo
  {pages} {204013} (\bibinfo {year} {2009})}\BibitemShut {NoStop}%
\bibitem [{\citenamefont {Mukherjee}\ \emph {et~al.}(2017-10-17)\citenamefont
  {Mukherjee}, \citenamefont {Messenger},\ and\ \citenamefont
  {Riles}}]{MukherjeeSpinWandering2016}%
  \BibitemOpen
  \bibfield  {author} {\bibinfo {author} {\bibfnamefont {A.}~\bibnamefont
  {Mukherjee}}, \bibinfo {author} {\bibfnamefont {C.}~\bibnamefont
  {Messenger}}, \ and\ \bibinfo {author} {\bibfnamefont {K.}~\bibnamefont
  {Riles}},\ }\bibfield  {title} {\enquote {\bibinfo {title} {Accretion-induced
  spin-wandering effects on the neutron star in {S}corpius {X}-1: implications
  for continuous gravitational wave searches},}\ }\href@noop {} {\  (\bibinfo
  {year} {2017-10-17})},\ \Eprint {http://arxiv.org/abs/1710.06185} {1710.06185
  [gr-qc]} \BibitemShut {NoStop}%
\bibitem [{\citenamefont {Prix}(2007)}]{PrixMultiMetric2007}%
  \BibitemOpen
  \bibfield  {author} {\bibinfo {author} {\bibfnamefont {R.}~\bibnamefont
  {Prix}},\ }\bibfield  {title} {\enquote {\bibinfo {title} {Search for
  continuous gravitational waves: metric of the multidetector {F}-statistic},}\
  }\href@noop {} {\bibfield  {journal} {\bibinfo  {journal} {{P}hys {R}ev {D}}\
  }\textbf {\bibinfo {volume} {75}},\ \bibinfo {pages} {023004} (\bibinfo
  {year} {2007})}\BibitemShut {NoStop}%
\bibitem [{\citenamefont {Whelan}\ \emph {et~al.}(2014)\citenamefont {Whelan},
  \citenamefont {Prix}, \citenamefont {Cutler},\ and\ \citenamefont
  {Willis}}]{WhelanNewAmplitude2014CQG}%
  \BibitemOpen
  \bibfield  {author} {\bibinfo {author} {\bibfnamefont {J.T.}\ \bibnamefont
  {Whelan}}, \bibinfo {author} {\bibfnamefont {R.}~\bibnamefont {Prix}},
  \bibinfo {author} {\bibfnamefont {C.J.}\ \bibnamefont {Cutler}}, \ and\
  \bibinfo {author} {\bibfnamefont {J.L}\ \bibnamefont {Willis}},\ }\bibfield
  {title} {\enquote {\bibinfo {title} {New coordinates for the amplitude
  parameter space of continuous gravitational waves},}\ }\href {\doibase
  10.1088/0264-9381/31/6/065002} {\bibfield  {journal} {\bibinfo  {journal}
  {{C}lass {Q}uant {G}rav}\ }\textbf {\bibinfo {volume} {31}},\ \bibinfo
  {pages} {065002} (\bibinfo {year} {2014})}\BibitemShut {NoStop}%
\bibitem [{\citenamefont {Blandford}\ and\ \citenamefont
  {Teukolsky}(1976)}]{BlandfordBinary1976}%
  \BibitemOpen
  \bibfield  {author} {\bibinfo {author} {\bibfnamefont {R.}~\bibnamefont
  {Blandford}}\ and\ \bibinfo {author} {\bibfnamefont {S.}~\bibnamefont
  {Teukolsky}},\ }\bibfield  {title} {\enquote {\bibinfo {title} {Arrival-time
  analysis for a pulsar in a binary system},}\ }\href@noop {} {\bibfield
  {journal} {\bibinfo  {journal} {Astrophys {J}}\ }\textbf {\bibinfo {volume}
  {205}},\ \bibinfo {pages} {580} (\bibinfo {year} {1976})}\BibitemShut
  {NoStop}%
\bibitem [{\citenamefont {Allen}\ and\ \citenamefont
  {Mendell}(2004)}]{AllenMendellSFT2004}%
  \BibitemOpen
  \bibfield  {author} {\bibinfo {author} {\bibfnamefont {B.}~\bibnamefont
  {Allen}}\ and\ \bibinfo {author} {\bibfnamefont {G.}~\bibnamefont
  {Mendell}},\ }\bibfield  {title} {\enquote {\bibinfo {title} {{SFT} {D}ata
  {F}ormat {V}ersion~2 {S}pecification},}\ }\href@noop {} {\bibfield  {journal}
  {\bibinfo  {journal} {LIGO DCC}\ }\textbf {\bibinfo {volume} {T040164}}
  (\bibinfo {year} {2004})},\ \bibinfo {note}
  {https://dcc.ligo.org/LIGO-T040164/public}\BibitemShut {NoStop}%
\bibitem [{\citenamefont {Allen}\ \emph {et~al.}(2002)\citenamefont {Allen},
  \citenamefont {Papa},\ and\ \citenamefont {Schutz}}]{Allen2002}%
  \BibitemOpen
  \bibfield  {author} {\bibinfo {author} {\bibfnamefont {B}~\bibnamefont
  {Allen}}, \bibinfo {author} {\bibfnamefont {M.A.}\ \bibnamefont {Papa}}, \
  and\ \bibinfo {author} {\bibfnamefont {B.F.}\ \bibnamefont {Schutz}},\
  }\bibfield  {title} {\enquote {\bibinfo {title} {Optimal strategies for
  sinusoidal signal detection},}\ }\href {\doibase 10.1103/PhysRevD.66.102003}
  {\bibfield  {journal} {\bibinfo  {journal} {{P}hys {R}ev {D}}\ }\textbf
  {\bibinfo {volume} {66}},\ \bibinfo {pages} {102003} (\bibinfo {year}
  {2002})}\BibitemShut {NoStop}%
\bibitem [{\citenamefont {Schutz}(2017)}]{SchutzCBCResamp2017}%
  \BibitemOpen
  \bibfield  {author} {\bibinfo {author} {\bibfnamefont {B.}~\bibnamefont
  {Schutz}},\ }\href@noop {} {}\bibinfo {howpublished} {private communication}
  (\bibinfo {year} {2017})\BibitemShut {NoStop}%
\bibitem [{\citenamefont {{The LIGO Scientific
  Collaboration}}()}]{LALAppsRepo}%
  \BibitemOpen
  \bibfield  {author} {\bibinfo {author} {\bibnamefont {{The LIGO Scientific
  Collaboration}}},\ }\href@noop {} {\enquote {\bibinfo {title} {{LALApps}
  repository},}\ }\bibinfo {howpublished} {Web:
  http://www.lsc-group.phys.uwm.edu/daswg/}\BibitemShut {NoStop}%
\bibitem [{\citenamefont {Prix}(2017)}]{PrixTimingModel2017}%
  \BibitemOpen
  \bibfield  {author} {\bibinfo {author} {\bibfnamefont {R.}~\bibnamefont
  {Prix}},\ }\bibfield  {title} {\enquote {\bibinfo {title} {Characterizing
  timing and memory-requirements of the {F}-statistic implementations in
  {LALSuite}},}\ }\href@noop {} {\  (\bibinfo {year} {2017})},\ \bibinfo {note}
  {https://dcc.ligo.org/LIGO-T1600531/public}\BibitemShut {NoStop}%
\bibitem [{\citenamefont {Prix}(2011)}]{PrixFStatModel2011}%
  \BibitemOpen
  \bibfield  {author} {\bibinfo {author} {\bibfnamefont {R.}~\bibnamefont
  {Prix}},\ }\bibfield  {title} {\enquote {\bibinfo {title} {The {F}-statistic
  and its implementation in {ComputeFStatistic\_v2}},}\ }\href@noop {}
  {\bibfield  {journal} {\bibinfo  {journal} {LIGO DCC}\ }\textbf {\bibinfo
  {volume} {T0900149-v3}} (\bibinfo {year} {2011})},\ \bibinfo {note}
  {https://dcc.ligo.org/LIGO-T0900149-v3/public}\BibitemShut {NoStop}%
\bibitem [{\citenamefont {Ming}\ \emph {et~al.}(2016)\citenamefont {Ming},
  \citenamefont {Krishnan}, \citenamefont {Papa}, \citenamefont {Aulbert},\
  and\ \citenamefont {Fehrmann}}]{MingSetup2015}%
  \BibitemOpen
  \bibfield  {author} {\bibinfo {author} {\bibfnamefont {J.}~\bibnamefont
  {Ming}}, \bibinfo {author} {\bibfnamefont {B.}~\bibnamefont {Krishnan}},
  \bibinfo {author} {\bibfnamefont {M.A.}\ \bibnamefont {Papa}}, \bibinfo
  {author} {\bibfnamefont {C.}~\bibnamefont {Aulbert}}, \ and\ \bibinfo
  {author} {\bibfnamefont {H.}~\bibnamefont {Fehrmann}},\ }\bibfield  {title}
  {\enquote {\bibinfo {title} {Optimal directed searches for continuous
  gravitational waves},}\ }\href {\doibase 10.1103/PhysRevD.93.064011}
  {\bibfield  {journal} {\bibinfo  {journal} {{P}hys {R}ev {D}}\ }\textbf
  {\bibinfo {volume} {93}},\ \bibinfo {pages} {064011} (\bibinfo {year}
  {2016})}\BibitemShut {NoStop}%
\bibitem [{\citenamefont {Suvorova}\ \emph {et~al.}(2017)\citenamefont
  {Suvorova}, \citenamefont {Clearwater}, \citenamefont {Melatos},
  \citenamefont {Sun}, \citenamefont {Moran},\ and\ \citenamefont
  {Evans}}]{SidebandBessel2017}%
  \BibitemOpen
  \bibfield  {author} {\bibinfo {author} {\bibfnamefont {S.}~\bibnamefont
  {Suvorova}}, \bibinfo {author} {\bibfnamefont {P.}~\bibnamefont
  {Clearwater}}, \bibinfo {author} {\bibfnamefont {A.}~\bibnamefont {Melatos}},
  \bibinfo {author} {\bibfnamefont {L.}~\bibnamefont {Sun}}, \bibinfo {author}
  {\bibfnamefont {W.}~\bibnamefont {Moran}}, \ and\ \bibinfo {author}
  {\bibfnamefont {R.J.}\ \bibnamefont {Evans}},\ }\bibfield  {title} {\enquote
  {\bibinfo {title} {Hidden {M}arkov model tracking of continuous gravitational
  waves from a binary neutron star with wandering spin. {II.} {B}inary orbital
  phase tracking},}\ }\href {\doibase 10.1103/PhysRevD.96.102006} {\bibfield
  {journal} {\bibinfo  {journal} {{P}hys {R}ev {D}}\ }\textbf {\bibinfo
  {volume} {96}},\ \bibinfo {pages} {102006} (\bibinfo {year}
  {2017})}\BibitemShut {NoStop}%
\bibitem [{\citenamefont {Cutler}\ \emph {et~al.}(2005)\citenamefont {Cutler},
  \citenamefont {Gholami},\ and\ \citenamefont {Krishnan}}]{CutlerSemi2005}%
  \BibitemOpen
  \bibfield  {author} {\bibinfo {author} {\bibfnamefont {C.}~\bibnamefont
  {Cutler}}, \bibinfo {author} {\bibfnamefont {I.}~\bibnamefont {Gholami}}, \
  and\ \bibinfo {author} {\bibfnamefont {B.}~\bibnamefont {Krishnan}},\
  }\bibfield  {title} {\enquote {\bibinfo {title} {Improved stack-slide
  searches for gravitational-wave pulsars},}\ }\href@noop {} {\bibfield
  {journal} {\bibinfo  {journal} {{P}hys {R}ev {D}}\ }\textbf {\bibinfo
  {volume} {72}},\ \bibinfo {pages} {042004} (\bibinfo {year}
  {2005})}\BibitemShut {NoStop}%
\bibitem [{\citenamefont {Prix}\ and\ \citenamefont
  {Shaltev}(2012)}]{PrixShaltev2012}%
  \BibitemOpen
  \bibfield  {author} {\bibinfo {author} {\bibfnamefont {R.}~\bibnamefont
  {Prix}}\ and\ \bibinfo {author} {\bibfnamefont {M.}~\bibnamefont {Shaltev}},\
  }\bibfield  {title} {\enquote {\bibinfo {title} {Search for continuous
  gravitational waves: optimal {S}tack{S}lide method at fixed computing
  cost},}\ }\href@noop {} {\bibfield  {journal} {\bibinfo  {journal} {{P}hys
  {R}ev {D}}\ }\textbf {\bibinfo {volume} {85}},\ \bibinfo {pages} {084010}
  (\bibinfo {year} {2012})}\BibitemShut {NoStop}%
\bibitem [{\citenamefont {Aasi}\ \emph {et~al.}(2013)\citenamefont {Aasi} \emph
  {et~al.}}]{EinsteinHomeS52013}%
  \BibitemOpen
  \bibfield  {author} {\bibinfo {author} {\bibfnamefont {J.}~\bibnamefont
  {Aasi}} \emph {et~al.},\ }\bibfield  {title} {\enquote {\bibinfo {title}
  {Einstein@home all-sky search for periodic gravitational waves in {LIGO S5}
  data},}\ }\href@noop {} {\bibfield  {journal} {\bibinfo  {journal} {{P}hys
  {R}ev {D}}\ }\textbf {\bibinfo {volume} {87}},\ \bibinfo {pages} {042001}
  (\bibinfo {year} {2013})}\BibitemShut {NoStop}%
\bibitem [{\citenamefont {Thrane}\ \emph {et~al.}(2009)\citenamefont {Thrane},
  \citenamefont {Ballmer}, \citenamefont {Romano}, \citenamefont {Mitra},
  \citenamefont {Talukder}, \citenamefont {Bose},\ and\ \citenamefont
  {Mandic}}]{ThraneStochastic2009}%
  \BibitemOpen
  \bibfield  {author} {\bibinfo {author} {\bibfnamefont {E.}~\bibnamefont
  {Thrane}}, \bibinfo {author} {\bibfnamefont {S.}~\bibnamefont {Ballmer}},
  \bibinfo {author} {\bibfnamefont {J.D.}\ \bibnamefont {Romano}}, \bibinfo
  {author} {\bibfnamefont {S.}~\bibnamefont {Mitra}}, \bibinfo {author}
  {\bibfnamefont {D.}~\bibnamefont {Talukder}}, \bibinfo {author}
  {\bibfnamefont {S.}~\bibnamefont {Bose}}, \ and\ \bibinfo {author}
  {\bibfnamefont {V.}~\bibnamefont {Mandic}},\ }\bibfield  {title} {\enquote
  {\bibinfo {title} {Probing the anisotropies of a stochastic
  gravitational-wave background using a network of ground-based laser
  interferomters},}\ }\href@noop {} {\bibfield  {journal} {\bibinfo  {journal}
  {{P}hys {R}ev {D}}\ }\textbf {\bibinfo {volume} {80}},\ \bibinfo {pages}
  {122002} (\bibinfo {year} {2009})}\BibitemShut {NoStop}%
\bibitem [{\citenamefont {Christensen}(1992)}]{ChristensenStochastic1992}%
  \BibitemOpen
  \bibfield  {author} {\bibinfo {author} {\bibfnamefont {N.}~\bibnamefont
  {Christensen}},\ }\bibfield  {title} {\enquote {\bibinfo {title} {Measuring
  the stochastic gravitational-radiation background with laser-interferometric
  antennas},}\ }\href@noop {} {\bibfield  {journal} {\bibinfo  {journal}
  {{P}hys {R}ev {D}}\ }\textbf {\bibinfo {volume} {46}},\ \bibinfo {pages}
  {5250} (\bibinfo {year} {1992})}\BibitemShut {NoStop}%
\bibitem [{\citenamefont {Flanagan}(1993)}]{FlanaganStochastic1993}%
  \BibitemOpen
  \bibfield  {author} {\bibinfo {author} {\bibfnamefont {E.E.}\ \bibnamefont
  {Flanagan}},\ }\bibfield  {title} {\enquote {\bibinfo {title} {Sensitivity of
  the {L}aser {I}nterferomter {G}ravitational {W}ave {O}bservatory to a
  stochastic background, and its dependence on the detector orientations},}\
  }\href@noop {} {\bibfield  {journal} {\bibinfo  {journal} {{P}hys {R}ev {D}}\
  }\textbf {\bibinfo {volume} {48}},\ \bibinfo {pages} {2389} (\bibinfo {year}
  {1993})}\BibitemShut {NoStop}%
\bibitem [{\citenamefont {Mitra}\ \emph {et~al.}(2008)\citenamefont {Mitra},
  \citenamefont {Dhurandhar}, \citenamefont {Souradeep}, \citenamefont
  {Lazzarini}, \citenamefont {Mandic}, \citenamefont {Bose},\ and\
  \citenamefont {Ballmer}}]{MitraRadiometer2008}%
  \BibitemOpen
  \bibfield  {author} {\bibinfo {author} {\bibfnamefont {S.}~\bibnamefont
  {Mitra}}, \bibinfo {author} {\bibfnamefont {S.}~\bibnamefont {Dhurandhar}},
  \bibinfo {author} {\bibfnamefont {T.}~\bibnamefont {Souradeep}}, \bibinfo
  {author} {\bibfnamefont {A.}~\bibnamefont {Lazzarini}}, \bibinfo {author}
  {\bibfnamefont {V.}~\bibnamefont {Mandic}}, \bibinfo {author} {\bibfnamefont
  {S.}~\bibnamefont {Bose}}, \ and\ \bibinfo {author} {\bibfnamefont
  {S.}~\bibnamefont {Ballmer}},\ }\bibfield  {title} {\enquote {\bibinfo
  {title} {Gravitational wave radiometry: mapping a stochastic gravitational
  wave background},}\ }\href@noop {} {\bibfield  {journal} {\bibinfo  {journal}
  {{P}hys {R}ev {D}}\ }\textbf {\bibinfo {volume} {77}},\ \bibinfo {pages}
  {042002} (\bibinfo {year} {2008})}\BibitemShut {NoStop}%
\bibitem [{\citenamefont {Keitel}\ \emph {et~al.}(2014)\citenamefont {Keitel},
  \citenamefont {Prix}, \citenamefont {Papa}, \citenamefont {Leaci},\ and\
  \citenamefont {Siddiqi}}]{KeitelRobust2014}%
  \BibitemOpen
  \bibfield  {author} {\bibinfo {author} {\bibfnamefont {D.}~\bibnamefont
  {Keitel}}, \bibinfo {author} {\bibfnamefont {R.}~\bibnamefont {Prix}},
  \bibinfo {author} {\bibfnamefont {M.A.}\ \bibnamefont {Papa}}, \bibinfo
  {author} {\bibfnamefont {P.}~\bibnamefont {Leaci}}, \ and\ \bibinfo {author}
  {\bibfnamefont {M.}~\bibnamefont {Siddiqi}},\ }\bibfield  {title} {\enquote
  {\bibinfo {title} {Search for continuous gravitational waves: improving
  robustness versus instrumental artifacts},}\ }\href {\doibase
  10.1103/PhysRevD.89.064023} {\bibfield  {journal} {\bibinfo  {journal}
  {{P}hys {R}ev {D}}\ }\textbf {\bibinfo {volume} {89}},\ \bibinfo {pages}
  {064023} (\bibinfo {year} {2014})}\BibitemShut {NoStop}%
\bibitem [{\citenamefont {Wette}(2016)}]{WettePRD2016}%
  \BibitemOpen
  \bibfield  {author} {\bibinfo {author} {\bibfnamefont {K.}~\bibnamefont
  {Wette}},\ }\bibfield  {title} {\enquote {\bibinfo {title} {Empirically
  extending the range of validity of parameter-space metrics for all-sky
  searches for gravitational-wave pulsars},}\ }\href@noop {} {\bibfield
  {journal} {\bibinfo  {journal} {{P}hys {R}ev {D}}\ }\textbf {\bibinfo
  {volume} {94}},\ \bibinfo {pages} {122002} (\bibinfo {year}
  {2016})}\BibitemShut {NoStop}%
\end{thebibliography}%

\end{document}